%% file: SDUpaper.tex
\documentclass[11pt]{article}

\input{preamble}

\begin{document}
	\title{The Infinite Horizon Investment-Consumption Problem for Epstein--Zin Stochastic Differential Utility\footnote{We would like to thank Frank Seifried for bringing Epstein--Zin stochastic differential utility to our attention and for discussing some of its subtleties with us. We are also grateful to Miryana Grigorova for a very helpful discussion on the topic of optional strong supermartingales, which inspired our proof that the paths of generalised utility processes are c\`adl\`ag.}}
	
	\author{Martin Herdegen, David Hobson, Joseph Jerome\thanks{All authors: University of Warwick, Department of Statistics, Coventry, CV4 7AL, UK; \{m.herdegen, d.hobson, j.jerome\}@warwick.ac.uk}}
	\date{\today}
	
	\maketitle
	
	\begin{abstract}
		In this article we consider the optimal investment-consumption problem for an agent with preferences governed by Epstein--Zin stochastic differential utility who invests in a constant-parameter Black\textendash Scholes\textendash Merton market.
		
		The paper has three main goals: first, to provide a detailed introduction to infinite-horizon Epstein--Zin stochastic differential utility, including a discussion of which parameter combinations lead to a well-formulated problem; second, to prove existence and uniqueness of infinite horizon Epstein--Zin stochastic differential utility under a restriction on the parameters governing the agent's risk aversion and temporal variance aversion; and third, to provide a verification argument for the candidate optimal solution to the investment-consumption problem among all admissible consumption streams.
		
		To achieve these goals, we introduce a slightly different formulation of Epstein--Zin stochastic differential utility to that which is traditionally used in the literature. This formulation highlights the necessity and appropriateness of certain restrictions on the parameters governing the stochastic differential utility function.
	\end{abstract}



	
\section{Introduction}
	The goal of this paper is to undertake a rigorous study of a Merton-style, infinite horizon, investment-consumption problem in the setting of stochastic differential utility (SDU). In particular the aim is to derive the optimal investment and consumption strategy, the value function and optimal utility process, and to decide when the problem is well-posed, for an agent investing in a Black--Scholes--Merton style frictionless stochastic market (consisting of a risk-free asset with constant interest rate, and a single risky asset whose price process follows a constant parameter exponential Brownian motion) for an agent whose preferences are given by Epstein--Zin stochastic differential utility (EZ-SDU). In the sense that SDU is a generalisation of additive utility, EZ-SDU preferences are a natural generalisation of constant-relative-risk-aversion (CRRA) preferences.
	
	The contributions of the paper come in two main directions. The first contribution is partly foundational and partly didactic. Within the economics literature, SDU (introduced by Duffie and Epstein~\cite{duffie1992stochastic} as the continuous-time analogue of recursive utility, (Epstein and Zin~\cite{epstein1989substitution}), and further developed by Duffie and Lions~\cite{duffie1992pde} and Schroder and Skiadas~\cite{schroder1999optimal}) is viewed as an extension to classical additive utilities, and recognised as having the potential to explain several of the inconsistencies between the predictions of the Merton model and agent behaviour (for example, the equity premium puzzle, Mehra and Prescott~\cite{mehra1985equity}). However, with several honourable exceptions (including Kraft and Seifried~\cite{kraft2014stochastic}, Seiferling and Seifried~\cite{seiferling2016epstein}, Xing~\cite{xing2017consumption}, Matoussi and Xing~\cite{matoussi2018convex} and Melnyk et al~\cite{melnyk2020lifetime}), SDU has not been widely studied in the mathematical finance literature. Given the deep connections with many areas of modern probability theory (for example backward stochastic differential equations (BSDEs)) this is in some ways surprising, but given the technical challenges involved it is also understandable. We introduce SDU and EZ-SDU for infinite horizon problems and give a clear interpretation of all the parameters, with a focus on the feasible ranges for these parameters. The fact that we concentrate on the infinite horizon brings several issues into focus. Over the infinite horizon it is not possible to work backwards from the terminal horizon and it is necessary to introduce some form of transversality condition as an alternative. Moreover, integrability (and uniform integrability) become much more significant challenges.
	
	The conventional wisdom (see for example Duffie and Epstein~\cite{duffie1992stochastic} and Melnyk et al~\cite{melnyk2020lifetime}) is that the best technical solution to these challenges is to replace the infinite horizon problem with a family of finite horizon problems (but note that this is not the way in which the candidate solution is found). We take a different approach. Key to the definition of SDU is an aggregator, and we introduce a slightly different aggregator to that which is traditionally used in the literature, the key point being that our aggregator takes only one sign. Where there exist utility processes associated with both our aggregator and the classical aggregator, then the utility processes agree, but crucially any utility process associated to the traditional aggregator is also a utility process associated with our modified aggregator, whereas the converse is not true. Moreover, when specialised to the case of additive utility, our aggregator corresponds to the classical formulation of the Merton problem, whereas the traditional aggregator has a non-standard specification in this context.
	
	Our reformulation of the problem brings significant new insights concerning the set of feasible parameters for the problem with Epstein--Zin preferences. In particular we conclude that the co-efficient of relative risk aversion (RRA) and the co-efficient of elasticity of intertemporal consumption (EIC)---see  Section~\ref{sec:EZ stochastic differential utility} for a definition of this latter quantity---must lie on the same side of unity for the problem to make sense, at least for infinite horizon problems. (In the classical Merton problem for power law utility the RRA and EIC are necessarily equal.) This seems to be a new finding. We argue that the putative solutions which have been found previously in the literature (in the case when the co-efficients of RRA and EIC are on opposite sides of one) correspond to a bubble-like behaviour, where the value associated with a consumption stream comes not from the utility of consumption in the short and medium term, but rather from a perceived and unrealisable value in the distant future.
	
	The second aim of the paper is to give a rigorous treatment of the Merton problem for Epstein--Zin stochastic differential utility. Our first results are existence results which show there exists a well-defined utility process for a large class of consumption streams. Then, under an important restriction on the parameters of the EZ-SDU (namely that the co-efficient of RRA is closer to unity than the co-efficient of EIC), we show how to extend the existence result further to give a well-defined (though not necessarily finite) utility process for {\em any} consumption stream. Again, key to our proofs is the fact that under our formulation the aggregator takes one sign.
	
	Then we turn to uniqueness. Under the same restriction on parameter values, we show that for EZ-SDU preferences the utility process associated to a consumption stream is unique.\footnote{When this condition fails, and despite claims to the contrary in the literature, there are simple examples showing non-uniqueness.} The main idea is to use a comparison theorem for (sub- and super-) solutions to a representation of the utility process.
	
	Finally, we turn to the identification of the optimal investment and consumption strategy, and the optimal utility process.  The candidate optimal strategy and candidate optimal utility process are known (see \cite{schroder1999optimal, melnyk2020lifetime, kraft2017optimal}), and the main techniques behind a verification argument are also well established in the literature. But, what distinguishes our results is the fact that we optimise over {\em all} admissible consumption streams, i.e., all consumption streams which can be financed from an initial wealth $x$. Typically in the extant literature optimisation only takes place over a sub-family of consumption streams for which the consumption stream and utility process posses certain regularity and integrability conditions. Further, since there are very few existence results in the literature, often the only strategies for which it can be verified that the utility process indeed satisfies the required regularity conditions are the constant proportional investment-consumption strategies. Since we optimise over \emph{all} admissible consumption streams, this is a significant advance.
	
	The paper comes in two parts. The first part focuses on characterising the set of parameter combinations for which the problem is well-founded. The second part takes a subset of these parameter combinations and discusses existence and uniqueness in this setting and gives a rigorous derivation of the value function and of the optimal investment-consumption strategy.
	
	Part I is structured as follows. In Sections \ref{sec:CRRA utility intro} and
	\ref{sec:general SDU}, we review the classical Merton-style investment-consumption problem for additive utility, and then we introduce the corresponding problem for SDU. In Section~\ref{sec:EZ stochastic differential utility}, we introduce Epstein--Zin SDU and carefully explain how the various parameters should be interpreted, and which parameter combinations lead to a well-founded problem. In Section~\ref{sec:BSM}, we embed EZ-SDU within a constant parameter financial market and derive the candidate value function, utility process and optimal strategy. In Sections~\ref{sec:difference form of EZ SDU} and \ref{sec:alternative}, we compare our formulation with the conventional formulation which has been used heretofore in the literature. We believe that our formulation has significant advantages; first in that it contributes to the understanding of when the problem is ill-founded, and second it makes possible in Part II an optimisation over {\em all} attainable consumption streams, and not just a restricted subclass of consumption streams as has been considered so far.
	
	Part II is concerned with a rigorous derivation of the value function and the optimal value process investment-consumption strategy. We mainly work in the case $\theta \in (0,1)$---here $\theta$ is defined in Section~\ref{sec:EZ stochastic differential utility} and depends on the coefficients of relative risk aversion and elasticity of intertemporal complementarity. Importantly, when $\theta \in (0,1)$, if the utility process exists then it is unique. (We defer to a subsequent paper the very interesting and very relevant case of $\theta>1$, in which uniqueness fails.) In Section~\ref{sec:existence}, we prove existence of EZ-SDU for a wide class of consumption streams, including all constant proportional consumption streams for which the problem is well-posed, and any strategies which are `close' to constant proportional streams, in a sense to be made precise. Still, this is not all consumption streams, so in Sections~\ref{sec:sub and supersolutions} and \ref{sec:rem bounds},
	we show how the utility process for an arbitrary attainable consumption stream can be obtained by approximation and taking limits. Finally, in Section~\ref{sec:verification of optimal strategy}, we prove optimality of the candidate optimal strategy (Theorem~\ref{thm:verification}) first derived in Section~\ref{ssec:candidate optimal utility process}, where the optimisation is taken over all attainable consumption streams and not just those satisfying regularity and integrability conditions. Key results along the way include a comparison result (Theorem~\ref{thm:comparison}), existence and uniqueness results (Theorem~\ref{thm: existence of a solution, first result}, Theorem~\ref{thm:existence of a F^epsilon solution, bounded case}) and an approximation result (Theorem~\ref{thm:approximation by evaluable processes is unique}).

	\part{Epstein--Zin stochastic differential utility: an introduction}\label{part:intro}

	\section{Constant relative risk aversion utility}\label{sec:CRRA utility intro}
	
	In this article our focus is on infinite-horizon, optimal investment-consumption problems for agents whose preferences are given under stochastic differential utility. Although the infinite-horizon problem brings potentially different (and greater) technical challenges when compared with the finite horizon problem, it can lead to a time-homogeneous problem and therefore to a dimension reduction and the greater prospect of closed-form solutions.
	
	Throughout we work on a filtered probability space $(\Omega, {\mathcal F},  (\cF_t)_{t \geq 0}, \P)$ satisfying the usual conditions and where $\cF_0$ is $\P$-trivial. Let $\sP$ be the set of progressively measurable processes, and let $\sP_+$ and $\sP_{++}$ be the restrictions of $\sP$ to processes that take non-negative and positive values, respectively. Moreover, denote by $\sS$ the set of all semimartingales. We identify processes in $\sP$ or $\sS$ that agree up to indistinguishability.
	
	Before we introduce the notion of stochastic differential utility, we first recall the definition of expected utility over the infinite horizon.
	We say $U : \R_+ \times \R_+ \mapsto \R$ is a \emph{utility function} if $U$ is increasing and concave in its second argument and $C$ is a consumption stream if $C \in \sP_+$. Then the utility associated to a consumption stream
	is given by
	$       J_U(C) = \EX{\int_0^\infty U(t,C_t) \dd t}$.
	Define the \textit{value process} or, as it is called in the SDU literature, the {\em utility process} $V=V^C \in \sS$ associated to the consumption stream $C$ by
	\begin{equation}
		\label{eq:UVadditive}
		V_t = V^C_t = \cEX[t]{\int_t^\infty U(s,C_s) \dd s}.
	\end{equation}
	Then, $J_U(C) = V^C_0$. The goal is to maximise $J_U(C)$ over an appropriate space of consumption streams.
	A specific example of a utility function is the discounted constant relative risk aversion (CRRA) utility function $U(t,c) = e^{-\delta t} \frac{c^{1-R}}{1-R}$. Under discounted CRRA utility, the utility process associated to $C$ is given by
	\begin{equation}\label{eq:CRRA utility}
		V_t = \cEX[t]{\int_t^\infty e^{-\delta s} \frac{C_s^{1-R}}{1-R} \dd s}.
	\end{equation}
	It is very well known that under CRRA preferences the parameter $R$ controls the agent's appetite for risk.
	In particular, since $R$ is a measure of the concavity of the utility function $U(t,c) = e^{-\delta t}\frac{c^{1-R}}{1-R}$, and more precisely $R= - c \frac{U'(t,c)}{U''(t,c)}$, $R$ captures the agent's aversion to variation of consumption over $\omega \in \Omega$. It is also known, though perhaps less well known, that the parameter $R$ also captures the agent's aversion to variation of consumption over time. (We will justify and explain this fact when we study EZ-SDU in Section~\ref{sec:EZ stochastic differential utility}.)
	
	There is no economic or mathematical justification (beyond mathematical tractability) for restricting attention to preferences in which the same parameter governs preferences over both fluctuations of consumption across sample paths and fluctuations of consumption across time. One of the motivations behind the introduction of SDU is to allow a disentanglement of preferences over these two types of fluctuations of consumption.
	
	\section{Stochastic differential utility}\label{sec:general SDU}
	Stochastic differential utility (SDU) is a generalisation of time-additive discounted expected utility and is designed to allow a separation of risk preferences from time preferences. The goal in this section is to explain how this statement should be interpreted.
	
	Under discounted expected utility the value or utility of a consumption stream is given by $J_U(C) = \EX{\int_0^\infty  U(t,C_t) \dd t}$ and the value or utility process is given by $V_t = \E[\int_t^\infty U(s,C_s) |{\mathcal F}_t]$. Under SDU the function $U=U(s,C_s)$ is generalised to become an \emph{aggregator} $g=g(s,C_s,V_s)$, and the stochastic differential utility process $V^C=(V^C_t)_{t\geq0}$ associated to a consumption stream $C$ solves
	(compare with \eqref{eq:UVadditive})
	\begin{equation}\label{eq:stochastic differential utility aggregator g}
		V^C_t = \cEX[t]{\int_t^\infty g(s,C_s,V^C_s)ds}.
	\end{equation}
	This creates a feedback effect in which the value at time $t$ may depend in a non-linear way on the value at future times. This feature leads to a separation of the two phenomena mentioned in the previous section: risk aversion and temporal variance aversion.
	
	Note that if $g$ takes positive and negative values, the conditional expectation on the right hand side of \eqref{eq:stochastic differential utility aggregator g} may not be well-defined. With this in mind, we introduce the following definitions.
	\begin{defn}\label{defn:integrable set I(g,C)}
		An \textit{aggregator} is a function $g:[0,\infty)\times \R_+ \times \R \to \R$.
		For $C\in\sP_+$, define $\II(g,C)\coloneqq\left\{V\in\sP:~ \E\int_0^\infty \left| g(s,C_s,V_s)\right| \dd s < \infty\right\}$.
		Further, let $\UU\II(g,C)$ be the set of elements of $\II(g,C)$ which are uniformly integrable. Then $V \in \II(g,C)$ is a \textit{utility process} associated to the pair $(g,C)$ if it has c\`adl\`ag paths and satisfies \eqref{eq:stochastic differential utility aggregator g} for all $t \in [0,\infty)$.
	\end{defn}
	
	\begin{rem}\label{rem:utility process UI}
		All utility processes are necessarily semimartingales and uniformly integrable. Indeed, let $M=(M_t)_{t \geq 0}$ be the   (c\`adl\`ag)  martingale      given by $M_t = \cEX[t]{\int_0^\infty g(s,C_s,V^C_s) ds}$ and $A=(A_t)_{t \geq 0}$ the continuous adapted process       given by $A_t = \int_0^t g(s,C_s,V^C_s) ds$. Then $V^C = M - A \in \sS$. Moreover, let $\tilde M=(\tilde M_t)_{t \geq 0}$ be the uniformly integrable martingale
		given by $\tilde M_t = \cEX[t]{\int_0^\infty |g(s,C_s,V^C_s)| ds}$. Then $V^C \in\UU\II(g,C)$ since
		\[|V^C_t| \leq \cEX[t]{\int_t^\infty |g(s,C_s,V^C_s)| ds}\leq \cEX[t]{\int_0^\infty |g(s,C_s,V^C_s)| ds} = \tilde M_t, \quad t\geq 0.\]
	\end{rem}
	
	\begin{defn}\label{def:evaluable}
		$C$ is \textit{$g$-evaluable} if there exists a utility process $V\in \II(g,C)$ associated to the pair $(g,C)$. The set of \textit{$g$-evaluable} consumption streams $C$ is denoted by $\sE(g)$.
		
		Furthermore, if the utility process is unique (up to indistinguishability), then $C$ is \textit{$g$-uniquely evaluable}. The set of \textit{$g$-uniquely evaluable} $C$ is denoted by $\sE_u(g)$.
	\end{defn}
	Throughout the first part of this paper (with a few exceptions where we explictly state otherwise), we will only consider uniquely evaluable consumption streams. Provided that $C$ is uniquely evaluable, we may therefore define the \textit{stochastic differential utility} of a consumption stream $C$ and aggregator $g$ by $J_g(C) \coloneqq V^C_0$ where $V^C$ satisfies \eqref{eq:stochastic differential utility aggregator g}.
	
	The restriction to evaluable or uniquely evaluable consumption streams is a very real restriction. For some parameter combinations for EZ-SDU there are consumption streams that are either not evaluable or not uniquely evaluable.
	
	\section{Epstein--Zin stochastic differential utility}\label{sec:EZ stochastic differential utility}
	
	The goals of this section are: to introduce Epstein--Zin stochastic differential utility, which is a generalisation of the discounted CRRA utility that was introduced in Section \ref{sec:CRRA utility intro}; to define the associated aggregator; to examine some of properties of EZ-SDU; and to justify any restrictions on coefficients that must be imposed to make EZ-SDU well-founded. We will see in Section \ref{ssec:EZ SDU risk aversion and EIC example} that EZ-SDU allows a disentanglement of risk preferences from temporal variance preferences.
	
	The Epstein--Zin aggregator corresponding to the vector of parameters $(b,\delta,R,S)$ is a function $g_{EZ}:\R_+ \times \R_+ \times \VV \to \VV$, given by
	\begin{equation}\label{eq:Epstein--Zin aggregator R and S}
		g_{EZ}(t,c,v) \coloneqq b e^{-\delta t}\frac{c^{1-S}}{{1-S}}\left( (1-R)v\right)^\frac{S-R}{1-R}.
	\end{equation}
	Here $\VV = (1-R)\R_+$ is the domain of the Epstein--Zin utility process and both $R$ and $S$ lie in $(0,1) \cup (1,\infty)$.
	It is convenient to introduce the parameters $\theta\coloneqq\frac{1-R}{1-S}$ and $\rho=\frac{S-R}{1-R} = \frac{\theta-1}{\theta}$, so that \eqref{eq:Epstein--Zin aggregator R and S} becomes
	\begin{equation}\label{eq:Epstein--Zin aggregator}
		g_{EZ}(t,c,v) = b e^{-\delta t}\frac{c^{1-S}}{{1-S}}\left( (1-R)v\right)^\rho.
	\end{equation}
	Note that when $S=R$ the aggregator reduces to the discounted CRRA utility function. This case corresponds to $\theta=1$ and $\rho=0$.
	
	\begin{rem}
		The expression in \eqref{eq:Epstein--Zin aggregator} is a reformulation of the classical Epstein--Zin stochastic differential utility. Other authors use the \textit{difference} form aggregator $g^\Delta_{EZ}$  given by
		\begin{equation}\label{eq:Epstein--Zin aggregator difference form}
			{g^\Delta_{EZ}}(c,v) \coloneqq b \frac{c^{1-S}}{1-S}((1-R)v)^\rho - \delta \theta v.
		\end{equation}
		When we want to emphasise the difference between the two formulations we will call \eqref{eq:Epstein--Zin aggregator} the {\em discounted} form of EZ-SDU. As might be expected there is a very close relationship between solutions of the two different forms, and we will discuss this further in Section~\ref{sec:difference form of EZ SDU}. Note immediately however, that the discounted form is easily recognised as the natural generalisation of CRRA utility as given in \eqref{eq:CRRA utility}. Indeed, when $R=S$ we recover \eqref{eq:CRRA utility} from \eqref{eq:Epstein--Zin aggregator} instantly.
	\end{rem}
	
	Let $g_{EZ}$ be the aggregator in \eqref{eq:Epstein--Zin aggregator}. We begin by trying to give interpretations of the various parameters and to show that (despite appearances) $R$ captures the agent's risk aversion whereas $S$ captures agent's elasticity of intertemporal complementarity, or temporal variance aversion. In addition, $\delta$ represent the agent's subjective discount rate, and $b$ is a scaling parameter which has no effect on the agent's preferences (as long as it is positive) - see Remark \ref{rem:b unimportant}. We have included $b$ to facilitate comparison with other forms of Epstein--Zin SDU used in the literature, but it may be set to $1$ without loss of generality (alternatively, sometimes it is set equal to $\delta$).
	
	\begin{sass}[Rational Parameter Assumption]\label{sass:rational} We assume $b>0$, $\delta \in \RR$ and $R \neq S \in \R_+ \setminus \{1\}$.\end{sass}
	
	The case $S=R$ corresponds to CRRA utility. We exclude the case $R=S$ as it has been extensively studied and is well understood.
	
	In addition to excluding $R=S$ we also exclude $R=1$ and $S=1$. Just as power law utility becomes logarithmic utility when $R=S = 1$, EZ-SDU also changes form. The parameter combination when $S=1$ is considered by Chacko and Viceira \cite{chacko2005dynamic}. (It is less clear how to extend EZ-SDU to the case $R=1$.) Rather than study these limiting cases we focus on the case $R \neq 1 \neq S$, where the issues are already substantial.
	
	Positivity of $b$ corresponds to monotone preferences which are increasing in consumption. We will show in Section~\ref{ssec:EZ SDU risk aversion and EIC example} via a pair of examples that
	the condition $R>0$ corresponds to the agent being risk averse (rather than risk seeking) to variance of consumption over $\omega$, and the condition $S>0$ corresponds to the agent being averse to variance (rather than variance seeking) in consumption over time.
	The parameter $\delta$ is left unrestricted. Whilst it is natural based on its interpretation as a discount factor to expect $\delta$ to be positive, when EZ-SDU is associated with a financial market model a deterministic change of consumption units leads to a change in the value of $\delta$ and potentially to a change in sign, see Section~\ref{ssec:change of numeraire}. Since typically the choice of accounting units is arbitrary there is no economic or mathematical reason to require or expect that $\delta \geq 0$.
	
	If $g_{EZ}$ is the Epstein--Zin aggregator given in \eqref{eq:Epstein--Zin aggregator} then the utility process $V^C = V = (V_t)_{t \geq 0}$ associated to consumption $C$ and aggregator $g_{EZ}$ solves
	\begin{equation}\label{eq:Epstein--Zin SDU}
		V_t = \E\left[\left.\int_t^\infty b e^{-\delta s}\frac{C_s^{1-S}}{{1-S}}\left( (1-R)V_s\right)^\rho \dd s\right| \cF_t \right].
	\end{equation}
	
	\begin{rem}\label{rem:b unimportant}
		The parameter $b$ has no effect on preferences, provided it is positive. To see this, suppose that $V$ is a solution to \eqref{eq:Epstein--Zin SDU} with $b=1$. For arbitrary $d>0$ it follows that $d^\theta V = (d^\theta V_t)_{t\geq0}$ is a solution to \eqref{eq:Epstein--Zin SDU} with $b=d$. Since preferences remain unchanged by a multiplicative scaling of the utility function, it does not matter which value of $b$ we choose.
	\end{rem}
	
	\subsection{Risk aversion and temporal variance aversion}\label{ssec:EZ SDU risk aversion and EIC example}
	
	Consider a deterministic consumption stream $c = (c(t))_{t\geq0}$. Then, $V^c = V = (V(t))_{t\geq0}$ can be found by solving the ordinary differential equation
	\begin{equation}
		\frac{\dd V(t)}{\dd t} =  - b e^{-\delta t}\frac{c(t)^{1-S}}{1-S} ((1-R) V(t))^\rho ,
	\end{equation}
	subject to $ \lim_{t\to\infty} V(t) = 0$. Making the change of variables to $W(t) = (1-R)V(t)$ and dividing through by ${W(t)^\rho}$, we find (recall $\theta = \frac{1-R}{1-S} = \frac{1}{1-\rho}$)
	\begin{equation} \label{eq:deterministic ODE for V}
		\frac{1}{W(t)^\rho}\frac{\dd W(t)}{\dd t} = - b e^{-\delta t} \theta c(t)^{1-S} , \hspace{10mm} \qquad \lim_{t\to\infty} W(t) = 0.
	\end{equation}
	Assuming that $e^{-\delta s} c(s)^{1-R}$ is integrable at infinity, a solution to \eqref{eq:deterministic ODE for V} is
	$       W(t) = \left(\int_t^\infty b e^{-\delta s} c(s)^{1-S} \dd s \right)^\theta. $
	Therefore, a utility process $V=V^c$ associated to $c$ is
	\begin{align}\label{eq:utility associated to a deterministic consumption plan}
		V(t) = \frac{1}{1-R}\left(b\int_t^\infty  e^{-\delta s} c(s)^{1-S} \dd s\right)^\theta.
	\end{align}
	In particular, when $C^{a,\gamma} = (C^{a,\gamma}_t)_{t \geq 0}$ is the deterministic, exponentially decaying consumption stream given by $C_t = C^{a,\gamma}_t = a e^{-\gamma t}$ and $\delta + \gamma(1-S) > 0$ we find
	\begin{equation}\label{eq:constant consumption stream evaluated}
		V(t) = V^{C^{a,\gamma}}_t = e^{-(\delta + \gamma(1-S)) \theta t}\left(\frac{b }{\delta + \gamma(1-S)} \right)^\theta \frac{a^{1-R}}{1-R}
	\end{equation}
	and $J_{g_{EZ}}(C^{a,\gamma}) \coloneqq V^{C^{a,\gamma}}_0 = \left( \frac{b}{\delta + \gamma(1-S)} \right)^{\theta} \frac{a^{1-R}}{1-R}$.
	
	Now consider a `purely random' consumption stream, whose paths have no variance over time, except for an exponential decay. Suppose that the non-negative random variable $Y$ is such that $Y$ and $Y^{1-R}$ are integrable.
	Let $\cF_t = \sigma(Y)$ for all $t>0$.\footnote{For the exposition, we temporarily drop the assumption that the filtration $(\cF_t)_{t \geq 0}$ is right-continuous.} Consider the (progressively measurable) consumption stream $C^{Y,\gamma}_t \equiv Y e^{-\gamma t}$ for $t>0$. All uncertainty is resolved instantaneously at $t=0$. The value of such a consumption stream is given by
	\begin{equation}
		J_{g_{EZ}}(C^{Y,\gamma}) = \left(\frac{b}{\delta+\gamma(1-S)} \right)^\theta\E\left[\frac{Y^{1-R}}{1-R} \right] \leq \left(\frac{b }{\delta+ \gamma(1-S)} \right)^\theta\frac{(\E[Y])^{1-R}}{1-R} = J_{g_{EZ}}(C^{\E[Y],\gamma}),
	\end{equation}
	where the inequality follows directly from Jensen's inequality. The loss in utility from the uncertainty is captured by the risk-aversion $R$ of the agent and the larger value of $R$, the stronger the agent's preference for certainty. Thus $R$ may interpreted as the agent's aversion to risk. Looking at \eqref{eq:Epstein--Zin SDU} or \eqref{eq:utility associated to a deterministic consumption plan} one might expect that the risk aversion comes from the value of $S$ but, contrary to naive intuition, this is not the case.
	
	Now consider the agent's preferences over deterministic consumption streams that vary over time.
	Assume temporarily and for the purposes of exposition that $\delta > 0$ and $\theta > 0$ and define a new (probability) measure $\QQ= \QQ_\delta$ on the Borel $\sigma$-algebra $\cB(\R_+)$ by
	\begin{equation}
		\QQ_\delta (A) = \int_A \delta e^{-\delta t} \dd t.
	\end{equation}
	The choice of $\delta$ accounts for the agent's temporal preferences for consumption in the sense that the higher the value of $\delta$ the greater the weighting on consumption which occurs earlier.
	
	Now compare a (deterministic) consumption stream $c = (c(t))_{t\geq0}$ with its $\QQ_\delta$-average value $\E^{\QQ_\delta}[c] = \int_0^\infty \delta e^{-\delta t} c(t) \dd t$ which we suppose finite. From \eqref{eq:utility associated to a deterministic consumption plan} we know that the value at time $0$ is given  is given by
	\[ V^c(0)=  \frac{1}{1-R} \left(\frac{b}{\delta}\right)^{\theta} \left( \int_0^\infty  \delta  e^{-\delta t} {c(t)^{1-S}} \dd t\right)^\theta
	= \theta \left(\frac{b}{\delta}\right)^{\theta} \frac{ \left( \E^{\QQ_\delta}\left[{c^{1-S}}\right]\right)^\theta}{1-S} .
	\]
	Again, Jensen's inequality (and $\theta>0$) gives $\frac{1}{1-S}(\E^{\QQ_\delta}[c^{1-S}])^\theta \leq \frac{1}{1-S} [(\E^{\QQ_\delta}[c])^{1-S}]^{\theta}$, which implies that $V^c_0 \leq V^{\E^{\QQ_\delta}[c]}_0$. Note that, all of the variance aversion (after changing the Lebesgue measure to an equivalent probability measure) comes from $S$. This justifies considering $S$ as the parameter governing aversion to variance over time. In the economics literature $S$ is named the elasticity of intertemporal complementarity (EIC).
	
	Note that if $(1-R) V_t < 0 $ then the integrand on the right hand side of \eqref{eq:Epstein--Zin SDU} is ill-defined for non-integer $\rho$. This justifies the choice $\VV = (1-R) \R_+$. Further, the integrand is either positive ($S<1$) or negative ($S>1$).  It is therefore necessary to impose a link between the co-efficient of RRA $R$ and co-efficient of EIC $S$ to ensure agreement in the sign of the left-hand-side of \eqref{eq:Epstein--Zin SDU} and the right hand side. Recall that $\theta = \frac{1-R}{{1-S}}$.
	\begin{thm}
		For EZ-SDU over the infinite horizon with generator given by \eqref{eq:Epstein--Zin aggregator} we must have $\theta > 0$ for there to exist solutions to \eqref{eq:Epstein--Zin SDU}.
	\end{thm}
	The condition $\theta > 0$, or equivalently $\rho \in (-\infty, 1)$ means that either both $R$ and $S$ are greater than unity, or both $R$ and $S$ are smaller than unity.
	
	In the finite time horizon problem the parity issue can be overcome by adding a bequest function so that \eqref{eq:Epstein--Zin SDU} is replaced by $V_t = \E[\int_t^T b e^{-\delta s} \frac{C_s^{1-S}}{1-S} ((1-R)V)^{\rho} + e^{-\delta T} \frac{B(X_T)}{1-R} | \cF_t]$ where $B: \R_+ \mapsto \R_+$ assigns a value to terminal wealth. But, even over the finite horizon this leads to conceptual issues: for example, when $S<1<R$ the utility process is negative at time $t$, even though the term corresponding to consumption over $(t,T)$ is everywhere positive, because this positive term is outweighed by the contribution from the bequest. Moreover if we let the terminal horizon tend to infinity the problem becomes even more stark---in order to outweigh the increasing (as terminal horizon $T$ increases) contribution from consumption the contribution from the bequest must also grow, and must become more (not less) influential as the terminal horizon increases.
	In Section \ref{ssec:difference form of EZ SDU} we argue that in the limit $T \nearrow \infty$ we end up with bubble-like behaviour which cannot be justified economically, and which is not consistent with any notion of transversality. This further justifies the requirement $\theta>0$.
	
	\section{Optimal investment and consumption in a Black--Scholes--Merton financial market}
	\label{sec:BSM}
	
	\subsection{The financial market and attainable consumption streams}
	\label{ssec:market}
	The Black--Scholes--Merton financial market consists of a risk-free asset with interest rate $r \in \RR$, whose price process $S^0=(S^0_t)_{t \geq 0}$ is given by $S^0_t = S^0_0\exp(r t)$, together with a risky asset whose price process $S= (S_t)_{t \geq 0}$ follows a geometric Brownian motion with drift $\mu \in \RR$ and volatility $\sigma > 0$, and whose initial value is $S_0 = s_0 > 0$. In particular,
	$S_t = {s_0} \exp ( \sigma B_t + (\mu - \frac{1}{2} \sigma^2)t )$, where $B = (B_t)_{t \geq 0}$ denotes a Brownian motion.
	
	The agent optimises over the controls variables \textit{the proportion of wealth invested in each asset} and the \textit{rate of consumption}.
	Let $\Pi_t$ represent the proportion of wealth invested in the risky asset at time $t$ and let $\Pi^0_t=1-\Pi_t$ represent the proportion of wealth held in the riskless asset at time $t$. Further, let $C_t$ denote the rate of consumption at time $t$. It then follows that the wealth process $X=(X_t)_{t\geq0}$ satisfies the SDE
	\begin{equation}
		\dd X_t =  X_t \Pi_t  \sigma \dd B_t + \left( X_t (r + \Pi_t (\mu - r))  - C_t\right)\dd t,
		\label{eqn:original wealth process}
	\end{equation}
	subject to initial condition $X_0=x$, where $x$ is the initial wealth.
	\begin{defn}
		Given $x>0$ an \emph{admissible investment-consumption strategy} is a pair $(\Pi,C)= (\Pi_t,C_t)_{t \geq 0}$ of progressively measurable processes, where $\Pi$ is real-valued and $C$ is nonnegative, such that the SDE \eqref{eqn:original wealth process} has a unique strong solution $X^{x, \Pi, C}$ that is $\as{\P}$ nonnegative. We denote the set of admissible investment-consumption strategies for $x > 0$ by $\sA(x; r,\mu,\sigma)$.
	\end{defn}
	
	The objective criteria by which the strategy is evaluated will depend only upon the consumption and not upon the investment portfolio in the financial assets. This motivates the following definition:
	
	\begin{defn}
		A consumption stream $C \in\sP_+$ is called \emph{attainable} for initial wealth $x > 0$ if there exists a progressively measurable process $\Pi = (\Pi_t)_{t\geq0}$ such that $(\Pi, C)$ is an admissible investment-consumption strategy. Denote the set of attainable consumption streams for $x  > 0$ by $\sC(x; r,\mu,\sigma)$.
	\end{defn}
	When it is clear which financial market we are considering, we simplify the notation and write $\sA(x) = \sA(x; r,\mu,\sigma)$ and $\sC(x) = \sC(x; r,\mu,\sigma)$.
	
	The goal of an agent with Epstein--Zin stochastic differential utility preferences
	is to maximise $J_{g_{EZ}}(C)$ over attainable consumption stream. However, $J_{g_{EZ}}(C)$ is currently only defined for $C \in \cE_u(g_{EZ})$ and therefore, we can currently only optimise over uniquely evaluable consumption streams. Thus, we seek to find
	\begin{align}\label{eq:Control problem uniquely formulation}
		V^*_{\sE_u(g_{EZ})}(x) = \sup_{C \in \sC(x)\cap \sE_u(g_{EZ})} V^C_0 ~=~ \sup_{C \in \sC(x) \cap \sE_u(g_{EZ})}J_{g_{EZ}}(C) .
	\end{align}
	This is very restrictive. For $\theta > 1$, one can show that $\sE_u(g_{EZ}) = \{0\}$ and so the problem \eqref{eq:Control problem uniquely formulation} is meaningless. Further, even when $\theta \in (0, 1)$, there are many attainable consumptions streams which are non-(uniquely)-evaluable and therefore to which we currently cannot assign them a utility. For example, when $S>1$,  the zero consumption stream is not evaluable. Since it might reasonably be argued that the zero consumption stream is clearly suboptimal (and when $S>1$ should give a utility process with negative infinite utility), we would like to eliminate this choice of consumption stream because it is suboptimal and not because we cannot evaluate it. The same applies to other non-evaluable consumption streams. Ideally, we would like \emph{every} attainable consumption stream to be considered, and not just the `nice' ones for which we can define a unique utility process. For $\theta \in (0, 1)$, this problem will be considered in Part II.
	
	\subsection{Changes of num\'eraire}\label{ssec:change of numeraire}
	
	One apparent advantage of the difference form $g^\Delta_{EZ}$ of the EZ-SDU aggregator given in \eqref{eq:Epstein--Zin aggregator difference form} over the discounted form $g_{EZ}$ given in \eqref{eq:Epstein--Zin aggregator} is that $g^\Delta_{EZ}$, unlike $g_{EZ}$, has no explicit time-dependence, i.e. $g^\Delta_{EZ}=g^\Delta_{EZ}(c,v)$ whereas $g_{EZ}=g_{EZ}(t,c,v)$. However, when we consider EZ-SDU in the constant parameter Black--Scholes--Merton model a simple change of accounting unit leads to a modification of the discount factor $\delta$, but leaves the problem otherwise unchanged. It follows that by an appropriate choice of units we can switch to a coordinate system in which the aggregator becomes time-independent. The change of accounting units has an effect upon the financial market model, but it remains a Black--Scholes--Merton financial market, albeit with modified interest rate and market drift.
	
	Let $C$ be a consumption stream with corresponding utility process $V$ for $g_{EZ}$. Let $\chi \in \RR$ and define the
	the \textit{discounted consumption stream} $\tilde{C}$ by $\tilde{C}_t = e^{-\chi t}C_t$. Then, $V$ satisfies
	\begin{equation}\label{eq:numeraire changed equation for V}
		V_t = \E\left[\left.\int_t^\infty b e^{-\delta s}\frac{C_s^{1-S}}{{1-S}}\left( (1-R) V_s\right)^\rho \dd s\right| \cF_t \right]
		= \E\left[\left.\int_t^\infty b e^{-(\delta - \chi(1-S))t} \frac{\tilde{C}_s^{1-S}}{{1-S}}\left( (1-R)V_s\right)^\rho \dd s\right| \cF_t \right].
	\end{equation}
	This implies $V$ is the utility process for $\tilde{C}$ with the aggregator ${g}_{\chi,EZ}$ defined by
	\begin{equation}\label{eq:discounted Epstein--Zin aggregator}
		{g}_{\chi,EZ}(t,c,v) = b  e^{-(\delta - \chi(1-S))t} \frac{c^{1-S}}{{1-S}}\left( (1-R)v\right)^\rho.
	\end{equation}
	Choosing $\chi = \frac{\delta}{1-S}$, we find that $V$ is the utility process for the time independent aggregator
	\[ f_{EZ}= f_{EZ}(c,v) = g_{\chi,EZ}(t,c,v) =  b \frac{c^{1-S}}{{1-S}}\left( (1-R)v\right)^\rho. \]
	Furthermore, $V \in \II(f_{EZ},\tilde{C}= (C_t e^{-\frac{\delta}{1-S}t})_{t \geq 0})$ if and only if $V \in \II(g_{EZ},C)=\II(g_{0,EZ},C)$ and $\tilde{C} \in \sE_u(f_{EZ})$ in and only if $X \in \sE_u(g_{EZ})$.
	
	If we consider the discounted wealth process $\tilde{X}^{\Pi,\tilde{C}}_t \coloneqq e^{-\frac{\delta}{{1-S}}t} X^{\Pi,C}_t$ then, by applying It\^o's lemma, we find that with $\tilde{r} = r - \frac{\delta}{1-S}$ and $\tilde{\mu} = \mu -\frac{\delta}{1-S}$,
	\begin{eqnarray}\label{eq:discounted wealth process}
		d \tilde{X}^{\Pi,\tilde{C}}_t & = & \tilde{X}^{\Pi,\tilde{C}}_t \Pi_t \sigma \dd B_t + \left(\tilde{X}^{\Pi,\tilde{C}}_t\left( \tilde{r} + \Pi_t (\tilde{\mu} - \tilde{r})\right) - \tilde{C}_t\right)\dd t, \quad \tilde{X}^{\Pi,\tilde{C}}_0 = x.
	\end{eqnarray}
	
	This means that our control problem \eqref{eq:Control problem uniquely formulation} admits the equivalent formulation,
	\begin{align}\label{eq:Control problem numeraire changed formulation}
		V^*_{\sE_u(g_{EZ})}(x)  = \sup_{C \in \sC(x; r,\mu,\sigma) \cap \sE_u(g_{EZ})} V^{C,g_{EZ}}_0 = \sup_{\tilde{C} \in \sC(x; \tilde{r},\tilde{\mu},\sigma) \cap \sE_u(f_{EZ})} V^{\tilde{C},f_{EZ}}_0 =   V^*_{\sE_u(f_{EZ})}(x).
	\end{align}
	In particular, by an appropriate change of accounting units the problem for EZ-SDU in discounted form reduces to an equivalent form with no discounting.
	This simplification result will be used extensively in Part II on existence and uniqueness, but whilst we are comparing and contrasting the discounting and difference forms we will continue to allow $\delta$ to be any real number.
	
	\subsection{The candidate optimal strategy}\label{ssec:candidate optimal utility process}
	Suppose now $\theta>0$. We seek to heuristically find an admissible (and uniquely evaluable) consumption stream $C$ that maximises the value of $V^C_0$, where
	\begin{equation}\label{eq:value process optimal for strategy derivation}
		V^C_t = \E\left[\left.\int_t^\infty b e^{-\delta s} \frac{C_s^{1-S}}{{1-S}}\left( (1-R)V^C_s\right)^\rho \dd s\right| \cF_t \right].
	\end{equation}
	As in the Merton problem with CRRA utility, it is reasonable to expect that the optimial strategy is to invest a constant proportion of wealth in the risky asset, and to consume a constant proportion of wealth. Consider the investment-consumption strategy $\Pi \equiv \pi\in\R$ and $C \equiv \xi X$ for $\xi\in\RR_{++}$. Then, solving \eqref{eqn:original wealth process}, the wealth process $X^{x,\pi,\xi} = X = (X_t)_{t \geq 0}$ is given by
	$       X_t  =  x\exp\left(\pi \sigma B_t + \left({r} + \pi ({\mu} - {r}) - \xi - \frac{\pi^2 \sigma^2}{2}\right)t \right)$, and then for $s>t$
	\begin{equation}\label{eq:X^1-R equation constant strategy}
		X_s^{1-R} = \ x^{1-R} \exp\left(\pi \sigma (1-R) B_t + (1-R)\left({r} + \lambda \sigma \pi - \xi - \frac{\pi^2\sigma^2}{2}\right)t\right).
	\end{equation}
	As in the Merton problem, consider a value process of the form $V_t = V(t,X_t) = A e^{- \beta t} \frac{X_t^{1-R}}{1-R}$ for some constant $\beta$ to be determined. Substituting this expression into \eqref{eq:value process optimal for strategy derivation}, and using $1-S + \rho(1-R) = 1-R$ yields
	\begin{equation}
		\label{eq:deriving the optimal value function step 1}
		V_t      = \E\left[\left.\int_t^\infty b e^{-\delta s} \frac{(\xi X_s)^{1-S}}{{1-S}} \left( A e^{-\beta s} X_s^{1-R}\right)^\rho \dd s\right| \cF_t \right]
		= b A^\rho \frac{\xi^{1-S}}{1-S}  \E\left[\left.\int_t^\infty  e^{-(\delta + \beta \rho)s}   X_s^{1-R} \dd s\right| \cF_t \right]
	\end{equation}
	Then, for $s>t$, $\E[e^{-(\delta + \beta \rho)s}X_{s}^{1-R} | \cF_t] = e^{-(\delta + \beta \rho)t} X_t^{1-R} e^{-H_{\delta + \beta\rho}(\pi,\xi)(s-t)}$,
	where for $\nu\in\R$, $H_\nu: \R \times \R_{++} \mapsto \R$ is given by
	\begin{equation}
		\label{eq:H_nu}
		H_\nu(\pi, \xi) ~=~ \nu + (R-1)\left({r} + \lambda \sigma \pi - \xi - \frac{\pi^2\sigma^2}{2}R\right).
	\end{equation}
	\begin{rem}
		If we consider the constant proportional investment-consumption $(\pi,\xi)$, then the drift of $(e^{-\nu t} X_t^{1-R})_{t \geq 0}$ is given by $-H_\nu(\pi,\xi)$. This means that $H_{\nu}(\pi,\xi)$ is a critical quantity for both the well-definedness of the integral
		$\E[\int_0^\infty e^{-\nu t}X_t^{1-R} \dd t]$ and the transversality condition $\lim_{t \to \infty} \E[e^{-\nu t} X_t^{1-R}] = 0$ which will feature heavily in Section \ref{sec:alternative}.
	\end{rem}
	Provided that $H_{\delta + \beta\rho}(\pi,\xi)>0$ so that the integral in \eqref{eq:deriving the optimal value function step 1} is well-defined, it follows that
	\begin{equation}\label{eq:valfungenstrat}
		V_t =  \frac{b e^{-(\delta + \beta \rho) t} A^\rho \xi^{1-S} }{H_{\delta + \beta\rho}(\pi,\xi)}\frac{X_t^{1-R}}{1-S} .
	\end{equation}
	Since $V$ was postulated to be of the form $V_t = A e^{- \beta t}\frac{X_t^{1-R}}{1-R}$, it must be the case that $\beta = \delta+ \beta \rho$ (i.e. $\beta = \delta\theta$) and $A = A(\pi,\xi) = \left(\frac{b \theta \xi^{1-S}}{H_\beta (\pi,\xi)}\right)^\theta>0$. Then, $\delta + \beta \rho = \delta \theta$ and $H \coloneqq H_{\delta\theta}$ satisfies
	\begin{equation}
		\label{eq:H(pi,xi)>0}
		H(\pi, \xi) = \delta\theta + (R-1)\left({r} + \lambda \sigma \pi - \xi - \frac{\pi^2\sigma^2}{2}R\right).
	\end{equation}
	It follows that any proportional investment strategy $(\Pi = \pi,~C=\xi X)$ is evaluable provided that $H(\pi,\xi)$ is positive.
	
	To find the optimal strategy amongst constant proportional strategies (and hence to find the candidate optimal strategy) it remains to maximise $\frac{A(\pi,\xi)}{1-R}$ over $(\pi,\xi) \in \R \times \R_{++}$ such that $H(\pi,\xi)>0$. There is a turning point of $\frac{A(\pi,\xi)}{1-R} = \frac{1}{1-R} \left( \frac{b \theta \xi^{1-S}}{H(\pi,\xi)}\right)^\theta$ at $(\hat{\pi},\hat{\xi})=
	(\frac{\lambda}{\sigma R}, \eta)$ where
	\begin{equation}
		\label{eq:etadef} \eta = \frac{1}{S} \left( \delta + (S-1)r + (S-1) \frac{\lambda^2}{2R} \right)
	\end{equation}
	and this point is such that $H(\pi,\xi)=H(\frac{\lambda}{\sigma R},\eta)>0$ provided $\eta > 0$. Under the condition $\eta>0$ it is easily checked that $(\pi=\frac{\lambda}{\sigma R},\xi=\eta)$ is a maximum of $(1-R)^{-1}A(\pi,\xi)$
	over $\{(\pi,\xi):H(\pi,\xi)>0 \}$;
	it then follows that $ \max_{\{\xi>0: H(\hat{\pi},\xi)>0 \} } V_0 = b^\theta\eta^{-\theta S }\frac{x^{1-R}}{1-R}$. Considering this as a function of the initial wealth, for $\eta>0$ the candidate value function is defined by
	\begin{equation} \label{eq:hat V candidate}
		\hat{V}(x) = b^\theta\eta^{-\theta S }\frac{x^{1-R}}{1-R}.
	\end{equation}
	The results of this section are summarised in the following proposition:
	\begin{prop}\label{prop:derivecandidate}
		Define $D= \{ (\pi, \xi)\in\R\times\R_+ : H(\pi, \xi)>0 \}$. Consider constant proportional strategies with parameters $(\pi,\xi)\in D$.        Suppose $\theta>0$ and $\eta  > 0$, where $\eta$ is given in \eqref{eq:etadef}.
		\begin{enumerate}[{\rm(}i{\rm)}]
			\item For $(\pi,\xi)\in D$, one solution $V=(V_t)_{t\geq0}$ to \eqref{eq:value process optimal for strategy derivation} is given by
			\begin{equation}\label{eq:valfungenstrat2}
				V_t =   e^{-\delta \theta t} \left( \frac{b \theta \xi^{1-S} }{H(\pi,\xi)} \right)^\theta \frac{X_t^{1-R}}{1-R} .
			\end{equation}
			\item The global maximum of $h(\pi,\xi) = \frac{1}{1-R}  \left( \frac{b \theta \xi^{1-S} }{H(\pi,\xi)} \right)^\theta$ over the set $D$ is attained at $(\pi, \xi) = (\frac{\lambda}{\sigma R}, \eta)$ and the maximum is $\frac{b^\theta \eta^{-\theta S}}{1-R}$.
			\item The optimal strategy for \eqref{eq:valfungenstrat2} is $(\hat{\pi},\hat{\xi}) = (\frac{\lambda}{\sigma R}, \eta)$ and satisfies $\hat V_0 = b^\theta\eta^{-\theta S }\frac{x^{1-R}}{1-R}= \hat{V}(x)$, where $x$ denotes initial wealth.
		\end{enumerate}
	\end{prop}
	
	The candidate well-posedness condition for the investment-consumption problem is $\eta>0$, where $\eta$ is given in \eqref{eq:etadef}. We shall see in Corollary \ref{cor:ill posed} that when $\theta\in(0,1)$ this is a necessary and sufficient condition for the well-posedness of the problem. The agent's (candidate) optimal investment in this case is a constant fraction $\hat{\pi} = \frac{\lambda}{\sigma R}$ of their wealth, a proportion which is independent of their EIC. The agent's investment preferences are controlled solely by the risk aversion coefficient $R$. The agent's (candidate) optimal consumption is a constant proportion $\eta$ of their wealth.
	
	To understand, the interpretation of $\eta$, it is insightful to perform a change of num\`eraire. As in \cite[Section 7]{herdegen2020elementary}, the problem may be rewritten in equivalent form as
	\begin{align}
		V_t =~ \E\left[\left.\int_t^\infty \frac{b e^{-(\delta + r(S-1))s}}{1-S}\left(\frac{C_s}{S^0_s}\right)^{1-S}\left( (1-R)V_s\right)^\rho \dd s\right| \cF_t \right].
	\end{align}
	With this in mind, it makes sense to call $\phi \coloneqq \delta + r(S-1)$ the \textit{impatience rate}. Then, the optimal proportional consumption rate is given by
	\begin{equation}
		\eta = \frac{\phi}{S} + \frac{S-1}{S}\frac{\lambda^2}{2R}.
	\end{equation}
	This is a linear (convex if $S>1$) combination of the impatience rate and (half of) the squared Sharpe ratio per unit of risk aversion, with the weights depending on the elasticity of intertemporal complementarity $S$.
	
	\begin{rem}
		The well-posedness condition $\eta>0$ is equivalent to $\delta > (1-S)\left( r + \frac{\lambda^2}{2R} \right)$ (or $\phi > (1-S)\tfrac{\lambda^2}{2R}$). This means that when  $S>1$ (or $r<0$), the problem can be well-posed even for negative values of $\delta$ (or $\phi$).
	\end{rem}
	
	\begin{rem} When $\theta>1$, uniqueness of a utility process fails (for example $V_t=0$ always solves \eqref{eq:value process optimal for strategy derivation}). In this case, the first issue is to decide which utility process to associate to a consumption stream; this in turn has implications for the optimal value function and optimal consumption stream, and ultimately for the well-posedness of the problem. Since this is a delicate issue and deserves a full discussion, we postpone it to a later paper covering the case $\theta>1$.
	\end{rem}
	
	\section{A comparison of the discounted and difference formulations}\label{sec:difference form of EZ SDU}
	
	The goal of this section is to compare the discounted and difference formulations of the aggregator for EZ-SDU. Despite the ubiquity of the latter in the literature, we will argue that the discounted form has many advantages. As demonstrated in Section~\ref{ssec:change of numeraire}, its main disadvantage, the fact that it has an explicit dependence on time, is easily overcome by a change in accounting unit.
	
	\subsection{The difference form of CRRA utility}\label{ssec:difference form of CRRA utility}
	
	Additive utilities such as CRRA may be thought of as special cases of SDU in which the aggregator has no dependence on $v$. In this sense CRRA utility may be indentified with the aggregator
	\begin{equation}\label{eq:Merton aggregator}
		g_{CRRA}(t,c,v) = g_{CRRA}(t,c) = e^{-\delta t} \frac{c^{1-R}}{1-R}.
	\end{equation}
	Note that provided $\E[ \int_0^\infty e^{-\delta s} |C_s^{1-R}| ds ] < \infty$ it follows that
	\begin{equation}\label{eq:Merton aggregatorV}
		V^C_t = \EX{\int_t^\infty e^{-\delta s} \frac{C_s^{1-R}}{1-R} \dd s}
	\end{equation}
	is the unique utility process associated with consumption $C$ for generator $g_{CRRA}$ and then $J_{g_{CRRA}}(C) = V^C_0$. Further, if $\E[\int_0^\infty e^{-\delta s} |C_s^{1-R}| ds ] = \infty$ we can set $J(C)=\infty$ if $R<1$ and $J(C)=-\infty$ if $R>1$.
	
	In particular, two subtle but important questions which are crucial to the study of SDU are absent from the additive utility setting: first, what value to assign to non-evaluable strategies, and second which utility process to assign to consumptions which are not uniquely evaluable.
	
	Suppose $C$ is such that $\E[ \int_0^\infty e^{-\delta s} |C_s^{1-R}| \dd s ] < \infty$. Then, the martingale
	$M=(M_t)_{0 \leq t \leq \infty}$ given by $M_t \coloneqq \cEX[t]{\int_0^\infty e^{-\delta s}\frac{C_s^{1-R}}{1-R} \dd s}$
	is uniformly integrable and satisfies
	$M_t = \int_0^t e^{-\delta s} \frac{C_s^{1-R}}{1-R} \dd s + V_t $ where $V$ is the utility process in \eqref{eq:Merton aggregatorV}.
	Using that $M_\infty = \int_0^\infty e^{-\delta s}\frac{C_s^{1-R}}{1-R} \dd s$ and rearranging, we find that $  V_t = \int_t^\infty e^{-\delta s} \frac{C_s^{1-R}}{1-R} \dd s  - \int_t^\infty \dd M_t. $
	Then, applying It\^o's formula to $V^\Delta$ given by $V^\Delta_t \coloneqq e^{\delta t}V_t$ and integrating yields $V^{\Delta}_t = \int_t^\infty \left( \frac{C_s^{1-R}}{1-R} - \delta V^\Delta_s\right)\dd s + \int_t^\infty e^{\delta s} \dd M_s,$
	provided such a solution is well-defined. Taking expectations, and assuming that $M^\delta= (M^\delta_t)_{t \geq 0}$ given by $M^\delta_t = \int_0^t e^{\delta s} dM_s$ is a uniformly integrable martingale we get the \textit{difference form} of discounted expected utility,
	\begin{equation}\label{eq:value process associated to infinite horizon difference CRRA utility}
		V^\Delta_t = \cEX[t]{\int_t^\infty \left( \frac{C_s^{1-R}}{1-R} - \delta V^\Delta_s \right)\dd s}.
	\end{equation}
	Modulo the technical issues, under CRRA preferences, it is possible to define the value associated to a consumption stream $C$ as the initial value $V^\Delta_0$ of the utility process ${V^\Delta} = (V^\Delta_t)_{t \geq 0}$ where $V^\Delta$ solves \eqref{eq:value process associated to infinite horizon difference CRRA utility}, rather than using \eqref{eq:Merton aggregatorV}.
	However, doing so brings several immediate disadvantages. It is no longer obvious if solutions to \eqref{eq:value process associated to infinite horizon difference CRRA utility} are unique or even exist. This may result in a smaller class of evaluable strategies. Indeed there are simple deterministic counter-examples to existence of a solution to \eqref{eq:value process associated to infinite horizon difference CRRA utility}, see Example~\ref{eg:CRRAdnotD}. The counterexamples arise because the integrand $\frac{C_s^{1-R}}{1-R} - \delta {V^\Delta_s} $ takes both signs and so the integral on the right hand side of \eqref{eq:value process associated to infinite horizon difference CRRA utility} may not be well-defined.
	(In contrast, $\E[ \int_0^\infty e^{-\delta s} \frac{C^{1-R}_s}{1-R}]$ is always well defined, at least in $[-\infty,\infty]$.)
	Further, whenever $\E[ \int_0^\infty e^{-\delta s} |C_s^{1-R}|ds ] < \infty$ we have that $M$ is a uniformly integrable martingale. But $M^\delta$ may not be uniformly integrable, and the representation \eqref{eq:value process associated to infinite horizon difference CRRA utility} may fail.
	
	\begin{example}
		\label{eg:CRRAdnotD}
		Suppose $\delta>0$ and let $A = \cup_{n \geq 0} [2n,2n+1)$.
		Consider the deterministic consumption stream $c = (c(t))_{t\geq0}$ which satisfies
		\begin{equation} \label{eq:non evaluable consumption plan}
			U(c(t))\coloneqq\frac{c(t)^{1-R}}{1-R} =
			\frac{2\delta}{1-R}e^{\delta(\lceil t \rceil-t)} \1_{A^c}(t).
		\end{equation}
		It is easily checked (consider the cases $t \in A$ and $t \in A^c$ separately) that $V^\Delta$ defined by
		$V^\Delta(t) = \frac{1}{1-R}e^{\delta(t-\lfloor t\rfloor)(\1_A(t) - \1_{A^c}(t))}$ satisfies $\dd V^\Delta(t) = \left[ \delta V^\Delta(t) - \frac{c(t)^{1-R}}{1-R} \right]\dd t$ (at least for non-integer $t$).
		
		Clearly, $\int_t^\infty \left( \frac{c(s)^{1-R}}{1-R} - \delta V^\Delta(s)\right)\dd s$ is \textit{not} well-defined since both the positive part and the negative part are infinite and hence it is {\em not} the case that $V^\Delta$ solves $V^\Delta = \int_t^\infty \left( \frac{c(s)^{1-R}}{1-R} - \delta V^\Delta(s)\right)\dd s$. On the other hand, $V(t) = e^{-\delta t} V^\Delta(t)$ is a solution to the discounted formulation
		$V(t) = \int_t^\infty e^{-\delta s} \frac{c(s)^{1-R}}{1-R} \dd s$. (Note that since $U(c(s))$ is bounded and $\delta>0$, $V(0)$ is finite.)
		
		Thus, if we set ${g}^\Delta_{CRRA}(t,c,v) = \frac{c^{1-R}}{1-R} - \delta v$ and $g_{CRRA}= e^{-\delta t} \frac{c^{1-R}}{1-R}$, then $\sE(g^\Delta_{CRRA})\subsetneq \sE(g_{CRRA})$. In particular, there are consumption streams which can be evaluated under the formulation \eqref{eq:Merton aggregatorV} but which cannot be evaluated using \eqref{eq:value process associated to infinite horizon difference CRRA utility}.
	\end{example}
	
	\subsection{The difference form of Epstein--Zin stochastic differential utility}\label{ssec:difference form of EZ SDU}
	In the previous section we argued that for additive CRRA preferences, the discounted form was better than the difference form for three reasons:
	first, existence and uniqueness of the utility process are guaranteed;
	second, there is a wider class of consumption streams to which it is possible to assign a (finite) value;
	and third, it is possible to assign a value (possibly infinite) to any consumption stream even when $\int_0^\infty g_{CRRA}(s,C_s) ds$ is not integrable.
	The goal in this section is to show that, although the first property in this list no longer applies, when we move to EZ-SDU preferences the second and third advantages of the discounted form remain. Indeed, much of the discussion is as in the additive case.
	
	Suppose that $C \in \sE_u(g_{EZ})$ and set $M_t \coloneqq \E[\int_0^\infty b e^{-\delta s}\frac{C_s^{1-S}}{{1-S}}\left( (1-R)V_s\right)^\rho \dd s| \cF_t ]$. After a re-arrangement, \eqref{eq:value process optimal for strategy derivation} becomes
	\begin{equation}
		V_t = M_t - \int_0^t b e^{-\delta s}\frac{C_s^{1-S}}{{1-S}}\left( (1-R)V_s\right)^\rho \dd s= \int_t^\infty b e^{-\delta s}\frac{C_s^{1-S}}{{1-S}}\left( (1-R)V_s\right)^\rho \dd s~ - \int_t^\infty \dd M_s.
	\end{equation}
	Furthermore, applying It\^o's lemma to the upcounted utility process ${V}^\Delta = (V^\Delta_t)_{t \geq 0}$ defined by $V^\Delta_t \coloneqq e^{\delta \theta t}V_t$, we find that ${V^\Delta}$ satisfies
	$V^\Delta_t = \int_t^\infty \left(b \frac{C_s^{1-S}}{{1-S}}\left( (1-R)V^\Delta_s\right)^\rho - \delta \theta V^\Delta_s \right)\dd s~ - \int_t^\infty e^{\delta \theta s} \dd M_s,
	$
	and we may reasonably hope to be able to define the (upcounted) utility process as the solution to
	\begin{equation}\label{eq:value process associated to infinite horizon difference SDU}
		V^\Delta_t = \cEX[t]{\int_t^\infty \left(b \frac{C_s^{1-S}}{{1-S}}\left( (1-R) V^\Delta_s \right)^\rho - \delta \theta V^\Delta_s \right)\dd s}.
	\end{equation}
	This is the utility process associated to the difference form of the Epstein--Zin aggregator, $g^\Delta_{EZ}$.
	
	As discussed in Section \ref{ssec:difference form of CRRA utility}, for some consumption streams \eqref{eq:value process associated to infinite horizon difference SDU} is not well defined because the integrand may be either positive or negative.
	If the utility process is defined via the difference aggregator $g^{\Delta}_{EZ}$ then it is necessary to
	restrict the class of consumption streams, when compared with those which may be evaluated under $g_{EZ}$.
	
	\begin{example}\label{eg:sducounterexample}
		This example is similar to Example~\ref{eg:CRRAdnotD}. Recall the definition of $A$, and consider the deterministic consumption stream $c = (c(t))_{t\geq0}$ such that
		$       \frac{c(t)^{1-S}}{1-S} \coloneqq 2\frac{\delta}{b(1-S)} e^{\delta(\lceil t \rceil-t)}\1_{A^c}(t) $.
		Let $V^\Delta=(V^\Delta(t))_{t \geq 0}$ be given by $V^\Delta(t) =  \frac{1}{1-R}\exp(\delta\theta(t-\lfloor t\rfloor)(\1_A(t) - \1_{A^c}(t)))$. Then, $$\dd V^\Delta(t) = \left[ \delta \theta V^\Delta(t) - b\frac{c(t)^{1-S}}{1-S}((1-R)V^\Delta(t))^\rho \right]\dd t.$$
		For this consumption stream, both the positive and negative part of the integral
		$$\int_t^\infty \left( b\frac{c(t)^{1-S}}{1-S}((1-R)V^\Delta(t))^\rho - \delta \theta V^\Delta(s)\right)\dd s = \int_t^\infty \delta\theta V^\Delta(s)\left[\1_A(s) - \1_{A^c}(s)\right]\dd s$$ are infinite for all $t\geq0$. Hence, it cannot be the case that $V^\Delta$ solves \eqref{eq:value process associated to infinite horizon difference SDU}. On the other hand, if $V(t)=e^{-\delta\theta t}V^\Delta(t)$, then
		\begin{equation}
			\int_0^\infty be^{-\delta t}\frac{c(t)^{1-S}}{1-S}((1-R)V(t))^\rho \dd t = \int_0^\infty 2 e^{-\delta t}\frac{\delta}{1-S}e^{\delta\theta(\lceil t \rceil -t)} \1_{A^c}(t) \dd t <\infty
		\end{equation}
		and $V=(V(t))_{t\geq0}\in\II(g_{EZ}, c)$. Furthermore, it can be shown that $V$ solves \eqref{eq:value process optimal for strategy derivation}. Thus, $\cE(g^\Delta_{EZ})\subsetneq \cE(g_{EZ})$.
	\end{example}
	
	\section{Alternative formulations of SDU}\label{sec:alternative}
	
	\subsection{A family of finite horizon problems}\label{ssec:finitefamily}
	Our approach to investment-consumption problems for EZ-SDU over the infinite horizon differs from the conventional approach in two important ways. First, we use the discounted aggregator given by \eqref{eq:Epstein--Zin aggregator} whereas the standard approach is to use the difference form. Second, we define the value function over the infinite horizon directly (with the natural transversality condition that the value process tends to zero in expectation following as a consequence), whereas the standard approach (formulated by Duffie, Epstein and Skiadas in the appendix to \cite{duffie1992stochastic}, and developed further by Melnyk et al~\cite{melnyk2020lifetime}) is to look for utility processes which solve a family of finite-horizon problems (where now the form of the transversality condition is not so clear, and may be part of the definition of a utility process). We have already compared the aggregators, so the goal in this section is to explain why we believe that it is better to define utility processes over the infinite horizon directly,
	and why, as a corollary, parameter combinations corresponding to $\theta<0$ cannot make economic sense.
	
	For the sake of exposition, we introduce some additional pieces of notation. Fix an aggegrator $g$ and $C \in \sP_+$. Then for  $T > 0$, let $\II_T(g,C) = \{ W \in \sP:  \int_0^T |g(s,C_s,W_s)| ds < \infty \}$ and $\JJ_T = \JJ_T(g,C)$ be a subset of $\II_T(g,C)$ such that elements of $\JJ_T$ have additional regularity and/or integrability properties. Let $\JJ \coloneqq \bigcap_{T > 0} \JJ_T$. Examples of suitable sets $\JJ_T$ will be given below.
	
	As an alternative to defining utility processes directly over the infinite horizon, \cite{duffie1992stochastic} and \cite{melnyk2020lifetime} define utility processes as solutions to a family of finite horizon problems.
	
	\begin{defn}\label{defn:family}
		$V$ is the $(\nu,\JJ)$-utility process associated to the consumption stream $C$ and generator $g$ if it has c\`adl\`ag paths, lies in $\JJ$, satisfies the transversality condition $\lim_{t \to \infty} e^{- \nu t} \E[| V_t |] = 0$, and for all $0\leq t \leq T < \infty$,
		\begin{equation}\label{eq:familyV}
			V_t = \E \left[  \left. \int_t^T  g(s, C_s,V_s) \dd s + V_T \right| \cF_t \right].
		\end{equation}
	\end{defn}
	
	\begin{rem}
		It follows as in Remark \ref{rem:utility process UI} that a $(\nu,\JJ)$-utility process is automatically a semimartingale.
	\end{rem}
	
	Let $\sE^{\nu,\JJ}(g)$ be the set of consumption streams $C$ such that there exists a $(\nu,\JJ)$-utility process associated to $C$ for aggregator $g$, and let $\sE_u^{\nu,\JJ}(g)$ be the subset of $\sE^{\nu,\JJ}(g)$, where there exists a exists a unique $(\nu,\JJ)$-utility process. Moreover, let $\sC_0(x)$ be some subset of $\sC(x)$, the set of attainable consumption streams from initial wealth $x$. Additional regularity conditions on the consumption streams may be encoded in $\sC_0$.
	
	In order to avoid the technical challenges of dealing with the infinite horizon problem directly, the idea in \cite{duffie1992stochastic, melnyk2020lifetime} is to replace the problem of finding $V(x)$ with the problem of finding $V_{\sC_0,\sE_u^{\nu, \JJ}(g)}(x) = \sup_{C \in \sC_0(x) \cap \sE_u^{\nu, \JJ}(g)} V^C_0$, for an appropriate transversality parameter $\nu$ and appropriate sets $\sC_0(x)$ and $\JJ$. But this immediately raises several issues. What exactly are the spaces $\sC_0(x)$, $\sE^{\nu,\JJ}(g)$ and $\sE_u^{\nu, \JJ}(g)$? How do we (easily) check whether $C \in \sC_0(x)$ and/or  $C \in \sE_u^{\nu,\JJ}(g)$?
	
	Regarding the choice of transversality condition, the issue crystalises as: first,
	how do we know that $\sE^{\nu,\JJ}(g)$ is non-empty?; second, how do we know that a utility process $V$ associated with a consumption $C$ makes economic sense?
	As regards the first issue, if $\nu < \nu'$, any $(\nu, \JJ)$-utility process is also a $(\nu', \JJ)$-utility process. Hence, $\sE^{\nu,\JJ}(g) \subseteq \sE^{\nu',\JJ}(g)$ and if $\nu$ is chosen too small, then it may easily follow that $\sE^{\nu,\JJ}(g)$ does not include the candidate optimal solution.
	As regards the second issue, in Section~\ref{ssec:transversalityandutilitybubbles} below we introduce the concept of a \textit{bubble solution} and argue that bubble solutions do not make economic sense.
	
	Duffie et al~\cite{duffie1992stochastic} impose Lipschitz-style conditions which exclude EZ-SDU. Melnyk et al~\cite{melnyk2020lifetime} do study EZ-SDU but the main focus of \cite{melnyk2020lifetime} is to understand the impact of market frictions on the investment-consumption problem for SDU-preferences. Nonetheless, in the frictionless case which is the subject of this paper, Melnyk et al prove some of the most complete results for Epstein--Zin preferences
	currently available in the literature. Melnyk et al~\cite{melnyk2020lifetime} only consider $R>1$ but this is mainly to limit the number of cases rather than because their methods do not extend to the general case. For the following definition, denote by
	
	\begin{defn}[Melnyk et al~\protect{\cite[Definition 3.1]{melnyk2020lifetime}}] Suppose $R>1$ and $\delta>0$.
		For $T > 0$, let
		\begin{eqnarray*} \bS^1_T & = & \{ V : V \in \sS {\mbox{ with $\E \left[ \sup_{0 \leq t \leq T} | V_t| \right] < \infty$}} \} \\
			\JJ^1_T & = & \bS^1_T  \cap \II_T(g^\Delta_{EZ},C). \\
			\JJ^2_T & = &  \big\{ V: V \in \JJ^1_T : V_t \leq - \tfrac{C_t^{1-R}}{R-1} \leq 0 \; {\mbox{ for all $0 \leq t \leq T$}} \big\}.
		\end{eqnarray*}
		For $k \in \{1, 2\}$ set $\JJ^k \coloneqq \bigcap_{T > 0} \JJ^k_T$  and let $\sC_0(x)$ be the set of $C \in \sC(x)$ for which there exists $\Pi$  such that $\Pi (X^{x, \Pi, C})^{1-R} \in \bS^1_T$ for all $T > 0$  and $\tfrac{1}{1-R}( X^{x, \Pi, C} )^{1-R} \in \JJ^1$. Moreover, if $0<\theta<1$, set $\JJ^{MMS} \coloneqq \JJ^1$ and $\sE^{MMS} = \sE^{MMS}(g^\Delta_{EZ}) \coloneqq \sE^{\delta \theta,\JJ^{MMS}}(g^\Delta_{EZ})$; if $\theta > 1$ or $\theta \in (-\infty,0)$, set $\JJ^{MMS} \coloneqq \JJ^2$ and $\sE^{MMS} = \sE^{MMS}(g^\Delta_{EZ}) \coloneqq \sE^{\delta,\JJ^{MMS}}(g^\Delta_{EZ})$.
		\label{def:MMS}
	\end{defn}
	Note that as we move from $\theta \in (0,1)$ to $\theta \notin (0,1)$ the transversality parameter $\nu$ changes from $\delta \theta$ to $\delta$. Moreover, an additional restriction that $V \leq -\frac{C^{1-R}}{R-1}$ is imposed.
	
	Melnyk et al~\cite{melnyk2020lifetime} take $b=\delta$. Then, from \eqref{eq:hat V candidate} we have that for $\eta>0$ the candidate value function is given by $\hat{V}(x)=\eta^{-\theta S} \delta^\theta \frac{x^{1-R}}{1-R}$.
	
	\begin{thm}[Melnyk et al~\protect{\cite[Corollary 2.3, Theorem 3.4]{melnyk2020lifetime}}] Suppose $R>1$ and $\delta>0$.
		Then $\sE^{MMS} = \sE_u^{MMS}$. Moreover, suppose $\frac{\mu- r}{R \sigma^2}\notin \{0, 1\}$ and $\eta > 0$.
		\begin{enumerate}[{\rm(}i{\rm)}]
			\item If $\theta \in (0,1)$ {\rm(}i.e. $1<R<S${\rm)}, then  $V_{\sC_0,\sE_u^{MMS}}(x) = \hat{V}(x)$.
			\item If  $\theta \in (1,\infty)$ {\rm(}i.e. $1<S<R${\rm)} and $ \frac{R-S}{R-1}\delta = \delta\rho < \eta < \delta$, then  $V_{\sC_0,\sE_u^{MMS}}(x) = \hat{V}(x)$.
			\item If  $\theta \in (-\infty,0)$ {\rm(}i.e. $S<1<R${\rm)}, then $\delta  < \eta < \delta\rho = \delta \frac{R-S}{R-1}$. Then,  $V_{\sC_0,\sE_u^{MMS}}(x) = \hat{V}(x)$.
		\end{enumerate}
		
	\end{thm}
	
	The results of Melnyk et al~\cite{melnyk2020lifetime} on the frictionless problem are amongst the few rigorous results on the investment-consumption problem over the infinite horizon. Nonetheless, they are incomplete in several respects. For all values of $\theta$, there is no existence result; although it is possible (at least under the conditions of the theorem) to verify that the candidate optimal consumption stream is a member of $\sC_0(x) \cap \sE_u^{MMS}$, in general little is said about which consumption streams are evaluable by Definition~\ref{def:MMS}, and it is unclear if the space of evaluable strategies goes beyond the set of constant proportional strategies.
	The fact that the wealth process must satisfy transversality and integrability conditions means that many plausible consumption streams are excluded by assumption, rather than because they are sub-optimal.
	
	When $\theta \notin (0,1)$ there are additional issues. In that case, the transversality condition in Definition~\ref{def:MMS} is that $\nu = \delta$. This condition leads to simple mathematics, but does not necessarily make economic sense---in Section~\ref{ssec:trans} we will argue that the economically-correct transversality condition is $\nu = \delta \theta$. Moreover, the restriction to consumption streams for which there exists a utility processes with $V \leq \frac{1}{1-R} C^{1-R}$  seems both hard to verify in general and hard to interpret.
	Finally, the analysis  in \cite{melnyk2020lifetime} leaves several parameter combinations uncovered, including the case $\{ \theta>1, \eta \in (0, \delta\rho] \cup [\delta , \infty) \}$.
	
	Although the space $\sE^{MMS}$ is difficult to describe, the following result, whose proof is given in Appendix~\ref{appendix:additional proofs}, says that if $C$ has an associated utility process in the sense of Melnyk et al, then automatically it has an associated utility process in the sense of a solution to \eqref{eq:stochastic differential utility aggregator g}. The converse is not true.
	
	\begin{prop}\label{prop:MMKS utility process is utility process}
		Suppose $\theta \in (0,1)$ or $\theta \in (1,\infty)$ and suppose $\delta>0$. Suppose $C \in \sE^{MMS}$ and let $V^\Delta$ be a $(\delta\theta, \JJ^{MMS})$-utility process associated to consumption stream $C$ and generator $g^\Delta_{EZ}$. Then, $V$ given by $V_t = e^{\delta\theta t} V^\Delta_t$ is a utility process associated to consumption stream $C$ and generator $g_{EZ}$ in the sense of Definition~\ref{defn:integrable set I(g,C)}. In particular, $\sE^{MMS}(g^\Delta_{EZ})\subset\sE(g_{EZ})$.
	\end{prop}
	
	Although Melnyk et al~\cite{melnyk2020lifetime} also define utility processes in the case $\theta<0$ we will argue that the solutions in this case do not make sense.
	
	\subsection{The transversality condition and utility bubbles in the additive case}\label{ssec:transversalityandutilitybubbles}
	
	Our goal is to show that, when coupled with the switch from the infinite horizon problem to the family of finite horizon problems approach, a mismatched transversality condition can lead to peculiar behaviour. We conclude that the modeller is not free to choose the transversality condition, at least in the framework of Definition~\ref{defn:family}, and electing to use the wrong condition can either rule out perfectly reasonable admissible strategies (and possibly rule out all strategies, including the candidate optimal strategy) or it can allow utility processes to be defined which have the characteristics of a bubble.
	
	In this section we consider the simpler case of time-additive CRRA utility. We will assume throughout this section that: the well-posedness condition $\eta_{a}\coloneqq\frac{\delta}{R} - \frac{1-R}{R} (r + \frac{\lambda^2}{2 R}) >0$ holds (see, for example, \cite[Corollary 6.4]{herdegen2020elementary}, for a discussion of the well-posedness of the Merton problem for additive utility); also, that $R>1$. The latter condition is only imposed to avoid case distinctions and similar behaviour is observed when $R<1$.
	
	In this case it is clear that for $g_{CRRA}$-evaluable consumption stream, the infinite horizon formulation
	\begin{align} \label{eq:value process infinite horizon discounted CRRA utility}
		V_t = & ~ \E \left[ \left. \int_t^\infty e^{-\delta s}\frac{C_s^{1-R}}{1-R} ds  \right| \cF_t \right], \quad  0\leq t  < \infty,
		\intertext{is equivalent to the finite horizon formulation:} \label{eq:value process finite horizon discounted CRRA utility}
		V_t = & ~ \cEX[t]{\int_t^T e^{-\delta s}\frac{C_s^{1-R}}{1-R} \dd s + V_T}, \quad 0\leq t \leq T < \infty,
	\end{align}
	if and only if the transversality condition $\lim_{T \to \infty}\E[V_T] = 0$ is met. Define $V^\Delta_t = e^{\delta t} V_t$.
	By arguing as in the proof of Proposition \ref{prop:MMKS utility process is utility process}
	(specialised to the case $\theta=1$), $V$ satisfies \eqref{eq:value process finite horizon discounted CRRA utility} if and only if ${V^\Delta}$ satisfies
	\begin{equation}
		\label{eq:value process finite horizon difference CRRA utility}
		V^\Delta_t = \cEX[t]{ \int_t^T \left( \frac{C_s^{1-R}}{1-R} - \delta V^\Delta_s\right)\dd s + V^\Delta_T }, \quad  0\leq t \leq T < \infty,
	\end{equation}
	where the transversality condition is $e^{-\delta t}\E[V^\Delta_t] \to 0$.
	
	The above observation suggests that the `correct' transversality condition for the problem with the difference aggregator is $e^{- \delta t}\E[V^\Delta_t] \rightarrow 0$. But, what happens if the transversality condition is modified to become  $e^{-\nu t}\E[V^\Delta_t] \to 0$ for some $\nu \neq \delta$?
	
	For $\hat{\pi} = \frac{\lambda}{\sigma R}$ and $\xi > 0$ with $H_\delta(\hat{\pi},\xi) = \delta + (R-1)({r} + \lambda \sigma \pi - \xi - \frac{\pi^2\sigma^2}{2}R)  >0$,  it follows from \eqref{eq:X^1-R equation constant strategy} that the constant proportional strategy with $\Pi \equiv \hat{\pi} $ and $C = \xi X$ satisfies
	$       \E[{C_t^{1-R}}] = \xi^{1-R} \E[{X_t^{1-R}}] = {\xi^{1-R}x^{1-R}} e^{(1-R)(r + \frac{\lambda^2}{2 R}  - \xi )t}$ and the solution to \eqref{eq:value process finite horizon discounted CRRA utility} is
	\begin{equation}\label{eq:VKX}
		V_t = V^\xi_t = \frac{K(\xi)}{1-R} e^{-\delta t}X_t^{1-R},
	\end{equation}
	where
	$       K(\xi) \coloneqq \frac{\xi^{1-R}}{H_\delta(\hat{\pi},\xi)} = \frac{\xi^{1-R}}{R \eta_{a} + (1-R) \xi}$. This implies that a solution to \eqref{eq:value process finite horizon difference CRRA utility} is given by
	\begin{equation}
		\label{eq:V^Delta CRRA}
		V^\Delta_t = V^{\Delta,\xi}_t = e^{\delta t} V_t = \frac{K(\xi)}{1-R} X_t^{1-R}.
	\end{equation}
	On the other hand, $e^{-\nu t} \E[V^{\Delta}_t] \to 0$ is equivalent to $e^{(\delta - \nu)t} \E[V_t] \to 0$, which in turn is equivalent to $H_\nu(\hat{\pi},\xi)>0$.
	We can therefore define the \textit{maximum} value of $\xi$ such that
	the transversality condition $e^{-\nu t} \E[V^{\Delta}_t] \to 0$ is satisfied. This is given by
	$$\xi^\nu_{\max} := \sup\{\xi > 0: \text{there is } {\pi\in\R} \text{ with } H_{\nu}(\pi, \xi)>0\}= \big(r +\tfrac{\lambda^2}{2} + \tfrac{\nu}{R-1}\big)_+<\infty.$$
	
\begin{figure}[ht]
		\begin{subfigure}{.5\textwidth}
			\centering
			\captionsetup{width=0.9\linewidth}
			\includegraphics[width=\linewidth]{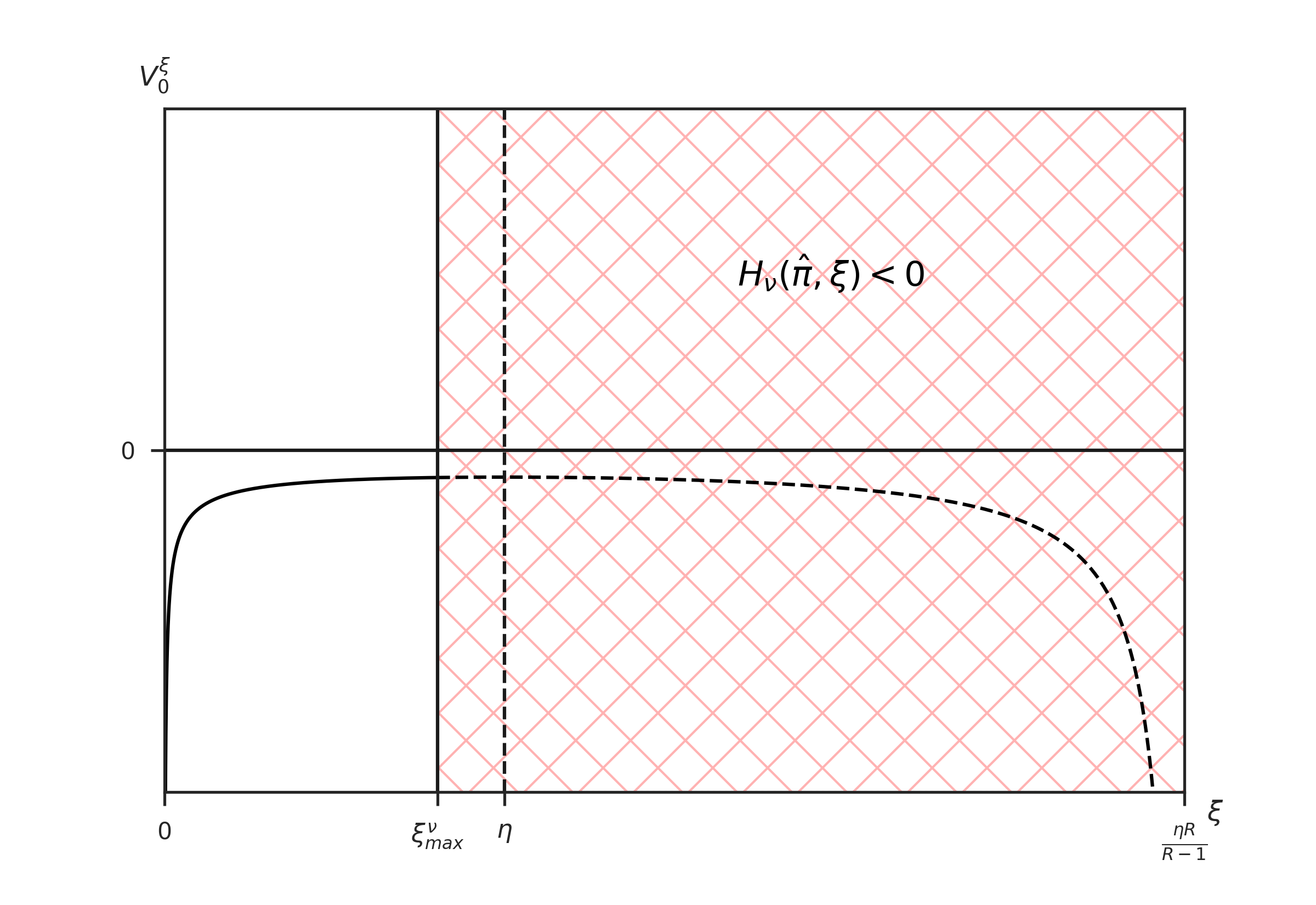}
			\caption{When the transversality condition is too small ($\nu<\delta$) the candidate optimal strategy may \emph{not be evaluable}.\\}
			\label{fig:wrong_nu_R<1}
		\end{subfigure}
		\begin{subfigure}{.5\textwidth}
			\centering
			\captionsetup{width=0.9\linewidth}
			\includegraphics[width=\linewidth]{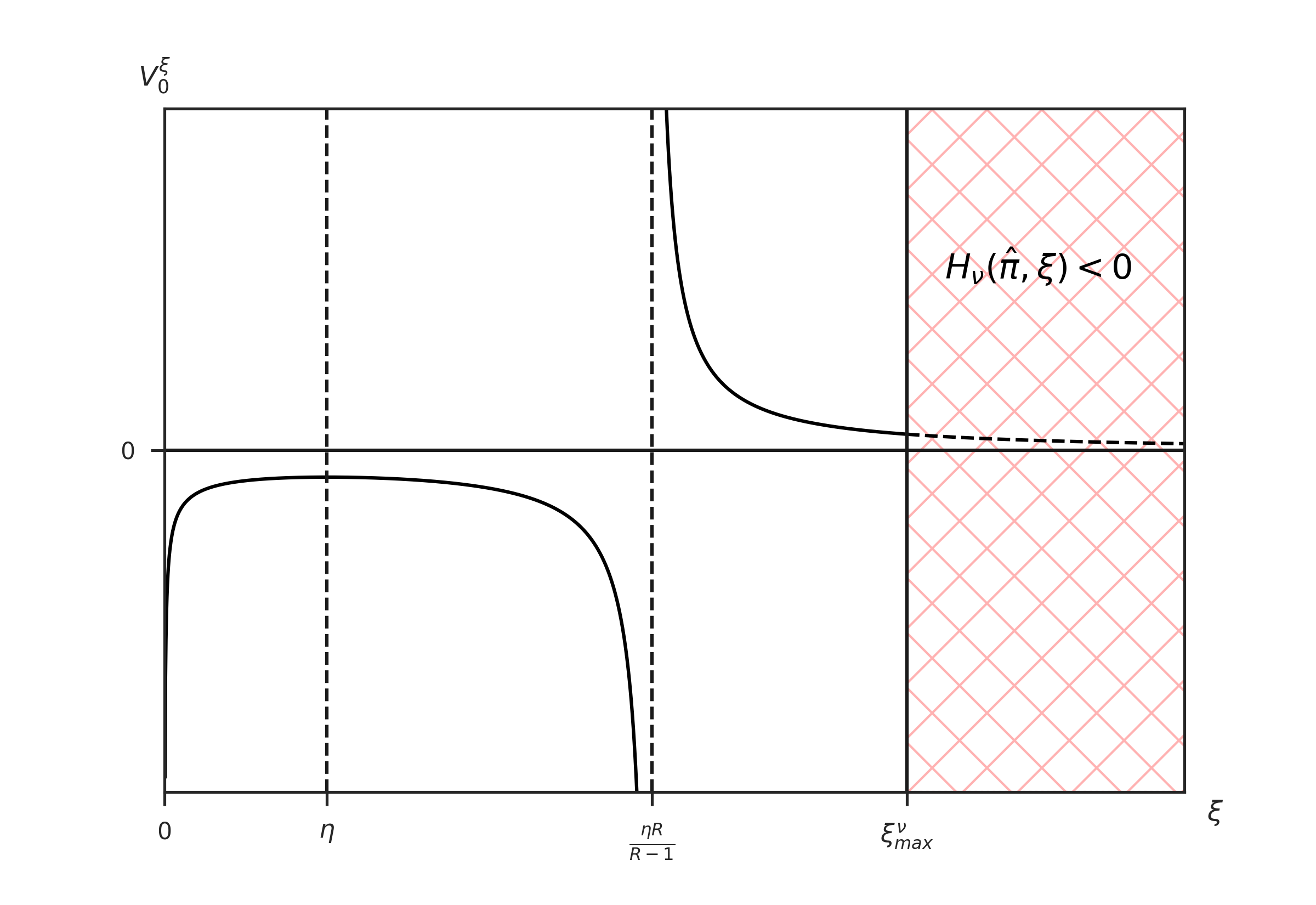}
			\caption{When the transversality condition is too large ($\nu>\delta$), the candidate optimal strategy is not optimal. Furthermore, some consumption streams lead to bubble-like utility processes.}
			\label{fig:wrong_nu_R>1}
		\end{subfigure}
		\caption{Plots of the solution to \eqref{eq:V^Delta CRRA} associated to the constant proportional investment-consumption strategy $(\hat{\pi},\xi)$ along with blocked out region where the transversality condition is not met ($H_{\nu}(\hat{\pi},\xi)\leq0$).}
		\label{fig:fig}
	\end{figure}

	First, consider a stronger transversality condition, $e^{-\nu t}\E[V^\Delta_t] \to 0$ for $\nu < \delta$. This means that $H_\delta(\hat{\pi},\xi)>H_\nu(\hat{\pi},\xi)$. In this case, if $H_\delta(\hat{\pi},\xi)>0 \geq H_\nu(\hat{\pi},\xi)$,
	or equivalently if $\xi$ is such that $R \eta_a > (R-1) \xi \geq \nu + (R-1) \big( r+\frac{\lambda^2}{2R}\big)$,
	then $V^\Delta$ defined in \eqref{eq:V^Delta CRRA} satisfies \eqref{eq:value process finite horizon difference CRRA utility} but it does not satisfy the transversality condition $e^{-\nu t}\E[V^\Delta_t] \to 0$. In particular, if $\eta_a > \xi^\nu_{\max}$
	then the candidate optimal strategy leads to a utility process which does not satisfy the transversality condition and hence does not lie in the set of consumption streams over which the optimisation takes place. This is illustrated in Figure \ref{fig:wrong_nu_R<1} for the case $R>1$ (but can also occur when $R<1$).
	
	Second, consider solving \eqref{eq:value process finite horizon difference CRRA utility} under a weaker transversality condition $e^{-\nu t}\E[V^\Delta_t] \to 0$ for $\nu > \delta$. In this case, $H_\nu(\hat{\pi},\xi)>H_\delta(\hat{\pi},\xi)$. Let $\xi\neq \frac{R \eta_{a}}{R-1}$ be such that $H_\nu(\hat{\pi},\xi)>0>H_\delta(\hat{\pi},\xi)$ (for example $\xi = \xi_\epsilon \coloneqq \frac{\delta + \epsilon}{R-1} + \big(r + \frac{ \lambda^2}{2 R}\big) = \frac{\epsilon+R \eta_{a}}{R-1}> 0$
	for $\epsilon \in (0, \nu - \delta)$). Again, it follows that $V^{\Delta,\xi_\epsilon}$ as defined \eqref{eq:V^Delta CRRA} solves \eqref{eq:value process finite horizon difference CRRA utility} for the constant proportional investment-consumption strategy $(\pi,\xi) = (\hat{\pi}, \xi_\epsilon)$. As $H_\nu(\hat{\pi},\xi_\epsilon)>0$, the transversality condition $e^{-\nu t}\E[V^{\Delta,\xi_\epsilon}_t] \to 0$ is met.
	
	Further $V^{\Delta,\xi_\epsilon}= - \frac{K(\xi_\epsilon)}{R-1} X^{1-R}$ where $K(\xi_\epsilon) = - \frac{\xi_\epsilon^{1-R}}{\epsilon}$.
	In particular, $V_0^{\xi_\epsilon} = \frac{\xi^{1-R}}{\epsilon}\frac{x^{1-R}}{R-1} > 0$. By comparison, $V_0^\eta = b \eta^{-\theta S} \frac{x^{1-R}}{1-R}<0$. Hence,
	the candidate optimal strategy does no longer maximise the initial value of the utility process over constant proportional strategies, in contradiction to the well-established theory for this case.
	
	In the case $R>1$ where we would expect to assign a negative utility, we may actually obtain an arbitrarily large \textit{positive} utility (see Figure \ref{fig:wrong_nu_R>1}). This can be done by letting $\epsilon\searrow0$ in the above. What is happening is that---whilst the integrand in \eqref{eq:value process finite horizon discounted CRRA utility} is always negative---the discounted expected future utility $\cEX[t]{V^\Delta_T}$ is diverging to positive infinity as $T \nearrow \infty$. The agent is always receiving a negative utility from consumption, but this is offset by an ever-increasing positive contribution from expectations of future utility. The endless optimism that things will always be better in the future creates bubble-like behaviour.
	
	Although there are special features in the additive case, the study of CRRA utility does show that some delicacy is needed when defining infinite horizon utility to be the solution to the finite horizon utilities paired with a transversality condition. If we wish to define stochastic differential utility in this manner, we must be very careful that we use the appropriate transversality condition.
	
	In preparation for the move beyond the additive case we record the following definition and proposition summarising the results of this section.
	\begin{defn}\label{def:bubbleCRRA}
		$V$ is a {\em CRRA-bubble} for a consumption stream $C$ if $V$ solves \eqref{eq:value process finite horizon discounted CRRA utility} for each $0 \leq t \leq T < \infty$ but $V$ and $U=U(t,C)$ are of opposite sign.
	\end{defn}
	\begin{prop}
		\begin{enumerate}[{\rm(i)}]
			\item For constant proportional strategies, there are no CRRA-bubbles which satisfy the transversality condition $e^{-\delta t} \E[V^\Delta_t] \rightarrow 0$.
			\item If $\nu < \delta$ then there is a financial market such that the candidate optimal investment-consumption strategy does not satisfy the transversality condition.
			\item If $\nu > \delta$, there is a financial market such that there is a consumption strean for which the associated utility process satisfies the transversality condition but is a CRRA-bubble. When $R>1$, the candidate optimal consumption stream does not maximise $V_0^C$ over attainable strategies.
		\end{enumerate}
	\end{prop}
	
	\subsection{Transversality, the case $\theta<0$, and the family of finite horizon problems.}\label{ssec:trans}
	
	For the EZ-SDU aggregator in discounted form over the infinite horizon it is not possible to define a utility process in the case $\theta<0$. However, several authors have attempted to define a utility process for $\theta<0$ using the difference form with the family of finite horizon problems approach or otherwise. Motivated by the analysis of the additive case, in this section we explain why the mathematical results they find may not have a sensible economic interpretation.
	
	The only strategies for which we can hope to find a non-trivial utility process in explicit form are constant proportional investment-consumption strategies. Moreover, the candidate optimal strategy is of this form. In consequence, and for this section only, we make the following assumption so we can see explicitly the issues which arise when $\theta<0$.
	
	\begin{tsass}
		\label{tsass:proportional}
		Consumption plans under consideration in this section are generated by constant proportional investment-consumption strategies $(\pi,\xi)$. If an associated utility process exists, then it is of the form $V^\Delta_t = B \xi^{1-R} \frac{X_t^{1-R}}{1-R}$ where $B=B(\pi,\xi)$ is a positive constant.
		If there is no solution of the form $V^\Delta_t = B \xi^{1-R} \frac{X_t^{1-R}}{1-R}$ for $B \in (0, \infty)$, then the consumption stream is not evaluable.
	\end{tsass}
	
	\begin{rem}
		Note that if $\theta \in (0,1)$, Corollary \ref{cor:uniqueness of utility process for g decreasing in v} below shows that if a utility process exists for a consumption stream  $C$, then it is unique. If $\theta \notin [0,1]$, then this need not be the case. In that case we must decide which utility process to assign to a given consumption stream. Typically the literature makes additional assumptions to ensure that the time-homogeneous solution $V^\Delta_t = B \xi^{1-R} \frac{X_t^{1-R}}{1-R}$ is the utility process
		associated with $C$, if such a solution exists. Without discussing what these assumptions might be, the impact of the temporary standing assumption is to assign the utility process $V^\Delta$ given by $V^\Delta_t = B \xi^{1-R} \frac{X_t^{1-R}}{1-R}$ to the constant proportional strategy.
	\end{rem}
	
	Consider $g^{\Delta}_{EZ}$ and a constant proportional investment-consumption strategy $(\pi,\xi)$. Suppose $V^\Delta = (V^{\Delta}_t)_{t \geq 0}$ is a solution to
	\begin{equation}\label{eq:t1}
		V^{\Delta}_t = \E \left[ \left. \int_t^T \left[ b \frac{\xi^{1-S} X_s^{1-S}}{1-S} \left((1-R) V^{\Delta}_s \right)^\rho - \delta \theta V^{\Delta}_s \right] \dd s + V^\Delta_T \right| \mathcal{F}_t \right]
	\end{equation}
	for all $0 \leq t \leq T < \infty$.
	We look for a solution of the form $V^\Delta_t = B \xi^{1-R} \frac{X_t^{1-R}}{1-R}$ where $B=B(\pi,\xi)$ is a positive constant which we seek to identify---we need $B \geq 0$ since we require $V^\Delta \in \VV$. For a constant proportional strategy $(\pi,\xi)$, we have  that $\E[ X_s^{1-R} | \cF_t] = X_t^{1-R} e^{-H_0(s-t)}$ where $H_0 = H_0(\pi,\xi)$ is as in \eqref{eq:H_nu} with $\nu = 0$. Then, substituting the candidate form for $V^\Delta$ into \eqref{eq:t1} and dividing by $\xi^{1-R} X_t^{1-R}$ yields
	\begin{equation}
		\label{eq:Bdef1}
		\frac{B}{1-R} = \int_t^T \left[ \frac{b}{1-S} B^\rho - \frac{\delta\theta B}{1-R} \right] e^{-H_0(s-t)} \dd s + \frac{B}{1-R} e^{-H_0(T-t)},
	\end{equation}
	and, provided $H_0(\pi,\xi) \neq 0$,
	\begin{equation}
		B = (b \theta B^\rho - \delta\theta B) \frac{1-e^{-H_0(T-t)}}{H_0(\pi,\xi)} + B e^{-H_0(T-t)}.
		\label{eq:Bdef2}
	\end{equation}
	It follows that there is a solution of the given form if there is a solution to
	\begin{equation}
		\label{eq:Bdef3}
		B H_{\delta\theta}(\pi,\xi) = B(\delta \theta + H_0(\pi,\xi)) = b \theta B^\rho,
	\end{equation}
	where $H_{\delta\theta}(\pi,\xi)$ is as in \eqref{eq:H_nu} with $\nu = \delta \theta$. (If $H_0(\pi,\xi)=0$, instead of \eqref{eq:Bdef2}, we get $B = (T-t) (b \theta B^\rho - \delta\theta B) + B$ which means that again $B$ solves \eqref{eq:Bdef3}.) Since $b>0$, there can only be a positive solution to \eqref{eq:Bdef3} if $\theta H_{\delta\theta}(\pi,\xi) > 0$.
	
	Note that already this is different to the additive case ($\rho=0$ and $\theta = 1$) in the way that it was presented in Section~\ref{ssec:transversalityandutilitybubbles}. In the additive case we (effectively) looked for solutions to $B(\delta + H_0(\pi,\xi)) = b$ but did not require that $B>0$; indeed we sometimes found (genuine) solutions with $B>0$ and sometimes bubble solutions with $B<0$. Solutions in the additive case with $B<0$ do not satisfy $V \in \VV$ and are automatically excluded when we consider utility processes in the EZ-SDU framework. We now argue that similar ideas mean that the case $\theta < 0$ does not make sense if bubble solutions are excluded.
	
	Suppose $\theta \neq 1$ (equivalently $\rho \neq 0$ or $R \neq S$) and consider non-negative solutions to \eqref{eq:Bdef3}.
	If $\theta \in (0,1)$ (equivalently $\rho < 0$), then this equation has a solution if and only if $H_{\delta\theta}(\pi,\xi)>0$ and then the solution is unique and given by $B = \big( \frac{b \theta}{H_{\delta\theta}(\pi,\xi)} \big)^{\theta}$. If $\theta > 1$, then $B=0$ is always a solution to \eqref{eq:Bdef3} (and so is $B = \infty$ if $H_{\delta\theta}(\pi,\xi)>0$) and there exists a strictly positive, finite solution if and only if $H_{\delta\theta}(\pi,\xi)>0$, whence again $B = \big( \frac{b \theta}{H_{\delta\theta}(\pi,\xi)} \big)^{\theta}$.  If $\theta < 0$, then $B=0$ is always a solution to \eqref{eq:Bdef3}, $B=\infty$ is a solution if $H_{\delta\theta}(\pi,\xi)<0$ and there exists a further solution if and only if $H_{\delta\theta}(\pi,\xi)<0$ whence $B = \left( \frac{b |\theta|}{|H_{\delta\theta}(\pi,\xi)|} \right)^{\theta}$. By the Temporary Standing Assumption, we exclude zero and infinity as solutions.
	
	For a constant proportional strategy $(\hat{\pi}=\frac{\lambda}{\sigma R},\xi)$, a change of accounting units will have the effect of changing the discount parameter. Fix $\delta$ and $g^\Delta_{EZ}$ but introduce also
	$g^\gamma = g^{\gamma}_{EZ}$ and $V^\gamma$ where $g^\gamma \coloneqq b \frac{c^{1-S}}{1-S} ((1-R)v)^\rho - \gamma \theta v$ and $V^\gamma = (V^\gamma_t)_{t \geq 0}$ is a solution to
	\begin{equation}\label{eq:t2}
		V^{\gamma}_t = \E \left[ \left. \int_t^T \left[ b e^{(\gamma - \delta)s}\frac{\xi^{1-S} X_s^{1-S}}{1-S} \left((1-R) V^{\gamma}_s \right)^\rho - \gamma \theta V^{\gamma}_s \right] \dd s + V^\gamma_T \right| \mathcal{F}_t \right]
	\end{equation}
	for all $0 \leq t \leq T < \infty$. (Then also $(g^\delta,V^\delta) \equiv (g^\Delta_{EZ},V^\Delta)$.)
	As before, we look for a solution of the form $V^\gamma_t = B_\gamma \xi^{1-R} \frac{X_t^{1-R}}{1-R}$ where $B_\gamma = B_\gamma(\pi,\xi) \in (0,\infty)$.

	\begin{lemma}\label{lem:accounting}
		Let  $(X^\gamma_t)_{t \geq 0}$ be given by $X^\gamma_t = X_t e^{-\frac{(\gamma-\delta)}{1-S} t}$ so that $X^\gamma$ is the wealth process which arises from a change of accounting unit.
		\begin{enumerate}[{\rm(}i{\rm)}]
			\item  $V^{\Delta}$ solves \eqref{eq:t1} if and only if $V^\gamma$ solves \eqref{eq:t2}.
			\item $V^\gamma$ solves \eqref{eq:t2} if and only if it also solves
			\begin{equation}\label{eq:t4}
				V^{\gamma}_t = \E \left[ \left. \int_t^T \left[ b \frac{\xi^{1-S} (X^\gamma_s)^{1-S}}{1-S} \left((1-R) V^{\gamma}_s \right)^\rho - \gamma \theta V^{\gamma}_s  \right] \dd s + V^\gamma_T \,\right|\, \mathcal{F}_t \right]
			\end{equation}
		\end{enumerate}
	\end{lemma}
	\begin{proof}
		The proof of {\rm(}i{\rm)} follows by a similar argument to the one used in the proof of Proposition~\ref{prop:MMKS utility process is utility process}. Statement {\rm(}ii{\rm)} is a simple renaming of variables.
	\end{proof}
	In particular, taking $\gamma = 0$, $V^0_t$ solves
	\begin{equation}\label{eq:t5}
		V^{0}_t = \E \left[ \left. \int_t^T  b \xi^{1-S} \frac{(X^0_s)^{1-S}}{1-S} \left((1-R) V^{0}_s \right)^\rho \dd s + V^0_T \,\right| \,\mathcal{F}_t \right].
	\end{equation}
	
	Considering solutions of \eqref{eq:t5} it is clear that the aggregator $g^0$ takes only one sign in the sense that (except possibly on the boundary where it may not be defined) either $g^0 : \R_+ \times \R_+ \times \VV \mapsto \R_+$ or $g^0 : \R_+ \times \R_+ \times \VV \mapsto \R_-$.
	\begin{defn}\label{def:bubbleEZ}
		$V$ is a \textit{bubble solution} for a consumption stream $C$ and generator $g$ if $V$ solves
		\begin{equation}\label{eq:bubbleg}
			V_t = \E \left[ \left. \int_t^T g(s,C_s,V_s) \dd s + V_T \,\right| \,\mathcal{F}_t \right]
		\end{equation}
		for each $0 \leq t \leq T < \infty$ and either $V \geq 0$ and $g \leq 0$ or $V \leq 0$ and $g\geq 0$ so that $V$ and $g=(g(s,C_s,V_s))_{s \geq 0}$ are of opposite sign.
	\end{defn}
	
	\begin{hypothesis} \label{hyp:H} There are no bubble solutions under any choice of accounting units.
	\end{hypothesis}
	
	\begin{thm}
		Under Hypothesis \ref{hyp:H} we must have $\theta>0$.
	\end{thm}
	
	\begin{proof}
		Consider the constant proportional strategy $(\pi,\xi)$.
		
		Suppose there exists a utility process $V^\Delta$ which solves \eqref{eq:t2}. Then, by Lemma \ref{lem:accounting}, we can switch accounting units so that $V^0$ solves \eqref{eq:t5}. There $g$ has one sign. Since there are no bubble solutions under any accounting units, $V^0$ is not a bubble and therefore has the same sign as $g^0$. Hence, $(1-S)V^0_t \geq 0$. Further, since the integral in \eqref{eq:t5} is monotonic in $T$ and $\E[V^0_T]$ always has exponential growth (or decay) for proportional investment-consumption strategies, we must have $\E[V^0_T] \rightarrow 0$.
		
		But, $\E[V^0_T] \rightarrow 0$ if and only if $e^{-\delta \theta t} \E[V^\Delta_t] \rightarrow 0$ which is equivalent to $H_{\delta\theta}(\pi,\xi)>0$.
		Since there exists a solution to \eqref{eq:Bdef3} if and only if $\theta H_{\delta\theta}(\pi,\xi) > 0$ it must be the case that $\theta>0$.
	\end{proof}
	
	Now we want to consider which transversality condition we should associate with \eqref{eq:t1}. Suppose the transversality condition is
	\begin{equation}
		\label{eq:T}
		e^{-\nu t} \E[ V^\Delta_t ] \rightarrow 0.
	\end{equation}
	It is easy to see that
	$\E[ e^{-\nu t} V^\Delta_t] \rightarrow 0$ if and only if $ e^{-(\nu - \delta\theta) t} \E[ e^{-\gamma \theta t} V^\gamma_t] \rightarrow 0$,
	and the transversality condition \eqref{eq:T} becomes $e^{-(\nu - \delta \theta)t} \E[V^0_t] \rightarrow 0$.
	
	\begin{hypothesis} \label{hyp:T}
		\begin{enumerate}[{\rm(}i{\rm)}]
			\item  The transversality condition associated with the aggregator $g$ should depend on the aggregator, but not on the financial market.
			\item Whenever the problem is well-posed, the utility process associated with the candidate optimal consumption stream satisfies the transversality condition \eqref{eq:T}.
		\end{enumerate}
	\end{hypothesis}
	
	\begin{prop}
		Under Hypothesis \ref{hyp:T} we must have that $\nu \geq \delta \theta$.
	\end{prop}
	
	\begin{proof}
		Suppose $\nu < \delta\theta$ and define $\epsilon = \delta\theta - \nu>0$.
		Then, the candidate optimal strategy $(\hat{\pi},\eta)$ satisfies the transversality condition $
		e^{-\nu t} \E[V^\Delta_t] \rightarrow 0$ if and only if it satisfies $e^{\epsilon t} \E[e^{-\delta\theta t} V^\Delta_t] \rightarrow 0$, which in turn is equivalent to $H_{\delta\theta}(\hat{\pi},\eta) > \epsilon $. Suppose the market parameters are such that $\eta \in (0,\frac{\epsilon}{\theta})$. Then, $H_{\delta\theta}(\hat{\pi},\eta) = \theta \eta < \epsilon$ and the candidate optimal utility process fails to satisfy the transversality condition.
	\end{proof}
	
	In general the larger the value of $\nu$, the weaker the admissibility condition and the more processes which will satisfy the transversality condition. However, for the Epstein--Zin aggregator, there is a point where increasing $\nu$ further makes no difference to the set of evaluable consumption streams.
	
	\begin{lemma} Fix $C$ and suppose that Hypothesis~\ref{hyp:H} holds.
		If there exists a solution $V^\Delta$ to \eqref{eq:t1}, then $V^\Delta$ satisfies \eqref{eq:T} for $\nu = \delta \theta$.
	\end{lemma}
	
	\begin{proof}
		$V^\Delta$ be a solution to \eqref{eq:t1}. Then
		by Lemma~\ref{lem:accounting}, $V^0$ solves \eqref{eq:t5}. Since $V^\Delta\in\VV$ and there are no bubble solutions, $\theta>0$ and $\E[V^0_t] \rightarrow 0$. Hence, $e^{-\delta \theta t} \E[V^\Delta_t] \rightarrow 0$.
	\end{proof}
	The final hypothesis says that we choose the smallest possible value for $\nu$ which allows us to evaluate all the strategies that we want.
	\begin{hypothesis} \label{hyp:T2}
		The transversality parameter should be the smallest parameter $\nu$ such that every solution to \eqref{eq:t1} satisfies \eqref{eq:T}.
	\end{hypothesis}
	
	\begin{prop}
		Under Hypotheses~\ref{hyp:H}, \ref{hyp:T} and \ref{hyp:T2}, the parameter $\nu$ in the transversality condition \eqref{eq:T} must take the value $\nu = \delta \theta$.
	\end{prop}
	
	\begin{rem}
		By construction there cannot be any bubble solutions in the infinite horizon discounted version. If $\E[|\int_t^\infty g(s,C_s,V_s) \dd s|]<\infty$ then
		$\E[V_T] \rightarrow 0$. Then, since $g$ has one sign, $V$ and $g$ must have the same sign.
	\end{rem}
	
	\begin{rem}
		For $\theta>1$, Melnyk et al~\cite{melnyk2020lifetime} take the transversality condition to be \eqref{eq:T} with $\nu=\delta < \delta \theta$. For some parameter values, the candidate optimal strategy may not be admissible because it fails the transversality condition. However, these parameter combinations are ruled out by the extra parameter restrictions imposed in \cite{melnyk2020lifetime}. In particular, \cite{melnyk2020lifetime} restrict attention to financial models for which $\eta > \delta \rho$. This is precisely enough to ensure that $e^{-\delta t} \E[ X_t^{1-R}] \rightarrow 0$ for the candidate optimal strategy. For $0 < \eta \leq \delta\rho$, the utility process for the candidate optimal strategy would fail the transversality condition. Further, both in the case $\eta > \delta \rho\geq 0$ and in the case $0 < \eta \leq \delta \rho$, many reasonable strategies are unnecessarily excluded because they fail the transversality condition, and not because they are suboptimal.
		
		For $\theta<0$ (and $R>1$), Melnyk et al~\cite{melnyk2020lifetime} define candidate solutions $V^\Delta$ as solutions to \eqref{eq:t1}. It follows that $V=(V_t)_{t \geq 0}$ given by $V_t = e^{-\delta \theta t}V^\Delta_t$ solves the family of finite horizon problems given in \eqref{eq:t5}. However, relative to the aggregator $g^0$, the solution $V$ is a bubble and would be ruled out by Hypothesis~\ref{hyp:T}.
		
		The same bubble feature can be observed without the switch in accounting units.
		For $\theta<0$, Melnyk et al~\cite{melnyk2020lifetime} define candidate solutions $V^\Delta$ of the form $V^\Delta_t =  - B \frac{1}{R-1} X_t^{1-R}$ where $B=B(\pi,\xi)$ solves \eqref{eq:Bdef2}. Since $H_{\delta\theta}(\hat{\pi},\hat{\xi}=\eta) = \eta\theta$, the condition $\eta>0$ implies that $\theta H_{\delta\theta}(\hat{\pi},\hat{\xi}) = \eta\theta^2>0$. Furthermore, the condition $\eta<\delta\rho$ ensures that $H_\delta(\hat{\pi},\hat{\xi}) = \eta\theta + \delta(1-\theta) = \theta (\eta - \delta\rho)>0$.
		Then, for $C_s = \eta X_s$, the proposed solution does indeed solve
		\begin{equation}\label{eq:SDUtTdifference}
			V^\Delta_t = \E \left[ \left. \int_t^T g^\Delta_{EZ}(C_s,V^\Delta_s) \dd s + V^\Delta_T \right| \cF_t \right]
		\end{equation}
		for all $0 \leq t \leq T < \infty$ together with the transversality condition $e^{-\delta t} \E[ X_t^{1-R}] \rightarrow 0$. However, \cite{melnyk2020lifetime} impose the additional admissibility condition $V^\Delta_s \leq -\frac{C_s^{1-R}}{R-1} \leq 0$ (which for the optimal strategy amounts to the condition $\eta>\delta$). This is precisely the condition under which $g^\Delta_{EZ}(C_s,V^\Delta_s) =\frac{\delta C_s^{1-S}}{1-S} ((1-R)V^\Delta_s)^\rho - \delta \theta V^\Delta_s \geq 0$ (recall that \cite{melnyk2020lifetime}  take $\delta = b$). Therefore, if $(C_s,V_s)$ is the candidate optimal strategy, it follows that $g^\Delta_{EZ}$ and $V^\Delta$ have the opposite sign, and so corresponds to a bubble, even in the original units.
	\end{rem}
	Due to the results in this section, we make the following standing assumption for the remainder of the paper.
	\begin{sass}\label{sass:theta>0} (Positive $\theta$ Assumption)
		The parameters $R$ and $S$ are such that $\theta= \frac{1-R}{1-S}>0$.
	\end{sass}
	
	\subsection{The dual approach}\label{ssec:dual}
	Dual methods have proved spectacularly successful for the Merton problem with additive utility. They work for general utility functions, and in principle they make it possible to move beyond the setting of constant parameter financial markets to non-Markovian settings and incomplete markets.
	However, it is not immediately clear how to extend dual methods to the SDU setting. One promising idea is based on stochastic variational utility as formulated by Dumas et al.~\cite{dumas2000efficient}.
	
	Building on work of Geoffard \cite{geoffard1996discounting} for deterministic consumption streams, \cite{dumas2000efficient} define the \textit{felicity function} $G$ to be the Fenchel--Legendre transform of the aggregator $g(c,v)$ in $v$, so that for $c>0$ and $(1-\theta) \nu>0$, $G(c,\nu) = \inf_{(1-R)u>0}(g(c,u) + \nu u)$. (\cite{dumas2000efficient} assume that $g$ is convex in its second argument, but a similar argument works if $g$ is concave.)
	Then, the \textit{stochastic variational utility} (SVU) is given by
	\begin{equation}
		\label{eq:stochastic variational utility dumas}
		U^C_t \coloneqq \sup_{(1-\theta)\nu > 0}\cEX[t]{\int_t^T e^{-\int_t^s \nu_u \dd u} G(C_s,\nu_s) \dd s + U_T(X_T)},
	\end{equation}
	where $U_T(\cdot)$ is a bequest function. \cite{dumas2000efficient} consider consumption streams $C$ that satisfy $\E[\int_0^T C_t^2 \dd t]<\infty$ and aggregators $g(c,v)$ that have linear growth in $c$ and are Lipschitz in $v$. Then, under these conditions, they show that $U$ is the stochastic variational utility associated to the pair $(g,c)$, if and only it is the finite horizon, stochastic differential utility associated to the pair $(g,c)$.
	
	Matoussi and Xing~\cite{matoussi2018convex} take the approach of \cite{dumas2000efficient} and extend it to the case of Epstein--Zin SDU in the finite horizon case. They show that if $\theta < 1$ and the consumption stream is such that a utility process exists and is uniformly integrable, then the solution to \eqref{eq:SDUtTdifference} is equal to the solution to \eqref{eq:stochastic variational utility dumas} for $G$ the Fenchel--Legendre transform of ${g}^\Delta_{EZ}$ and $V_T = U_T(X_T)$.
	
	Exploiting the equivalence of \cite{dumas2000efficient} between SDU and SVU, \cite{matoussi2018convex} show that if the bequest function is of an appropriate power law form, the maximisation problem of finding $\sup_{C \in \sC(x)\cap\sE_u(g)}V^C_0$ where $V^C$ solves \eqref{eq:SDUtTdifference} becomes that of finding $\sup_{C \in \sC(x)\cap\sE_u(g)}U^C_0$, where $U^C$ solves \eqref{eq:stochastic variational utility dumas}. Exchanging the order of suprema, the problem becomes to find
	\begin{equation}\label{eq:inner prob Matoussi-Xing}
		\sup_{(1-\theta)\nu > 0}\sup_{C \in \sC(x)\cap\sE_u(g)}\cEX[t]{\int_t^T e^{-\int_t^s \nu_u \dd u} G(C_s,\nu_s) \dd s + U_T(X_T)}.
	\end{equation}
	
	For EZ-SDU both $G(\cdot, \nu)$ and $U_T$ are power law functions, and hence standard duality techniques can be applied to the inner problem in \eqref{eq:inner prob Matoussi-Xing} with fixed $\nu$. Finally, by taking the dual with respect to the second argument again, the dual stochastic variational problem can be transformed back into what Matoussi and Xing call the \textit{stochastic differential dual}. They then prove that
	\begin{equation}\label{eq:duality inequality Matoussi-Xing}
		\sup_{C \in \sC(x)\cap\sE_u(g)}V_0^C \leq \inf_{k>0}\left(\inf_{D\in\sD_a} Y_0^{kD} + xk\right).
	\end{equation}
	where $\sD_a$ is the class of state-price densities and $Y^{kD}$ is the stochastic differential dual associated to a state-price density $D$ and a positive real number $k$.
	Matoussi and Xing show that under certain restrictions on the financial market (for example, bounded market price of risk) there is no duality gap and that \eqref{eq:duality inequality Matoussi-Xing} is satisfied with equality. Finally, they show that the optimal strategy is defined in terms of a BSDE and in particular it exists.
	
	The papers of Dumas et al~\cite{dumas2000efficient} and especially Matoussi and Xing~\cite{matoussi2018convex} provide great insights and a potential route-map describing how dual methods might be extended to the investment-consumption problem for SDU. However, there are several obstacles which make it difficult to apply these ideas to the infinite horizon problem. First, at present, the dual method has little to say about existence of solutions, and typically for existence it relies on results from the primal approach---in turn these have traditionally involved imposing restrictive assumptions on the consumption stream which are not satisfied in the infinite horizon problem.  Second, the equivalence between the SDU and SVU formulations may be challenging to prove in the infinite horizon setting, without imposing substantive technical assumptions. Third, we shall see that there are major issues of non-uniqueness when $\theta>1$; these issues do not disappear simply by a change of viewpoint.
	
	\subsection{Summary}\label{ssec:summary}
	The conclusions from Part~\ref{part:intro} are twofold.
	
	First, for Epstein--Zin stochastic differential utility over the infinite horizon combined with a constant parameter Black--Scholes--Merton frictionless financial model, certain restrictions on the parameters are necessary to have a well-founded problem. In particular, in addition to $b>0$, for the problem to make sense it must the case that the
	coefficient of relative risk aversion and the coefficient of elasticity of intertemporal complimentarity both lie on the same side of unity, i.e. $\theta>0$. (However, the condition that the discount parameter $\delta$ must be positive can sometimes be weakened. Indeed, since this parameter depends on the accounting units it is sometimes natural to consider a case where it takes a negative value.)
	
	Second, for the infinite horizon problem, it is preferable  to consider a discounted aggregator rather than a difference aggregator. The one-sign property of the discounted-form EZ-SDU aggregator means that the integral $\int_0^\infty g(s,C_s,V_s) \dd s$ and its expectation are always well defined in $[-\infty,\infty]$ whereas this is not always the case for the difference aggregator. Then, in addition to the fact that the discounted aggregator is the natural generalisation of the standard form of the Merton problem for additive utility, for the discounted aggregator there are no issues over bubble solutions. In the second part of this paper we shall strengthen this result further by showing that, at least when $\theta \in (0,1)$, for the aggregator of discounted form it is possible to define a (generalised) utility process for {\em every} consumption stream. This means that we can prove the optimality of the candidate optimal strategy within the class of {\em all} admissible investment-consumption strategies, and not just a subclass satisfying certain integrability properties.
	
	\part{Existence and uniqueness results}\label{part:existence and uniqueness results}
	
	Our goal in Part \ref{part:existence and uniqueness results} of the paper is to prove well-posedness of the investment-consumption problem under Epstein--Zin stochastic differential utility and verify that the candidate optimal investment-consumption strategy we derived in
	Section \ref{ssec:candidate optimal utility process} is optimal. There are three main issues which we must address: first, the existence of a utility process associated to a general consumption stream; second, the uniqueness of such a utility process; and third, optimality of the \textit{candidate} optimal investment-consumption strategy.
	
	Our results and approach are as follows. From the arguments in Section \ref{sec:BSM}, we have existence of a utility process for admissible consumption stream where the investment and consumption processes are proportional to wealth (provided that $H_{\delta\theta}(\pi,\xi)>0$) in a Black--Scholes--Merton financial market. The first major contribution is an extension of the existence result to all strictly positive consumption streams $C=(C_t)_{t \geq 0}$ which satisfy $kC^{1-R}_t \leq
	\cEX[t]{\int_t^\infty e^{-\delta \theta (s-t)}{C}_s^{1-R} ds}\leq K C_t^{1-R}$, for some constants
	$0 < k \leq K < \infty$. In particular, we may evaluate strategies that are, in a very precise sense, within a multiplicative constant of a constant proportional investment-consumption strategy. Moreover, for each such $C$ there is a unique utility process $V=(V^C_t)_{t\geq 0}$ such that $k_V C_t^{1-R} \leq V_t \leq K_V C^{1-R}_t$ for a different pair of constants $(k_V,K_V)$. (Note that this does not preclude the existence of other utility processes which do not satisfy such bounds.) The proof relies on the construction of a contraction mapping and a fixed point argument.
	
	To make further progress, we assume that $\theta \in (0,1)$ (equivalently, $\rho < 0$). In this case, we can show that any utility process is unique (in fact we show uniqueness for a wide class of aggregators, the main restriction being that they are decreasing in $v$). The key idea is to use concepts from the theory of BSDEs to extend the concept of a solution to \eqref{eq:Epstein--Zin SDU} to include subsolutions and supersolutions, depending (roughly speaking) on whether the equality in \eqref{eq:Epstein--Zin SDU} is replaced by $\leq$ or $\geq$. Then, again under the assumption that the aggregator is decreasing in $v$, we prove a comparison theorem which tells us that any subsolution always lies below any supersolution. Uniqueness of solutions follows---any solution is simultaneously both a sub-solution and a super-solution so if $V^1$ and $V^2$ are solutions then $V^1 \leq V^2$ and $V^2 \leq V^1$ and hence $V^1 = V^2$.
	
	For EZ-SDU, when $\theta>1$ the comparison argument fails and the uniqueness argument does not hold. Note that it is not merely that we need to look for a different strategy of proof---instead, it is simple to give examples for which there are multiple solutions to \eqref{eq:Epstein--Zin SDU}. In this case, a different comparison theorem and a modification of the definition of the utility process is required. For these reasons, we defer discussion of this case to a later paper.
	
	Returning to the case of $\theta \in (0,1)$, in order to remove the constraints $k>0$ and $K<\infty$ we again exploit the comparison theorem to obtain a monotonicity property for solutions. Provided we allow utility processes to take values in the extended real line, we can exploit the fact that the aggregator takes one sign to show that it is possible to define a unique, possibly infinite, utility process for \textit{any} admissible consumption stream. Here we make use of the notion of generalised supermartingales.
	
	Finally, still under the assumption that $\theta\in(0,1)$, we turn to the verification argument. By the arguments of the previous paragraphs, for any attainable consumption stream $C=(C_t)_{t \geq 0}$, we can define a utility process $V^C = (V_t^C)_{t \geq 0}$ and time-zero value $J(C) = V^C_0$. Our goal is to find $\sup_{C \in \sC(x)} J(C) $. Note that here the supremum is taken over \textit{all} admissible consumption stream; not just over consumption streams for which there exists a finite value function, or consumption/utility process pairs lying in some special set as is common in much of the literature. (In many cases, the only strategies/utility processes known to lie in this special set are those derived from constant proportional investment and consumption.)
	
	From the results of Section \ref{ssec:candidate optimal utility process}, we have candidates for the optimal strategy and value function, but several issues remain.
	The key is proving that $\hat{V}(X^{\Pi,C})= (\hat{V}(X^{\Pi,C}_t))_{t\geq0}$ is a supersolution for any admissible $C$ where $X^{\Pi, C}$ is the wealth process arising from the investment-consumption strategy $(\Pi, C)$. Then, by the comparison theorem $V^C_t \leq \hat{V}(X^{\Pi,C}_t)$ and $J(C) = V^C_0 \leq \hat{V}(x)$. (Further, for $(\hat{\Pi},\hat{C})$ the candidate optimal investment-consumption strategy, $V^{\hat{C}}_0 = \hat{V}(x)$ and so $\sup_{C \in \sC(x)} J(C) = \hat{V}(x)$.) However, as in the case of rigorous primal verification arguments for the Merton problem, there are several challenges to overcome. First $X_t^{\Pi,C} \in \sP_+$ but is not necessarily a member of $\sP_{++}$ and so we cannot naively apply It\^{o}'s formula to $\hat{V}(X^{\Pi,C}_t)$. Second, for general $(\Pi, C)$, $\hat{V}(X^{\Pi, C})$ does not (always) satisfy a transversality condition (and we do not want to artificially restrict the class of admissible $C$ by requiring that it does). Third, the local martingale term arising from applying It\^{o}'s formula to $\hat{V}(X^{\Pi, C}_t)$ is in general not a true martingale and cannot be assumed to have constant expectation. Nonetheless, as we show, these challenges can all be overcome. The key idea is a perturbation argument applied to the Merton problem in \cite{herdegen2020elementary}.
	
	Where proofs are not given in the main text, they are given in the appendices.
	
	\section{Existence of Epstein--Zin SDU}\label{sec:existence}
	
	For the Epstein--Zin aggregator $g_{EZ}$ we showed in Section \ref{ssec:candidate optimal utility process} that the candidate optimal strategy---along with many other proportional consumption streams---is evaluable.
	The goal of this section is to prove existence for a much larger class of consumption streams. The authors are not aware of any results on the existence of infinite horizon Epstein--Zin stochastic differential utility, so this is an essential result that is currently missing from the literature.
	
	A transformation of the coordinate system leads to a simplified problem. Define the $[0, \infty]$-valued processes $W = (W_t)_{t\geq0}$ and $U = (U_t)_{t\geq0}$ by\footnote{Here, we agree that $U_t \coloneqq \infty$ if $C_t = 0$ and $S > 1$.}
	\begin{equation}\label{eq:W and U from V and C}
		W_t = (1-R)V_t \quad \text{and} \quad U_t = u(t,C) = b \theta e^{-\delta t} C_t^{1-S}.
	\end{equation}
	Let $h_{EZ}(u,w): [0, \infty) \times (0, \infty) \to [0, \infty)$ be defined by $h_{EZ}(u,w) = uw^\rho$ and extend the definition of $h_{EZ}$ to the domain $[0,\infty]^2$ and co-domain $[0, \infty]$ as follows:
	\begin{equation}
		h_{EZ}(u,w) \coloneqq \left\{
		\begin{array}{rl}
			u w^\rho, & (u,w) \in (0,\infty)\times(0,\infty),
			\\
			w^\rho,   & (u,w) \in (0,\infty) \times \{0, \infty\}, \\
			u,        & (u,w) \in \{0,\infty \} \times [0,\infty],
		\end{array}
		\right.
	\end{equation}
	with the standard convention $0^\rho \coloneqq \infty$ and $\infty^\rho = 0$ for $\rho < 0$.
	The motivation behind the definition on the boundary is to ensure continuity in $w$ for fixed $u$.
	
	Note that $V \in \II(g_{EZ},C)$ if and only if $ W \in \II(h_{EZ}, U)$. Consequently, $V^C$ is a utility process associated to consumption stream $C$ with aggregator $g_{EZ}$ if and only if $W^{U}$ is a utility process associated to consumption stream $U$ with aggregator $h_{EZ}$.
	
	We next aim to define an operator $F_{U}$ from an appropriate subset of $\sP_{++}$ to itself satisfying\footnote{Here, we always choose a  c\`adl\`ag version for the right-hand side of \eqref{eq:fixed point operator F}.}
	\begin{equation}\label{eq:fixed point operator F}
		F_{U}(W)_t \coloneqq \E\left[\left.\int_t^\infty h_{EZ}(U_s,W_s) \dd s \right| \cF_t \right].
	\end{equation}
	Note that $V$ is a solution to \eqref{eq:stochastic differential utility aggregator g} with aggregator $g_{EZ}$ and consumption $C$ if and only $W$ is a fixed point of the operator $F_{U}$ for the transformed consumption $U$. In particular, every fixed point of the operator $F_{U}$ has c\`adl\`ag paths.
	
	\begin{defn}
		Suppose that $U=(U_t)_{t\geq0} \in \sP_+$ and $Y=(Y_t)_{t\geq0} \in \sP_+$. We say that $U$ has the same order as $Y$ if there exist constants $k,K\in(0,\infty)$ such that
		\(      0\leq k Y \leq U \leq K Y.\)
		Denote the set of processes with the same order as $Y$ by $\OO(Y)$.
	\end{defn}
	
	\begin{defn}
		Define $L^\theta_{++}$ to be the subset of all $\Lambda\in\sP_{++}$ such that $\EX{\int_0^\infty \Lambda^\theta_s \dd s}<\infty$.
		For $\Lambda\in L^\theta_{++}$, we may define the c\`adl\`ag process $I^\Lambda = (I^\Lambda_t)_{t\geq0}$ by
		\(I^\Lambda_t \coloneqq \cEX[t]{\int_t^\infty \Lambda_s^\theta \dd s}.
		\)
		Further, define $\hat{L}^\theta_{++}\subseteq L^\theta_{++}$ by $\hat{L}^\theta_{++} = \{\Lambda\in L^\theta_{++}:~ \Lambda^\theta \in \OO(I^\Lambda)\}$.
	\end{defn}
	
	\begin{example}\label{rem:geometric Brownian motion in Ltheta++}
		Geometric Brownian motion raised to a power remains a geometric Brownian motion. Let $Z = (Z_t)_{t\geq0}$ be a geometric Brownian motion such that $Z^\theta$ has drift $\gamma<0$. Then, $Z^\theta = \frac{1}{\gamma}I^Z$. Hence, $Z\in\hat{L}^\theta_{++}$.
		
		If $\eta>0$ and if $\hat{C}$ is the candidate optimal strategy, then $\hat{U} = u(t,\hat{C})$ is a geometric Brownian motion, and $(\hat{U})^\theta$ has drift $-\eta<0$. Hence, $\hat{U}\in\hat{L}^\theta_{++}$. Similarly, all the constant proportional investment-consumption strategies $(\pi,\xi)$ with $H_{\delta\theta}(\pi,\xi)>0$ lie in $\hat{L}^\theta_{++}$ (after a suitable transformation). Roughly speaking, the same holds true for any strategy which is close to a constant proportional strategy (for which $H_{\delta \theta}(\pi,\xi)>0$).
	\end{example}
	
	\begin{lemma}
		Let $\Lambda\in\hat{L}^\theta_{++}$ and $U\in \OO(\Lambda)$.    Then, $F_U(\cdot)$ maps from $\OO(\Lambda^\theta)$ to itself.
	\end{lemma}
	\begin{proof}
		This is follows from the more general Lemma \ref{lem:F^eps maps from O(Delta^theta) to itself} in Appendix \ref{app:existence}.
	\end{proof}
	
	We may now state a first existence result. Whilst it is not the strongest existence result we prove in this paper, (Theorem~\ref{thm: existence of a solution, first result} is a special case of Theorem~\ref{thm:existence of a F^epsilon solution, bounded case}) it forms the backbone of further existence arguments. The idea of the proof is to transform the problem to an alternative space where the transformed form of $F_{U}$ is a contraction mapping. The existence of a fixed point then follows from the Banach Fixed Point Theorem.
	\begin{thm}\label{thm: existence of a solution, first result}Let $\Lambda\in\hat{L}^\theta_{++}$ and $U\in\OO(\Lambda)$.        Then, $F_U$ defined by \eqref{eq:fixed point operator F} has a unique fixed point $W \in \OO(\Lambda^\theta) \subseteq \II(h_{EZ}, U)$, which has c\`adl\`ag paths.
	\end{thm}
	
	\begin{proof}
		This is a specific version of the more general Theorem \ref{thm:existence of a F^epsilon solution, bounded case}.
		For a stand-alone proof, one just needs to set $\epsilon=0$ in the proof of Theorem \ref{thm:existence of a F^epsilon solution, bounded case}.
	\end{proof}
	The following theorem is a direct corollary to Theorem \ref{thm: existence of a solution, first result} and the definitions of $W$ and $U$ in terms of $V$ and $C$ given in \eqref{eq:W and U from V and C}.
	\begin{thm}
		\label{cor:evaluableconsumptionsbycloseness}
		Suppose $C\in\sP_{++}$ satisfies $\E[\int_0^\infty e^{-\delta \theta s} C_s^{1-R} ds] < \infty$, and for some $0<k<K<\infty$,
		\begin{equation}\label{eq:C bounded existence}
			k\cEX[t]{\int_t^\infty e^{-\delta \theta s} C_s^{1-R} ds} \leq e^{-\delta\theta t}C_t^{1-R} \leq K\cEX[t]{\int_t^\infty e^{-\delta \theta s} C_s^{1-R} ds}
		\end{equation}
		for all $t\geq0$. Then, there exists a utility process $V=(V^C_t)_{t \geq 0}$ associated with $g_{EZ}$ and $C$. Moreover this utility process is unique in the class of processes with the property that $V_t/\cEX[t]{\int_t^\infty e^{-\delta \theta s} C_s^{1-R} ds}$ is bounded above and below by strictly positive constants.
	\end{thm}
	\begin{proof}
		Take $U_t=\Lambda_t=e^{-\delta t}C_t^{1-S}$. Then,  $U$ satisfies the conditions of Theorem \ref{thm: existence of a solution, first result} and so there exists a utility process $W$ associated to $(h_{EZ},U)$ which is unique in $\OO(\Lambda^\theta)$. Therefore, $V = \frac{W}{1-R}$ is a utility process associated to $(g_{EZ},C)$; uniqueness in the appropriate class is also inherited.
	\end{proof}
	
	Relative to the extant literature, Theorem \ref{cor:evaluableconsumptionsbycloseness} massively expands the set of consumption streams which are known to be evaluable. However, it still does not allow us to assign a utility to every consumption stream. For example, the zero consumption stream is excluded. Note also that Theorem \ref{cor:evaluableconsumptionsbycloseness} does not exclude the possibility of other utility processes which do not satisfy the condition that
	$V_t/\cEX[t]{\int_t^\infty e^{-\delta \theta s} C_s^{1-R} ds}$ is bounded.
	
	\section{Subsolutions and supersolutions}\label{sec:sub and supersolutions}
	
	The aim of this section is to introduce the notions of subsolutions and supersolutions and then to prove a comparison theorem for aggregators that take only one sign and are nonincreasing in $v$. As a consequence, all evaluable consumption streams for such aggregators are \emph{uniquely} evaluable.
	
	Let $\VV\subseteq  [-\infty,\infty]$ denote the set in which $V$ may take values.
	For EZ-SDU we have that either $\VV\subseteq \ol \R_+$ or $\VV\subseteq \ol \R_-$. This one-sign property ensures that integrals are always well defined. From now on we make this a standing assumption.
	\begin{sass}[One-sign property of the aggregator]\label{sass:VV pos or neg}
		Either $\VV\subseteq \ol \R_+$ or $\VV\subseteq \ol \R_-$.
	\end{sass}
	
	The following definition extends the notion of an aggregator, allowing it also to depend on the state of the world $\omega\in\Omega$.
	\begin{defn}
		An \textit{aggregator random-field} $g:[0,\infty) \times \Omega \times \R_+ \times \VV \to \VV$ is a product measurable mapping such that
		$g(\cdot,\omega,\cdot,\cdot)$ is an aggregator for fixed $\omega\in\Omega$, and for progressively-measurable processes $C=(C_t)_{t\geq0}$ and $V=(V_t)_{t\geq0}$, the process $(g(t,\omega,C_t(\omega),V_t(\omega)))_{t\geq0}$ is progressively-measurable.
		
	\end{defn}
	
	\begin{example}
		Let $G:\R_+\times\VV\times \R\to\VV$ be continuous and $Y:[0,\infty)\times\Omega\to\R$ a progressively measurable process.
		Then \(   g(t,\omega,c,v) \coloneqq G(c,v,Y(t,\omega))\) is an aggregator random field.
	\end{example}
	
	Let $g$ be an aggregator random field. The definitions of $\II(g,C)$, $\UU\II(g,C)$, the utility process associated to the pair $(g,C)$, and the sets of evaluable and uniquely evaluable consumption streams $\sE(g)$ and $\sE_u(g)$ follow verbatim from Definitions \ref{defn:integrable set I(g,C)} and \ref{def:evaluable}.
	
	We now introduce the notion of subsolutions and supersolutions.
	
	\begin{defn}
		Let $C \in \cP_+$ and $g$ be an aggregator random field. A $\VV$-valued, l\`ad, optional process $V$ is called
		\begin{itemize}
			\item  a \textit{subsolution} for the pair $(g,C)$  if $\limsup_{t\to\infty}~ \EX{V_{t+}} \leq 0$ and for all bounded stopping times ${\tau_1}\leq{\tau_2}$,
			\begin{equation}\label{eq:subsolution equation}
				V_{\tau_1} ~\leq~ \cEX[{\tau_1}]{V_{{\tau_2}{+}} +  \int_{\tau_1}^{{\tau_2}} g(s,\omega, C_s,V_s)\dd s}.
			\end{equation}
			\item   a \textit{supersolution} for the pair $(g,C)$  if $\liminf_{t\to\infty}~ \EX{V_{t+}} \geq 0$ and for all bounded stopping times ${\tau_1}\leq{\tau_2}$,
			\begin{equation}\label{eq:supersolution equation}
				V_{\tau_1} ~\geq~ \cEX[{\tau_1}]{V_{{\tau_2}{+}} +  \int_{\tau_1}^{{\tau_2}} g(s,\omega, C_s,V_s)\dd s}.
			\end{equation}
			\item a \emph{solution} for the pair $(g,C)$ if it is both a subsolution and a supersolution and $V\in\II(g,C)$.
		\end{itemize}
	\end{defn}
	\begin{rem}
		(a) $V$ is a supersolution associated to the pair $(g,C)$ if and only if $\tilde V \coloneqq -V$ (which is valued in $\tilde \VV \coloneqq-\VV$) is a subsolution for the pair $(\tilde g, C)$, where $\tilde g (t,\omega, c, \tilde v) = - g (t,\omega, c, -\tilde v)$.        \\
		(b) While we do not to require sub- or supersolutions to be in $\II(g,C)$, we require this integrability for solutions. \\
		(c) It might be expected that the definition would require subsolutions and supersolutions to be c\`adl\`ag. However, we will construct the utility process for a general consumption stream by taking limits and the monotone limit of c\`adl\`ag processes is not necessarily c\`adl\`ag. In contrast, optionality is preserved in the limit.
	\end{rem}
	
	If $V$ is a utility process for the pair $(g,C)$, then $V\in\II(g,C)$ by definition. By Remark \ref{rem:utility process UI} it then follows that $V$ is uniformly integrable. Similar results hold for sub- and supersolutions.
	
	\begin{lemma} \label{lem:UIofsupersolutions}
		Suppose that $\VV \subseteq \ol \R_+$ and $V$ is a subsolution or $\VV \subseteq \ol \R_-$ and $V$ is a supersolution for the pair $(g, C)$. If $V\in\II(g,C)$ then $V\in\UU\II(g,C)$.
	\end{lemma}
	
	\begin{proof}
		We only consider the case that $\VV \subseteq \ol \R_+$ and $V \in\II(g,C)$ is a subsolution. The other case is symmetric.
		Define the UI martingale $M = (M_t)_{t \geq 0}$ by $M_t \coloneqq \cEX[t]{\int_0^\infty g(s,\omega, C_s,V_s)\dd s}$. Since $\VV\subseteq\ol \R_+$, setting $\tau_1 \coloneqq t$ and $\tau_2 \coloneqq u$  in \eqref{eq:subsolution equation} and taking the limsup as $u \to \infty$ gives $0\leq V_t ~\leq~ \cEX[t]{  \int_t^\infty g(s,\omega, C_s,V_s)\dd s }\leq M_t$. Hence, $V$ is uniformly integrable.
	\end{proof}
	
	It is useful to introduce two monotonicity conditions on an aggregator random field.
	\begin{defn}
		Let      $g:[0,\infty) \times \Omega \times \R_+ \times \VV \to \VV$ be an aggregator random field. Then $g$ is said to satisfy
		\begin{center}
			\begin{itemize}
				\item (c$\uparrow$) if it is nondecreasing in $c$, its third argument, $\P\otimes\mathrm{d}t$-a.e.
				\vspace{-10pt}
				\item (v$\downarrow$) if it is nonincreasing in $v$, its fourth argument, $\P\otimes\mathrm{d}t$-a.e.
			\end{itemize}
		\end{center}
	\end{defn}
	\begin{rem}
		For EZ-SDU, (v$\downarrow$) is satisfied if and only if $\theta \in (0, 1]$;  if $\theta>1$ then the aggregator is increasing in its fourth argument.
	\end{rem}
	The following result shows that under condition (v$\downarrow$), a comparison result holds for sub- and supersolutions.
	
	\begin{thm}[Comparison Theorem for Subsolutions and Supersolutions] \label{thm:comparison}
		Let  $C \in \sP_+$ and let $g$ be an aggregator random field satisfying \textup{(v$\downarrow$)}. If $V^1$ is a subsolution and $V^2$ is a supersolution to the pair $(g,C)$, and either $V^1$ or $V^2$ is in $\UU\II(g,C)$, then $V^1_\tau \leq V^2_\tau$ $\as{\P}$ for all finite stopping times $\tau$.
	\end{thm}
	
	We deduce two simple but important corollaries. The first one shows that under condition (v$\downarrow$), all $g$-evaluable strategies are $g$-\emph{uniquely} evaluable. The second one shows that for aggregators $g$ satisfying (c$\uparrow$) and (v$\downarrow$), the utility associated to $(g,C)$ is increasing in $g$ and~$C$.
	
	\begin{cor}\label{cor:uniqueness of utility process for g decreasing in v}
		Let $g$ be an aggregator random field
		satisfying \textup{(v$\downarrow$)}. Then, $\sE(g) = \sE_u(g)$.
	\end{cor}
	\begin{proof}
		Clearly, $\sE(g) \supseteq \sE_u(g)$. For the converse inclusion, fix $C\in\sE(g)$. Suppose there are two utility processes $V^1$ and $V^2$ for the pair $(g,C)$. Since $V^1$ and $V^2$ are both solutions, they are in $\UU\II(g,C)$ by Lemma~\ref{lem:UIofsupersolutions}. Since they are both sub- and supersolutions, we may apply Theorem \ref{thm:comparison} twice to show $V^1_\tau \geq V^2_\tau$ $\as{\P}$ and $V^2_\tau \geq V^1_\tau$ $\as{\P}$ for all finite stopping times $\tau\geq0$. Thus, $V^1_\tau = V^2_\tau$ $\as{\P}$ for all finite stopping times $\tau$. Since $V^1$ and $V^2$ are both optional, this implies that they are indistinguishable (see e.g.~\cite[Theorem 3.2]{nikeghbali2006essayon}).
	\end{proof}
	
	\begin{cor}\label{cor: associated value process increasing in C and g}
		Let  $C^1, C^2 \in \sP_+$ and $g^1,g^2:[0,\infty) \times \Omega \times \R_+\times \VV  \to \VV$ be aggregator random fields satisfying \textup{(c$\uparrow$)} and \textup{(v$\downarrow$)}.  Suppose that  $C^2 \geq C^1$ $\P\otimes\mathrm{d}t$-a.e.~and $g^2(\cdot, \cdot, c, v) \geq g^1(\cdot, \cdot, c, v)$ $\P\otimes\mathrm{d}t$-a.e.~for $(c,v) \in \R_+ \times \VV$. Moreover suppose there exists a utility process $V^i\in\II(g^i,C^i)$ for the pair $(g^i,C^i)$, $i \in \{1, 2\}$. Then, $V^1_\tau \leq V^2_\tau$ for all finite stopping times $\tau$.
	\end{cor}
	
	\begin{rem}
		\label{rem: associated value process increasing in C and g}
		If $g_1,g_2$ are both nonincreasing rather than nondecreasing in $c$ but otherwise the hypotheses of the corollary are unchanged, then $V^1_\tau \geq V^2_\tau$.
	\end{rem}
	
	\section{Removing the bounds on evaluable strategies when $\theta \in (0,1)$}
	\label{sec:rem bounds}
	
	The goal of this section will be to show that if $\theta \in (0,1)$ we may: first, remove the lower bound restriction from Theorem \ref{thm: existence of a solution, first result}; and second, generalise the notion of a utility process, allowing us to evaluate the Epstein--Zin stochastic differential utility of \textit{any}
	consumption
	stream.
	
	\begin{sass}\label{sass:rho negative} Henceforth we assume that $\rho<0$, or equivalently $\theta \in (0,1)$.
	\end{sass}
	
	\begin{thm}\label{thm:existence when U<K Lambda}
		Let $\Lambda\in \hat{L}^\theta_{++}$, and suppose that $U\in\sP_+$ is such that there exists $K\in\R_+$, with $0 \leq  U \leq  K \Lambda$.
		Then, $F_U$ defined by \eqref{eq:fixed point operator F} has a unique fixed point $W \in \II(h_{EZ}, U)$.
	\end{thm}
	
	Recall that $\hat{X} = X^{\hat{C},\hat{\Pi}}$ is the candidate optimal wealth process---the solution to \eqref{eqn:original wealth process} under the candidate optimal strategy $\hat{\Pi}\equiv\frac{\mu - r}{\sigma R}$ and $\hat{C}=\eta \hat{X}$---and that $\hat{C} = \eta \hat{X}$ is the associated candidate optimal consumption.
	\begin{cor}\label{cor:C<K*hat(C) existence}
		Suppose that $C\in\sP_+$ is such that there exists $K\in\R_+$ with $C^{1-S}\leq K (\hat{C})^{1-S}$. Then, $C\in\sE_u(g_{EZ})$.
	\end{cor}
	\begin{proof}
		Since, $C^{1-S}\leq K (\hat{C})^{1-S}$, it follows that $U = u(t,C_t) \leq Ku(t,\hat{C}_t) = K \hat{U}_t$ where $\hat{U}_t \coloneqq u(t,\hat{C}_t)$. Furthermore, $\hat{U}\in\hat{L}^\theta_{++}$ by Example \ref{rem:geometric Brownian motion in Ltheta++}. Finally, using Theorem \ref{thm:existence when U<K Lambda} we may deduce that $U\in\sE_u(h_{EZ})$ and consequently that $C\in\sE_u(g_{EZ})$.
	\end{proof}
	
	Corollary \ref{cor:C<K*hat(C) existence} gives us a large class of evaluable consumption streams. The rest of this section is dedicated to generalising the notion of a utility process. In particular, for any aggregator $g$ satisfying (c$\uparrow$) and (v$\downarrow$), the results of this section make it possible to assign a utility to any process $C\in\sP_+$ that we can express as the monotone limit of processes $C^n\in\sE_u(g)$. For the Epstein--Zin aggregator this includes \emph{all} consumption streams.
	
	\begin{defn}\label{def:closure of evaluable}
		For a general aggregator $g:[0,\infty) \times \Omega \times \R_+ \times \VV \to \VV$, let $\overline \sE(g)$ denote the set of consumption streams $C\in\sP_+$ that are monotone limits of a sequence $(C^n)_{n\in\N}$ of processes in $\sE(g)$ and either
		1) $\VV \subseteq \ol \R_+$ and $(C^n)_{n\in\N}$ is non-decreasing, or
		2) $\VV \subseteq \ol \R_-$ and $(C^n)_{n\in\N}$ is non-increasing.
	\end{defn}
	We now state the central result of this section---that we may extend the notion of a utility process and evaluate processes in $\overline{\sE}(g)$.
	\begin{thm}\label{thm:approximation by evaluable processes is unique}
		Let $g$ be an aggregator random field satisfying \textup{(c$\uparrow$)} and \textup{(v$\downarrow$)}, and let $C\in\overline\sE(g)$. Let $(C^n)_{n\in\N}$ be a monotone approximating sequence. Let $V^n$ be the utility process associated to $C^n$ for each $n\in\N$. Then, there exists an adapted c\`adl\`ag process $V^\dagger = \lim_{n\to\infty}V^n$ that is independent of the approximating sequence. Moreover, if $\VV\subseteq \ol \R_+$, then $V^\dagger$ is the minimal supersolution and if $\VV\subseteq \ol \R_-$, then $V^\dagger$ is the maximal subsolution.
	\end{thm}
	\begin{defn}
		We call the unique process $V^\dagger = (V^\dagger_t)_{t\geq0}$ constructed in Theorem \ref{thm:approximation by evaluable processes is unique} the \textit{generalised solution} or the \textit{generalised utility process} associated to $(g,C)$.
	\end{defn}
	
	The following theorem tells us that the notion of a generalised solution \textit{extends} the notion of a solution, in the sense that if a solution exists, then it is equal to the generalised solution.
	\begin{thm}\label{thm:minimal supersolution agrees with solution when defined}
		Let $g$ satisfy \textup{(c$\uparrow$)} and \textup{(v$\downarrow$)}. If there exists a solution $V$ associated to the pair $(g,C)$ then it agrees with the generalised solution $V^\dagger$.
	\end{thm}
	\begin{proof}
		We only prove the result in the case $\VV \subseteq \ol \R_+$. The case $\VV \subseteq \ol \R_-$ follows by a symmetric argument. By Theorem \ref{thm:approximation by evaluable processes is unique}, $V^\dagger$ is the minimal supersolution. Let $\tau$ be an arbitrary finite stopping time. Since $V\in\UU\II(g,C)$ is a subsolution and $V^\dagger$ is a supersolution, $V_\tau\leq V^\dagger_\tau$ by Theorem \ref{thm:comparison}. Since $V$ is a supersolution and $V^\dagger$ is minimal in the class of supersolutions, $V^\dagger_\tau\leq V_\tau$. Hence, $V^\dagger_\tau=V_\tau$. Since $V^\dagger$ and $V$ are both optional ($V^\dagger$ by Theroem~\ref{thm:approximation by evaluable processes is unique}, and $V$ by defintion)
		and they agree for all bounded stopping times, $V^\dagger$ is equivalent to $V$ up to indistinguishability (see, for example, \cite[Theorem 3.2]{nikeghbali2006essayon}).
	\end{proof}
	We therefore drop the superscript$~^\dagger$ and denote the \textit{generalised utility process} by $V$. The next proposition shows that the generalised solution is increasing in $C$.
	\begin{prop}\label{prop:generalised supersolution increasing in C and g}
		Let $g$ be an aggregator random field satisfying \textup{(c$\uparrow$)} and \textup{(v$\downarrow$)} and let $C^1,C^2\in\overline{\sE}(g)$. Suppose further that $C^2$ dominates $C^1$ $\P\otimes\mathrm{d}t$-a.e. For $i=1,2$, let $V^i$ be the generalised solution associated to the pair $(g,C^i)$. Then, $V^2_\tau \geq V^1_\tau$ for all bounded stopping times $\tau$.
	\end{prop}
	
	If we consider Epstein--Zin aggregator $g_{EZ}$, we may assign a generalised utility process to \textit{any} consumption stream.
	\begin{thm}
		Let $C\in\sP_+$. There exists a unique generalised utility process      associated to the pair $(g_{EZ},C)$.
	\end{thm}
	\begin{proof}
		Suppose that $\VV \subseteq \ol \R_+$. We therefore want to find a non-decreasing sequence of consumption streams $(C^n)_{n\in\N}$ such that $C^n \in \sE_u(g_{EZ})$ for all $n\in\N$ and $C^n\nearrow C$. Let $\hat{C}=\eta\hat{X}$ be the candidate optimal strategy.
		Let $C^n = C\wedge n\hat{C}$. Then, $(C^n)_{n\in\N}\in\sE_u(g_{EZ})$ by Corollary \ref{cor:C<K*hat(C) existence} and $C^n \nearrow C$. Therefore, by
		Theorem \ref{thm:approximation by evaluable processes is unique} there exists a unique generalised utility process associated to $C$.
		
		The proof in the case $\VV \subseteq \ol \R_-$ goes through in exactly the same manner if we consider the sequence of processes $C^n = C \vee \frac{1}{n}\hat{C}$.
	\end{proof}
	
	We can now extend the definition of Epstein--Zin utility to any consumption stream.
	\begin{defn}
		Let $C\in\sP_+$. Define the Epstein--Zin utility process associated to $C$ to be the generalised utility process $V^{C,g_{EZ}}$ associated to the pair $(g_{EZ},C)$. Define the Epstein--Zin utility of the consumption stream to be $J_{g_{EZ}}(C) \coloneqq V^{C,g_{EZ}}_0$.
	\end{defn}
	
	This allows us to consider the infinite-horizon investment-consumption problem for EZ-SDU over all admissible strategies:
	\begin{align}\label{tilde Control problem}
		\sup_{C \in \sC(x) }J_{g_{EZ}}(C) ~=~ \sup_{C \in \sC(x) } V^{C,g_{EZ}}_0.
	\end{align}
	This definition of the stochastic control problem is different to that considered by Schroder and Skiadas \cite{schroder1999optimal}, Xing \cite{xing2017consumption}, Matoussi and Xing \cite{matoussi2018convex}, Melnyk et {al.} \cite{melnyk2020lifetime} and the rest of the literature on the Merton problem for Epstein--Zin SDU in the fact that it optimises over all consumption streams and does not impose any regularity conditions beyond attainability.
	
	\section{The verification argument for the candidate optimal strategy}\label{sec:verification of optimal strategy}
	
	The goal of this final section is to verify that the candidate optimal strategy is indeed optimal. The general structure of a primal verification argument for recursive optimal investment problems is as follows: first, apply It\^o's lemma to $\hat V(X^{\Pi,C})$ for a general strategy $(\Pi,C)$; next, use the HJB equation to show that $\hat V(X^{\Pi,C})$ is a supersolution associated to the pair $(g_{EZ},C)$; finally, the Comparison Theorem (Theorem \ref{thm:comparison}) for sub- and supersolutions implies $\hat V(x) \geq V^C_0$ for any admissible strategy $C\in\sC(x)$. Optimality follows since we showed in Section \ref{ssec:candidate optimal utility process} that $V^{\hat{C}}_0 = \hat V (x)$.
	
	Unfortunately, there are at least three difficulties with this approach. The first difficulty is that the candidate value function $\hat V(x)$ defined in \eqref{eq:hat V candidate} does not have a well-defined derivative at zero, meaning that we cannot apply It\^o's lemma to $\hat V(X^{\Pi,C}_t)$ for a general admissible wealth process $X^{\Pi,C}_t$. The second difficulty is that for a general strategy $(\Pi,C)$, the standard proof that $\hat V(X^{\Pi,C}_t)$ corresponds to a supersolution involves showing that the local martingale part of $\hat V(X^{\Pi,C})$ is a supermartingale, and in the case $R>1$ this is not true in general. The third difficulty is that $V^C$ might fail to exist.
	
	The first two issues arise also in the case of CRRA utility. In \cite{herdegen2020elementary}, the current authors show how they may be overcome using a stochastic perturbation of the value function. We now extend the ideas in \cite{herdegen2020elementary} to the setting of EZ-SDU. The third issue has been dealt with in Section \ref{sec:rem bounds}.
	
	\begin{thm}[Verification Theorem]\label{thm:verification}
		Suppose that $\eta>0$ and $\theta \in (0,1)$.
		If $V^C$ is the (generalised) utility process associated to the pair $(g_{EZ},C)$ and $\hat{V}(x)$ is the candidate optimal utility given in \eqref{eq:hat V candidate} then $\sup_{C \in \sC(x)} V^C_0 = V^{\hat{C}}_0 = \hat{V}(x)$, and the optimal investment-consumption strategy is given by $(\hat{\Pi},\hat{C})$.
	\end{thm}
	
	\begin{proof} We showed in Section \ref{ssec:change of numeraire} that
		$               \sup_{C \in \sC(x; r,\mu,\sigma)} V^{g_{EZ},C}_0 = \sup_{C \in \sC(x; \tilde{r},\tilde{\mu},\sigma)} V^{f_{EZ},C}_0$
		for $\tilde{r}=r - \frac{\delta}{1-S}$ and $\tilde{\mu}=\mu - \frac{\delta}{1-S}$. Hence, without loss of generality we may assume $\delta=0$.
		It follows from Section \ref{ssec:candidate optimal utility process} that $V^{f_{EZ},\hat{C}}_0 = \hat{V}(x)$, so it only remains to prove that $\hat{V}(x) \geq \sup_{c\in \sC(x)} V^{f_{EZ},C}_0$.
		
		Let $Y$ denote the candidate optimal wealth process started from unit wealth, i.e.
		\begin{equation}
			\frac{dY_t}{Y_t } = \frac{\lambda}{R} \dd B_t + \left( r + \frac{\lambda^2}{R} - \eta \right)\dd t, \hspace{20mm} Y_0 = 1.
		\end{equation}
		Fix $\epsilon > 0$, and let
		$               f^\epsilon_{EZ}(c,y,v) = f_{EZ}(c+\epsilon y,v) = b \frac{(c + \eta \epsilon y)^{1-S}}{1-S}((1-R)v)^\rho$.
		Fix an arbitrary admissible strategy $(\Pi, C)\in \sC(x)$. The dynamics of $X + \epsilon Y = X^{\Pi, C} + \epsilon Y$ are given by
		\[ d(X_t + \epsilon Y_t) = \left(\sigma \Pi_t X_t + \frac{\lambda \epsilon}{R}Y_t\right) \dd B_t + \left(X_t({r} + \Pi_t({\mu} - {r})) - C_t + \left({r} + \frac{\lambda^2}{R} - \eta\right)\epsilon Y_t\right) \dd t.
		\]
		Let $\sL^{c,\pi}$ denote the infinitesimal generator of the diffusion $X + \epsilon Y$ when the instantaneous rates of investment and consumption are, respectively, $\pi$ and $c$: for $h=h(x,y)$,
		\begin{equation}
			\sL^{c,\pi}h \coloneqq~ \left[x \left( {r} + \pi \sigma \lambda \right) - c + \left({r} + \frac{\lambda^2}{R} - \eta \right)\epsilon y \right]h' + \frac{1}{2}\left(\sigma \pi x + \frac{\lambda}{R}\epsilon y\right)^2 h''.
		\end{equation}
		The first aim is to show that $\hat{V}$ satisfies a perturbed HJB equation
		\begin{equation}\label{eq:perturbed HJB equation}
			\sup_{c \in \R_+,\pi \in \R}\left[\sL^{c,\pi} \hat{V}(x + \epsilon y) + f^\epsilon_{EZ}(c,y, \hat{V}(x + \epsilon y))\right] = 0.
		\end{equation}
		This follows from the fact that for general $c \in \R_+$ and $\pi \in \R$
		\begin{equation}
			\sL^{c,\pi} \hat{V}(x + \epsilon y) + f^\epsilon_{EZ}(c,y, \hat{V}(x + \epsilon y)) =  A^1(c,x,y) + A^2(\pi,x,y) + A^3(x,y),
		\end{equation}
		where
		\begin{align}
			A^1(c,x,y) =   & ~ b\frac{(c + \eta \epsilon y)^{1-S}}{1-S}((1-R)\hat{V}(x + \epsilon y))^\rho - \hat{V}'(x + \epsilon y) \left(c + \eta \epsilon y +\eta \frac{S}{1-S}(x + \epsilon y)\right),
			\\
			A^2(\pi,x,y) = & ~ \hat{V}'(x+\epsilon y) \left(x \pi \sigma \lambda + \frac{\lambda^2}{R} \epsilon y \right) + \frac{1}{2}\hat{V}''(x+\epsilon y) \left(\pi \sigma x + \frac{\lambda}{R}\epsilon y \right)^2 + \frac{\lambda^2}{2}\frac{(\hat{V}'(x + \epsilon y))^2}{\hat{V}''(x + \epsilon y)},
			\\
			A^3(x,y) =     & ~ (x + \epsilon y) \tilde{r}\hat{V}'(x + \epsilon y) - \frac{\lambda^2}{2}\frac{(\hat{V}'(x + \epsilon y)))^2}{\hat{V}''(x + \epsilon y)} + \eta \frac{S}{1-S}(x + \epsilon y)\hat{V}'(x + \epsilon y),
		\end{align}
		and the trio of inequalities $A^1 \leq 0$, $A^2 \leq 0$, $A^3=0$.
		Taking the derivative with respect to $c$ we find that the maximum of $A^1(c,x,y)$ is attained when
		$               c = \left(\frac{b((1-R)\hat{V}(x + \epsilon y))^\rho}{\hat{V}'(x + \epsilon y)}\right)^{\frac{1}{S}} - \eta \epsilon y$
		and then using the explicit form of $\hat{V}$ we find that the maximising value of $c$ is $c=\eta x$ and that $A^1(\eta x, x, y) = 0$. Similarly, by taking the derivative with respect to $\pi$, the maximum of $A^2(\pi,x,y)$ is attained when
		$               \pi = \frac{\lambda}{\sigma x}\left(\frac{\epsilon y}{R} -\frac{\hat{V}'(x + \epsilon y)}{\hat{V}''(x + \epsilon y)} \right) = \frac{\lambda}{\sigma R}$
		and then $A^2(\frac{\lambda}{\sigma R}, x, y) = 0$. Finally, by using the definition of $\hat{V}$ and $\eta$ we find that $A^3(x,y) = 0$.
		Consequently, \eqref{eq:perturbed HJB equation} is satisfied and the supremum is attained.
		Note that, since $\epsilon Y$ is just a scaling of the wealth process under the optimal strategy, it follows that $(\hat{V}(\epsilon Y_t))_{t\geq0} \in \UU\II(f_{EZ},\eta\epsilon Y)$ is the utility process associated to the consumption stream $\eta \epsilon Y$. Consequently, $\lim_{t \to \infty} \E[\hat{V}(\epsilon Y_{t+})] = 0$.
		
		Fix arbitrary bounded stopping times $\tau_1\leq\tau_2$, define $N=(N_t)_{t \geq 0}$ by $$N_t = \int_0^t \hat{V}'(X_u + \epsilon Y_u) \left(\sigma \Pi_u X_u + \frac{\lambda}{R}\epsilon Y_u\right) d W_u$$ and for $n\in\N$, set  $\zeta_n\coloneqq\inf\{s\geq \tau_1:  \langle {N} \rangle_s - \langle {N} \rangle_{\tau_1} \geq n\}$. It follows by It\^o's lemma, \eqref{eq:perturbed HJB equation} and the definition of $f^\epsilon_{EZ}$ that
		\begin{align*}
			\hat{V}(X_{\tau_1} + \epsilon Y_{\tau_1}) = & ~ \hat{V}(X_{{\tau_2}\wedge\zeta_n} + \epsilon Y_{{\tau_2}\wedge\zeta_n}) - \int_{\tau_1}^{{\tau_2}\wedge\zeta_n} \sL^{C_s,\Pi_s} \hat{V}(X_s + \epsilon Y_s) \dd s + N_{\tau_1} - N_{{\tau_2}\wedge\zeta_n}               \\
			\geq                                        & ~ \hat{V}(X_{{\tau_2}\wedge\zeta_n} + \epsilon Y_{{\tau_2}\wedge\zeta_n}) + \int_{\tau_1}^{{\tau_2}\wedge\zeta_n} f^\epsilon_{EZ}(C_s, Y_s, \hat{V}(X_s + \epsilon Y_s)) \dd s + N_{\tau_1} - N_{{\tau_2}\wedge\zeta_n}
			\\
			=                                           & ~ \hat{V}(X_{{\tau_2}\wedge\zeta_n} + \epsilon Y_{{\tau_2}\wedge\zeta_n}) + \int_{\tau_1}^{{\tau_2}\wedge\zeta_n} f_{EZ}(C_s+\eta\epsilon Y_s,\hat{V}(X_s + \epsilon Y_s)) \dd s + N_{\tau_1} - N_{{\tau_2}\wedge\zeta_n}.
		\end{align*}
		Taking conditional expectations and using that $(N_{t \wedge \zeta_n} - N_{t \wedge \tau_1})_{t \geq 0}$ is an $L^2$-bounded martingale, the Optional Sampling Theorem gives
		\begin{equation}\label{eq:localised verification}
			\hat{V}(X_{\tau_1} + \epsilon Y_{\tau_1})\geq \cEX[{\tau_1}]{\hat{V}(X_{{\tau_2}\wedge\zeta_n} + \epsilon Y_{{\tau_2}\wedge\zeta_n}) + \int_{\tau_1}^{{\tau_2}\wedge\zeta_n} f_{EZ}(C_s+\eta\epsilon Y_s,\hat{V}(X_s + \epsilon Y_s)) \dd s}.
		\end{equation}
		Since $\hat{V}$ is increasing and wealth is non-negative, $\hat{V}(X_{{\tau_2}\wedge\zeta_n} + \epsilon Y_{{\tau_2}\wedge\zeta_n}) \geq \hat{V}(\epsilon Y_{{\tau_2}\wedge\zeta_n})$. Using that $(\hat{V}(\epsilon Y_t))_{t \geq 0}$ is uniformly integrable, taking the liminf as $n \to \infty$, the generalised conditional version of Fatou's Lemma and the conditional Monotone Convergence Theorem yield
		
		\begin{equation}\label{eq:hat V supersol}
			\hat{V}(X_{\tau_1} + \epsilon Y_{\tau_1})\geq \cEX[{\tau_1}]{\hat{V}(X_{\tau_2} + \epsilon Y_{\tau_2}) + \int_{\tau_1}^{\tau_2} f_{EZ}(C_s+\eta\epsilon Y_s,\hat{V}(X_s + \epsilon Y_s)) \dd s}.
		\end{equation}
		Furthermore, $\liminf_{t\to\infty}\E[\hat{V}(X_{t+}+\epsilon Y_{t+})]\geq \lim_{t\to\infty}\E[\hat{V}(\epsilon Y_{t+})] = 0$. Consequently, $\hat{V}(X + \epsilon Y)$ is a supersolution associated to the pair $(f_{EZ},C+\eta\epsilon Y)$.
		
		Suppose $R<1$. Since $C+\eta\epsilon Y > C$ and $f_{EZ}$ is increasing in its first argument, $\hat{V}(X+\epsilon Y)$ is a supersolution associated to the pair $(f_{EZ}, C)$ by \eqref{eq:hat V supersol}. Thus, the (generalised) utility process $V^{f_{EZ},C}$ associated to $(f_{EZ},C)$ is the minimal supersolution by Theorem \ref{thm:approximation by evaluable processes is unique}. Consequently, $\hat{V}(X+\epsilon Y)\geq V^{f_{EZ},C}$.
		
		Suppose $R>1$, and hence also $S>1$ by Standing Assumption~\ref{sass:theta>0}. Then, since $(C+\eta\epsilon Y)^{1-S} \leq (\eta\epsilon)^{1-S} Y^{1-S}$, $C+\eta\epsilon Y\in\sE_u(f_{EZ})$ by Corollary \ref{cor:C<K*hat(C) existence}. Hence, there exists a utility process $V^{f_{EZ},C+\eta\epsilon Y}\in\UU\II(f_{EZ}, C+\eta\epsilon Y)$ associated to $C+\eta\epsilon Y$. Since also $\hat{V}(X+\epsilon Y)\leq0$, applying Theorem \ref{thm:comparison}  and then Proposition \ref{prop:generalised supersolution increasing in C and g} gives $\hat{V}(X+\epsilon Y)\geq V^{f_{EZ},C+\eta\epsilon Y}\geq V^{f_{EZ},C}$.
		
		In both cases, taking the supremum over attainable consumption streams at time zero gives $\hat V (x + \epsilon) \geq \sup_{c\in \sC(x)} V_0^{f_{EZ},C}$. Letting $\epsilon \searrow 0$ gives the result.
	\end{proof}
	
	We conclude this section by showing that the correct well-posedness condition of the investment-consumption problem is $\eta>0$.
	\begin{cor}     \label{cor:ill posed}
		Suppose that $\theta \in (0,1)$. Then, the infinite-horizon investment consumption problem for EZ-SDU is well-posed if and only if $\eta > 0$.
		
		In particular, suppose that $\eta\leq0$ and let $V^C$ be the (generalised) utility process associated to the pair $(g_{EZ},C)$.
		If $R<1$, then $\sup_{C \in \sC(x)}V_0^C  = \infty$. If $R > 1$, then, $\sup_{C \in \sC(x)}V_0^C  = -\infty$.
	\end{cor}
	
	\begin{proof}
		When $\eta>0$ the investment-consumption problem is well-posed by Theorem \ref{thm:verification}.
		
		Now suppose $\eta\leq0$. Since $\theta\in(0,1)$, the utility process is unique, and if $H(\pi,\xi)>0$ then $V$ given by \eqref{eq:valfungenstrat2} is the utility process for a constant proportional strategy.
		
		Suppose $R<1$ and then also $S<1$. Let $f(\pi, \xi) = \frac{\xi^{1-R}}{1-R}\left( \frac{b \theta}{H_{\delta\theta}(\pi,\xi)} \right)^{\theta}$ and $D=\{ (\pi, \xi) \in \R \times (0,\infty): H_{\delta\theta}(\pi,\xi) > 0 \}$. Note that $\theta (H_{\delta\theta}(\hat{\pi}, \xi))^{-1} = (\eta S + (1-S)\xi)^{-1}$.
		Letting $\xi\searrow -\eta \frac{S}{1-S}$ yields $\theta (H_{\delta\theta}(\hat{\pi}=\frac{\mu-r}{\sigma R}, \xi))^{-1}\nearrow\infty$.
		It follows that $f(\pi,\xi)\nearrow\infty$ and the supremum of $V_0^C$ over constant proportional strategies is $+\infty$. Hence, $\sup_{C \in \sC(x)}V_0^C  = \infty$.
		
		Now suppose $R>1$ and fix an arbitrary $C\in\sC(x; r,\mu,\sigma)$ with associated wealth process $X$. Denote by $V$ the generalised utility process associated to the pair $(g_{EZ},C)$. It suffices to show that $V_0 = -\infty$. For $n\in\N$, let $\alpha_n \coloneqq \frac{S}{S-1}(\frac{1}{n} - \eta) > 0$, $r_n\coloneqq r+\alpha_n$ and $\mu_n\coloneqq\mu+\alpha_n$. Consider the modified consumption stream $C^n$, given by $C^n_t\coloneqq e^{\alpha_n t}C_t$. Then, by calculating the dynamics of $X^n_t\coloneqq e^{\alpha_n t}X_t$ as in Section \ref{ssec:change of numeraire} it can be shown that $C^n\in\sC(x; r_n,\mu_n,\sigma)$.
		Furthermore, $\eta_n = \frac{1}{S}[\delta - (1-S)(r_n + \frac{\lambda^2}{2R})] = \frac{1}{n} >0$. Then, considering the Black--Scholes--Merton financial market with parameters $(r_n,\mu_n,\sigma)$ and applying Theorem \ref{thm:verification} gives $V^n_0 \leq \hat{V}^n(x) = \eta_n^{-\theta S } b^\theta\frac{x^{1-R}}{1-R}$. It follows from Proposition \ref{prop:generalised supersolution increasing in C and g} that if $V^n$ is the (generalised) solution associated for the pair $(g_{EZ},C^n)$, then $C\leq C^n$ implies $V\leq V^n$. Combining the inequalities and taking limits yields $V_0 \leq \lim_{n\to\infty} n^{\theta S } b^\theta\tfrac{x^{1-R}}{1-R} = -\infty$.
	\end{proof}
	
	\bibliographystyle{plain}
	\bibliography{SDU}
	{\small
	\include{bibliography}
}
	\appendix
	
	\section{Proof of the Comparison Theorem}
	
	\begin{lemma}\label{lem:right accumulation point of uncountable set}
		Let $-\infty < a <  b < \infty$. Every uncountable set $U\subseteq[a,b)$ contains at least one of its right accumulation points.
	\end{lemma}
	\begin{proof}
		Seeking a contradiction, suppose $U$ contains none of its right accumulation points. Then, for each $x\in U$, we may find $\epsilon_x>0$ such that $[x,x+\epsilon_x)\cap U = \{x\}$. Let $U_n \coloneqq \{x\in U:~ \epsilon_x>\frac{1}{n}\}$. Then, each $U_n$ is finite since the pairwise disjoint union $\bigcup_{x\in U_n} [x,x+\frac{1}{n})$ is contained in the interval $[a,b+\frac{1}{n})$. Hence, $U=\bigcup_{n\in\N}U_n$ is countable, and we arrive at a contradiction.
	\end{proof}
	
	\begin{proof}[\textbf{Proof of Theorem \ref{thm:comparison}}]
		We prove the result when $\VV\subseteq \ol \R_+$. The case $\VV\subseteq \ol \R_-$ is symmetric.
		
		Suppose for contradiction that there exists a finite stopping time $\tau$ and a set of positive measure $A \in \mathcal{F}_\tau$ such that $V^1_\tau(\omega) > V^2_\tau(\omega)$ for $\omega \in A$, whence $\EX{\1_A\left(V^1_\tau - V^2_\tau \right)} > 0$. Since $V^1$ and $V^2$ are l\`ad, the processes $(V^1_{s+})_{t\geq0}$ and $(V^2_{s+})_{t\geq0}$ exist and are right-continuous. Moreover, $\sigma \coloneqq \inf\{s\geq \tau : V^1_{s{+}} - V^2_{s{+}} \leq 0\}$ is a stopping time. The right continuity of $(V^1_{t{+}})_{t\geq0}$ and $(V^2_{t{+}})_{t\geq0}$ gives $( V^1_{\sigma{+}} - V^2_{\sigma{+}} )\1_{\{\sigma < \infty\}}  \leq 0$ $\as{\P}$
		
		For each $\omega\in A$, we have  $V^1_s(\omega) \geq V^2_s(\omega)$ for almost all $s\in[\tau(\omega),\sigma(\omega))$. Indeed, seeking a contradiction suppose there is $\omega\in A$ and a set of positive Lebesgue measure $U$ such that $V^1_s(\omega) < V^2_s(\omega)$ for $s\in U\subseteq[\tau(\omega),\sigma(\omega))$. Since $U$ is uncountable, it has a right accumulation point $q\in U$ by Lemma \ref{lem:right accumulation point of uncountable set}. Then, $q < \sigma(\omega)$ and $V^1_{q{+}}(\omega) \leq V^2_{q{+}}(\omega)$, and we arrive at a contradiction.
		
		Next, fix $n \in \NN$. By subtracting \eqref{eq:supersolution equation} from \eqref{eq:subsolution equation} for the bounded stopping times $\tau_1 \coloneqq \tau\wedge n$ and $\tau_2 \coloneqq \sigma \wedge n$, noting that the expectations are well defined since either $V^1$ or $V^2$ is in $\UU\II(g,C)$, and using the fact that $g$ is a.s.~decreasing in $v$ and $V^1_s(\omega) \geq V^2_s(\omega)$ for almost all $s\in[\tau(\omega),\sigma(\omega))$ for $\omega \in A$, we obtain
		\begin{eqnarray}
			\lefteqn{\EX{\1_A\1_{\{\tau\leq n\}}\left(V^1_\tau - V^2_\tau \right)} } \\
			& \leq & \EX{\1_A\1_{\{\tau\leq n\}}\left(V^1_{(\sigma\wedge n){+}} - V^2_{(\sigma\wedge n){+}}  +  \int_{ \tau\wedge n}^{\sigma\wedge n} g(s,\omega,C_s,V^1_s) - g(s,\omega,C_s,V^2_s)\dd s\right)} \\
			& \leq  & \ \EX{\1_A\1_{\{\tau\leq n\}}\left(V^1_{(\sigma\wedge n){+}} - V^2_{(\sigma\wedge n){+}}\right)}.
			\label{eq:Stopped W a submartingale}
		\end{eqnarray}
		Finally, taking the limsup as $n \to \infty$, monotone convergence, the fact that $(V^1_{t{+}})_{t\geq0}$ and $(V^2_{t{+}})_{t\geq0}$ are $\ol \R_+$-valued, the transversality condition for subsolutions and $( V^1_{\sigma{+}} - V^2_{\sigma{+}} )\1_{\{\sigma < \infty\}}  \leq 0$ $\as{\P}$ give
		\begin{align}
			\EX{\1_A(V^1_\tau - V^2_\tau) } & \leq \limsup_{n\to\infty}\EX{\1_A \1_{\{\tau\leq n<\sigma \}}( V^1_{n{+}} - V^2_{n{+}} )}+ \limsup_{n\to\infty} \EX{\1_A \1_{\{\sigma \leq n\}}( V^1_{\sigma{+}} - V^2_{\sigma{+}})} \\
			& \leq \limsup_{n\to\infty}\EX{ V^1_{n{+}}}+ \EX{\1_A \1_{\{\sigma < \infty \}} (V^1_{\sigma{+}} - V^2_{\sigma{+}})} \leq 0.
		\end{align}
		We arrive at a contradiction.
	\end{proof}
	
	\begin{proof}[\textbf{Proof of Corollary \ref{cor: associated value process increasing in C and g}}]
		Suppose that $\VV = \ol \R_+$; the proof for $\VV\subseteq \ol \R_-$ is symmetric. As $g_2(\cdot, \cdot, c, v) \geq g_1(\cdot, \cdot, c, v)$ $\P \otimes \dd t$-a.e.~and $g_1$ and $g_2$ are increasing in $c$, we have
		$g_2(s,\omega,C^2_s,V^2_s) \geq g_1(s,\omega,C^2_s,V^2_s) \geq g_1(s,\omega,C^1_s,V^2_s)\geq0$ for $\P \otimes \dd t$-a.e.~$(s, \omega)$.
		It then follows that, for all bounded stopping times $\tau\leq \sigma$,
		\[
		V^2_\tau =  \cEX[\tau]{V^2_{\sigma{+}} + \int_\tau^\sigma g_2(s,\omega,C^2_s,V^2_s)\dd s}
		\geq  \cEX[\tau]{V^2_{\sigma{+}} + \int_\tau^\sigma g_1(s,\omega,C^1_s,V^2_s)\dd s}.
		\]
		Since $\lim_{t \to \infty} V^2_{t+} = 0$ $\as{\P}$, $V^2$ satisfies the definition of a supersolution associated to the pair $(g_1,C^1)$. As $V^2\in\UU\II(g_2,C^2)\subseteq\UU\II(g_1,C^1)$ and $V^1$ is a (sub)solution associated to $(g_1,C^1)$, it follows that $V^1_\tau \leq V^2_\tau$ for all finite stopping times $\tau$ by Theorem \ref{thm:comparison}.
	\end{proof}
	
	\section{Proving Existence and Uniqueness of a Utility Process}\label{app:existence}
	
	For $\Lambda\in\hat{L}^\theta_{++}$, define the $\epsilon$-perturbed operator $F^\epsilon_{U,\Lambda}:\II(h_{EZ},U)\to \sP_+$ by\footnote{Here, we always choose a  c\`adl\`ag version for the right-hand side of \eqref{eq:perturbed fixed point operator}.}
	\begin{equation}\label{eq:perturbed fixed point operator}
		F^\epsilon_{U,\Lambda}(W)_t =  \cEX[t]{\int_t^\infty (U_s W_s^\rho + \epsilon \Lambda_s^\theta) \dd s}.
	\end{equation}
	A key property of $F^\epsilon_{U,\Lambda}$ is, when $\epsilon>0$ and $\Lambda\in\hat{L}^\theta_{++}$, $F^\epsilon_{0,\Lambda}$ is bounded away from zero. Another property is the following.
	\begin{lemma}\label{lem:F^eps maps from O(Delta^theta) to itself}
		Let $\epsilon\geq0$, $\Lambda\in\hat{L}^\theta_{++}$ and $U\in \OO(\Lambda)$.   Then, $F^\epsilon_{U,\Lambda}(\cdot)$ maps from $\OO(\Lambda^\theta)$ to itself.
	\end{lemma}
	
	\begin{proof}
		Fix arbitrary $W\in\OO(\Lambda^\theta)$. It follows that there exist $k_W,K_W \in (0,\infty)$ such that $k_W \Lambda^\theta \leq W \leq K_W \Lambda^\theta$. Similarly, since $U\in\OO(\Lambda)$ and $\Lambda^\theta \in \OO(I^\Lambda)$, there exist $k_U,K_U,k_\Lambda,K_\Lambda \in (0,\infty)$ such that
		$k_U \Lambda \leq U \leq K_U \Lambda$ and $k_\Lambda I^\Lambda \leq \Lambda^\theta \leq K_\Lambda I^\Lambda$.
		We only prove that $F^\epsilon_{U,\Lambda}(W) \geq \kappa \Lambda^\theta$ for $\rho<0$; the argument for $\rho>0$ involves $W^\rho \geq (k_W \Lambda)^{\theta \rho}$ and the argument for the upper bound is symmetric.
		By the definition of $F^\epsilon_{U,\Lambda}(\cdot)$ in \eqref{eq:perturbed fixed point operator} and since $U \geq k_U \Lambda$, $W \leq K_W \Lambda^\theta$ and $\Lambda^\theta \leq K_\Lambda I^\Lambda$, and $1+\theta\rho = \theta$, we see that
		\begin{align*}
			F^\epsilon_{U,\Lambda}(W)_t \geq & ~ \cEX[t]{\int_t^\infty k_U \Lambda_s (K_W \Lambda^\theta)^\rho + \epsilon \Lambda_s^\theta \dd s}
			\\
			=                                & ~ (k_U K_W^\rho + \epsilon)\cEX[t]{\int_t^\infty \Lambda_s^\theta \dd s} \geq \left(\frac{k_U K_W^\rho + \epsilon}{K_\Lambda}\right)\Lambda^\theta. \qedhere
		\end{align*}
	\end{proof}
	
	The subsequent theorem is the preliminary existence result and includes Theorem~\ref{thm: existence of a solution, first result} as a special case.
	
	\begin{thm}\label{thm:existence of a F^epsilon solution, bounded case}Let $\epsilon\geq0$, $\Lambda\in\hat{L}^\theta_{++}$ and $U\in\OO(\Lambda)$.      Then, $F^\epsilon_{U,\Lambda}$ defined by \eqref{eq:perturbed fixed point operator}
		has a unique fixed point $W \in \OO(\Lambda^\theta) \subseteq \II(h_{EZ}, U)$, which has c\`adl\`ag paths.
	\end{thm}

	For the proof of Theorem \ref{thm:existence of a F^epsilon solution, bounded case}, we use the following sufficient condition for an operator $T$ from a Banach space $\mathscr{B}$ to intself to be a contraction; see \cite[Theorem 3.3]{stokey1989recursive} for a proof.
	
	\begin{lemma}[Blackwell's sufficient conditions for a contraction]\label{lemma: Blackwell's sufficient conditions for a contraction}
		Let $\mathscr{B}$ be a Banach space and $T: \mathscr{B} \to \mathscr{B}$ an operator that is nonincreasing.  Suppose there exists $\beta \in (0,1)$ with
		\begin{equation}\label{eq:Blackwell decreasing condition for contraction mapping}
			T(X+a) \geq T(X) - \beta a \qquad \text{for all }X \in \mathscr{B}, \ a > 0.
		\end{equation}
		Then $T$ is a contraction with constant $\beta$.
	\end{lemma}
	
	\begin{proof} [\textbf{Proof of Theorem \ref{thm:existence of a F^epsilon solution, bounded case}}]
		Let $\text{Prog}$ denote the progressive $\sigma$-algebra on $\Omega \times \R_{+}$ and set
		$\mathscr{B} = L^\infty(\Omega \times \RR_{+}, \text{Prog}, \P \otimes \diff t)$.
		Consider the change of variables
		\begin{equation}
			P_t = \log(U_t) - \log\left(\Lambda_t\right), \qquad Q_t = \log(W_t) - \theta \log\left(\Lambda_t\right).
		\end{equation}
		Then $U \in \OO(\Lambda)$ if and only if $P \in \sB$ and $W \in \OO(\Lambda^\theta)$ if and only if $Q \in \sB$.
		
		The fixed point condition $W = F^\epsilon_{U,\Lambda}(W)$ is equivalent to the fixed point condition $Q = G^\epsilon_{P,\Lambda}(Q)$ where
		\begin{equation}\label{eq: log fixed point operator}
			G^\epsilon_{P,\Lambda}(Q)_t \coloneqq \log\left(\cEX[t]{\int_t^\infty \Lambda_s^\theta \exp(P_s + \rho Q_s) + \epsilon \Lambda_s^\theta \dd s }\right)  - \theta \log\left(\Lambda_t\right).
		\end{equation}
		Note that since the first term on the right-hand side of \eqref{eq: log fixed point operator} has c\`adl\`ag paths, every fixed point $Q$ to \eqref{eq: log fixed point operator} corresponds to a $W$ with c\`adl\`ag paths.
		
		Since $G^\epsilon_{P,\Lambda}(Q)$ is the difference of two continuous functions of progressive processes, it is progressive. Furthermore, as a consequence of Lemma \ref{lem:F^eps maps from O(Delta^theta) to itself}, $G^\epsilon_{P,\Lambda}$ maps $\B$ to itself.
		
		Suppose $\rho \in (-1,0)$ and let $a > 0$. Then, $G^\epsilon_{P,\Lambda}(Q)$ is decreasing. Furthermore,
		\begin{eqnarray*}
			G^\epsilon_{P,\Lambda}(Q+a)_t &=& \log\left(\exp(\rho a)\cEX[t]{\int_t^\infty \Lambda_s^\theta \exp(P_s + \rho Q_s) + \epsilon \frac{\Lambda_s^\theta}{\exp(\rho a)} \dd s }\right)  - \theta \log\left(\Lambda_t\right)
			\\
			&\geq& \log\left(\cEX[t]{\int_t^\infty \Lambda_s^\theta \exp(P_s + \rho Q_s) + \epsilon \Lambda_s^\theta \dd s }\right)  - \theta \log\left(\Lambda_t\right) + \rho a
			\\
			&=& G^\epsilon_{P,\Lambda}(Q)_t + \rho a.
		\end{eqnarray*}
		By Lemma~\ref{lemma: Blackwell's sufficient conditions for a contraction},      this implies that $G^\epsilon_{P,\Lambda}$ is a contraction with constant $\rho$.
		Hence, by the Contraction Mapping Theorem, there exists a unique $Q \in \B$ satisfying \eqref{eq: log fixed point operator}.
		
		If $\rho \in (0,1)$, then $G^\epsilon_{P,\Lambda}(Q)$ is increasing and one can show that $G^\epsilon_{P,\Lambda}(Q+a)_t \leq G^\epsilon_{P,\Lambda}(Q)_t + \rho a$. Again the result follows from  Lemma~\ref{lemma: Blackwell's sufficient conditions for a contraction} and the Contraction Mapping Theorem.
		
		Finally, to extend the result to ${\rho \in (-\infty, -1]}$, we borrow an idea from Schroder and Skiadas \cite{schroder1999optimal} and show by induction that the following holds for each $k \in \NN$:
		\begin{equation}
			\label{eq:schroder:ind}
			\text{For $0 > \rho > -k$ and $P\in\sB$, there exists a unique fixed point $Q \in \mathscr{B}$ of $G^\epsilon_{P,\Lambda}(Q)$.}
		\end{equation}
		The induction hypothesis ($k=1$) holds by the above. For the induction step, suppose that \eqref{eq:schroder:ind} holds true for some $k \geq 1$. In order to show that \eqref{eq:schroder:ind} holds true for $k +1$, it suffices to consider $\rho \in (-(k+1),k]$.
		So fix $\rho \in (-(k+1),k]$ and choose  $\chi \in (0,1)$ small enough that $ -k <\rho +  \chi < 0$. Now define the map $\tilde G^\epsilon_{P,\Lambda}: \B \times  \B \to \B$ by\footnote{Here, we always choose a  c\`adl\`ag version for the conditional expectation in the right-hand side of \eqref{eq: breaking the equation into Delta and Z}.}
		\begin{equation}\label{eq: breaking the equation into Delta and Z}
			\tilde G^\epsilon_{P,\Lambda}(Q, Z)_t = \log\left(\cEX[t]{\int_t^\infty \Lambda_s^\theta \exp(P_s - \chi Q_s + (\rho  + \chi) Z_s) + \epsilon \Lambda_s^\theta \dd s }\right)  - \theta \log\left(\Lambda_t\right).
		\end{equation}
		If suffices to show that there exists a unique $Q \in \B$ such that $Q=\tilde G^\epsilon_{P,\Lambda}(Q, Q)$. Note that since the first term on the right-hand side of \eqref{eq: breaking the equation into Delta and Z} has c\`adl\`ag paths, every $Q \in \B$ satisfying $Q=\tilde G^\epsilon_{P,\Lambda}(Q, Q)$ corresponds to a $W$ with c\`adl\`ag paths.
		By the induction hypothesis, for each fixed $Q \in \mathscr{B}$, and since $P - \chi Q \in \sB$, there exists a unique $Z \in \mathscr{B}$ such that $Z = \tilde G^\epsilon_{P,\Lambda}(Q, Z)$. So, we can define the operator $Z^\epsilon_{P,\Lambda}: \B  \to \B$ implicitly by
		\begin{equation}
			\label{eq: Z epsilon:op}
			Z^\epsilon_{P,\Lambda}(Q) = \tilde G^\epsilon_{P,\Lambda}(Q,Z^\epsilon_{P,\Lambda}(Q)).
		\end{equation}
		If we can show that $Z^\epsilon_{P,\Lambda}$ has a unique  fixed point, we are done. To this end, arguing as above, it suffices to show that $Z^\epsilon_{P,\Lambda}$ is nonincreasing and satisfies \eqref{eq:Blackwell decreasing condition for contraction mapping} for $\beta := \chi$.
		
		To argue that $Z^\epsilon_{P,\Lambda}$ is nonincreasing, let $Q^1,Q^2 \in \sB$ with $Q^1 \leq Q^2$ $\P \otimes \diff t$-a.e. For $i \in \{1, 2\}$, set $\tilde C^i \coloneqq \Lambda^\theta \exp(Q^i)$ and $\tilde V^i \coloneqq \Lambda^\theta \exp(Z^\epsilon_{P,\Lambda}(Q^i))$. Then \eqref{eq: Z epsilon:op} implies that
		\begin{align}
			\tilde V_t^i = & ~ \cEX[t]{\int_t^\infty \left( \Lambda_s^\theta \left(\frac{U_s}{\Lambda_s} \right) \left(\frac{\tilde C_s^i}{\Lambda_s^\theta}\right)^{-\chi}\left(\frac{\tilde V_s^i}{\Lambda_s^\theta}\right)^{\rho +\chi} + \epsilon \Lambda_s^\theta \right) \dd s }
			\\
			= & ~ \cEX[t]{\int_t^\infty \left( U_s \left(\tilde C_s^i\right)^{-\chi}\left(\tilde V_s^i\right)^{\rho + \chi} + \epsilon \Lambda_s^\theta \right) \dd s}.
		\end{align}
		Since $\tilde h(t,\omega,c,v) = U_t(\omega) c^{-\chi} v ^{\rho + \chi} + \epsilon (\Lambda_t(\omega))^\theta$ satisfies (c$\downarrow$) and (v$\downarrow$), by Remark~\ref{rem: associated value process increasing in C and g} it follows that $\tilde V^1 \geq\tilde V^2$, and consequently $Z^1 \geq Z^2$.
		
		Finally, to show that $Z^\epsilon_{P,\Lambda}$ satisfies  \eqref{eq:Blackwell decreasing condition for contraction mapping} for $\beta := \chi$, let $a > 0$ and set $\Psi = (Z^\epsilon_{P,\Lambda}(Q+a) - Z^\epsilon_{P,\Lambda}(Q))/a \leq 0$. It suffices to show that  $\Psi \geq -\chi$. Let $L \coloneqq \Lambda^\theta\exp(Z^\epsilon_{P,\Lambda}(Q))$. Then
		\begin{align*}
			L_t \exp(\Psi_t a)
			& = \Lambda_t^\theta\exp(Z^\epsilon_{P,\Lambda}(Q)_t)\exp(\Psi_t a) = \Lambda_t^\theta \exp(Z^\epsilon_{P,\Lambda}(Q+a)_t)                                                                                       \\
			& = \cEX[t]{\int_t^\infty \left( \Lambda_s^\theta \exp(P_s - \chi (Q_s+a) + (\rho +  \chi) Z^\epsilon_{P,\Lambda}(Q+a)_s) + \epsilon \Lambda_s^\theta \right) \dd s }                                            \\
			& = \mathbb{E}_t\left[\int_t^\infty \left( \Lambda_s^\theta e^{-\chi a + (\rho + \chi)a \Psi_s} e^{P_s - \chi Q_s + (\rho + \chi) Z^\epsilon_{P,\Lambda}(Q)_s} + \epsilon \Lambda_s^\theta \right) \dd s \right] \\
			& \geq L_t \exp(-\chi a ),
		\end{align*}
		where in the last line we have used that $(\rho + \chi)
		\Psi \geq 0 $. Hence, $ \exp(\Psi a) \geq \exp(-\chi a )$ and consequently $\Psi \geq -\chi$.
	\end{proof}
	
	We may now prove Theorem \ref{thm:existence when U<K Lambda}.
	\begin{proof}[\textbf{Proof of Theorem \ref{thm:existence when U<K Lambda}}]
		The proof is formed of two parts. The first part removes the lower bound on $U$ for $\epsilon>0$; the second part shows that we may remove the restriction $\epsilon>0$.
		
		Let $U^n = \max\{U,\frac{1}{n}\Lambda\}$. Then, $U^n\in\OO(\Lambda)$ for every $n\in\N$. Hence, by Theorem \ref{thm:existence of a F^epsilon solution, bounded case}, for each $n\in\N$, there exists $W^n$ that satisfies
		$
		W^n_t = \ \cEX[t]{\int_t^\infty U^n_s (W^n_s)^\rho + \epsilon \Lambda_s^\theta \dd s}.
		$
		Since $\Lambda\in\hat{L}^\theta_{++}$, there exists $\kappa$ such that $\Lambda^\theta\leq \kappa I^\Lambda$. Hence,
		$W^n \geq \epsilon I^\Lambda \geq \frac{\epsilon}{\kappa} \Lambda^\theta$ and
		\begin{align}\label{eq:bound on U^n (W^n)^rho}
			U^n (W^n)^\rho \leq (K \Lambda) (\epsilon^\rho \kappa^{-\rho} \Lambda^{\theta-1}) = K \kappa^{-\rho} \epsilon^\rho \Lambda^\theta.
		\end{align}
		Since $\rho<0$, $g$ satisfies (v$\downarrow$). Hence, by Corollary \ref{cor: associated value process increasing in C and g}, the sequence $(W^n)_{n\in\N}$ is decreasing (and positive) so it converges almost surely. Therefore, applying the Dominated Convergence Theorem with the bound in \eqref{eq:bound on U^n (W^n)^rho} and the condition $\Lambda\in\hat{L}^\theta_{++}$, we find that ${W^* \coloneqq \lim_{n\to\infty}W^n}$  satisfies
		\[
		W^*_t = \lim_{n\to\infty}\E\left[\left.\int_t^\infty U^n_s (W^n_s)^\rho + \epsilon \Lambda^\theta  \dd s \right| \cF_t \right]
		= \E\left[\left.\int_t^\infty U_s (W^*_s)^\rho + \epsilon \Lambda_s^\theta \dd s \right| \cF_t \right],
		\]
		so that $W^*$ is a fixed point of $F^\epsilon_{U,\Lambda}(\cdot)$. Uniqueness follows from Corollary \ref{cor:uniqueness of utility process for g decreasing in v} since $h^\epsilon(t,\omega,u,v) = u v^\rho + \epsilon (\Lambda(t,\omega))^\theta$ satisfies (v$\downarrow$). This concludes the first part of the proof.
		
		Let $U$ be a non-negative progressively measurable process such that $0\leq U \leq K\Lambda$. Define the aggregator random field $h^\epsilon$ by $h^\epsilon(t,\omega,u,v) = u v^\rho + \epsilon (\Lambda(t,\omega))^\theta$. By the preceding argument, for each $\epsilon>0$ there exists a utility process associated to the pair $(h^\epsilon,U)$.
		
		It follows from Corollary \ref{cor: associated value process increasing in C and g} that the fixed point $W^\epsilon$ to the operator $F^\epsilon(\cdot)$ given in \eqref{eq:perturbed fixed point operator} is decreasing as $\epsilon\searrow0$. Define $W_t = \lim_{\epsilon\to0}W^\epsilon_t$. Then,
		\begin{align}
			W_t = & \lim_{\epsilon\to0}\cEX[t]{\int_t^\infty U_s (W^\epsilon_s)^\rho + \epsilon \Lambda_s^\theta \dd s}
			\\
			=     & \lim_{\epsilon\to0}\cEX[t]{\int_t^\infty h_{EZ}(U_s, W^\epsilon_s) \dd s} + \lim_{\epsilon\to0}\cEX[t]{\int_t^\infty \epsilon\Lambda_s^\theta \dd s}
			\\
			=     & ~ \cEX[t]{\int_t^\infty h_{EZ}(U_s, W_s) \dd s},
		\end{align}
		where the last line follows from the Monotone Convergence Theorem and the fact that $h_{EZ}$ was chosen so that $\lim_{w \rightarrow w_0} h_{EZ}(u,w) = h_{EZ}(w,w_0)$ even for $(u,w_0)=(0,0)$ and $(u,w_0)=(\infty,\infty)$.
		Furthermore, $W\in\II(h_{EZ},U)$ since $\EX{\int_0^\infty U_s W_s^\rho \dd s} = W_0 \leq W^\epsilon_0 < \infty$. Uniqueness follows from Corollary \ref{cor:uniqueness of utility process for g decreasing in v} since $h_{EZ}$ satisfies (v$\downarrow$).
	\end{proof}
	
	\section{Existence and Uniqueness of a Generalised Utility Process}
	To prove Theorem \ref{thm:approximation by evaluable processes is unique} we must first introduce generalisations of some well-known concepts. We focus on the supermartingale case, but the submartingale case is symmetric.
	
	\begin{defn}[Generalised supermartingale, Doob \cite{doob1953stochastic}, Snell \cite{snell1952applications}]
		A $(-\infty, \infty]$-valued process $M = (M_t)_{t\geq0}$ is called a \emph{generalised supermartingale}  if, $M^-_t\in L^1$ for all $t\geq0$, $M$ is adapted and $M_s \geq \cEX[s]{M_t}$ for all $t\geq s \geq0$.
	\end{defn}
	
	\begin{rem}
		Since $M^-_t\in L^1$ ($M_t$ is quasi-integrable), the conditional expectation $\cEX[s]{M_t}$ exists and is unique, even if $M_t \notin L^1$.
	\end{rem}
	
	Compared to an (ordinary) supermartingale, a generalised supermartingale does not require $M_t \in L^1$ for all $t \geq 0$. So it is possible to have $M_s = +\infty \geq \cEX[s]{M_t}$.
	We next need to generalise this notion even further.\footnote{In \cite{mertens1972theory}, Mertens referred to such processes simply as \textit{supermartingales}.}
	\begin{defn}[Generalised Optional Strong Supermartingale,  Mertens \cite{mertens1972theory}]
		A generalised supermartingale is called a \textit{generalised optional strong supermartingale} if it is optional and  for all bounded pairs of stopping times $\tau_1\leq{\tau_2}$, $M_{\tau_2}^-\in L^1$ and $\cEX[{\tau_1}]{M_{\tau_2}}  \leq M_{\tau_1}$.
	\end{defn}
	
	\begin{rem}
		Note that every c\`adl\`ag supermartingale is an optional strong supermartingale by the Optional Sampling Theorem.
	\end{rem}
	
	\begin{prop}\label{prop:generalised supermartingale is ladlag}
		A generalised optional strong supermartingale $M$ that is either bounded above or below is almost surely l\`adl\`ag and for a.e. $\omega$, the path $t \mapsto M_t(\omega)$ is right-continuous outside a countable set.
	\end{prop}
	\begin{proof} Suppose first that $M$ is bounded below by a constant $K$ and define the continuous bijection $f: [K,\infty] \to [1-e^{-K},1]$ by  $f(x) \coloneqq 1-e^{-x}$ with the convention that $e^{-\infty} = 0$.
		It follows from Jensen's inequality (note that $f^{-1}$ is convex) that
		\(              M_{\tau_1} \geq \cEX[{\tau_1}]{M_{\tau_2}} = \cEX[{\tau_1}]{(f^{-1}\circ f)(M_{\tau_2})} \geq f^{-1}(\cEX[{\tau_1}]{f(M_{\tau_2})}).
		\)
		Consequently, if $\widetilde{M} = f(M)$, then for all bounded pairs of stopping times ${\tau_1} \leq {\tau_2}$ we have $\widetilde{M}_{\tau_1} = f(M_{\tau_1}) \geq \cEX[{\tau_1}]{f(M_{\tau_2})} = \cEX[{\tau_1}]{\widetilde{M}_{\tau_2}}$ and $\widetilde{M}$ is a bounded optional strong supermartingale. Hence, it is l\`adl\`ag (see for example \cite[Theoreom A1.4]{dellacherie1982}). Moreover, it has a Mertens decomposition (see, for example \cite[Theorem A1.20]{dellacherie1982}) given by $\tilde M = \tilde N - \tilde A$, where $\tilde N = (\tilde N_t)_{t \geq 0}$ is a c\`adl\`ag local martingale and $\tilde A = (\tilde A_t)_{t \geq 0}$ is a nondecreasing adapted  l\`adl\`ag  process. Since a noncreasing  l\`adl\`ag function is  (right-)continuous up to a countable set, it follows that for for a.e. $\omega$, the path $t \mapsto \tilde M_t(\omega)$ is right-continuous outside a countable set. Then, using that $f^{-1}$ is continuous, it follows that $M$ is l\`adl\`ag and for a.e.~$\omega$, the path $t \mapsto M_t(\omega)$ is right-continuous outside a countable set.
		
		For the case when $M$ is bounded above, we may use the concave function $g(x) = 1-e^x$.
	\end{proof}
	
	The following results are generalised versions of the Backwards Martingale Convergence Theorem (BMCT) and Hunt's Lemma. For lack of an easy reference, we provide proofs.
	
	\begin{prop}[Generalised Backwards Martingale Convergence Theorem]\label{prop:GBMCT}
		Suppose that $X$ is a $[0, \infty]$-valued random variable and let $\cF \supseteq \cF_0 \supseteq\cF_{-1} \supseteq \cF_{-2} \supseteq \cdots$ be a decreasing sequence of sub-$\sigma$-algebras and $\cF_{-\infty} \coloneqq \bigcap_{k = 1}^\infty \cF_{-k}$. Then $\lim_{n \to \infty}\cEX[-n]{X} = \cEX[-\infty]{X}$ $\as{\P}$
	\end{prop}
	\begin{proof}
		For  $n\in\N$,  set  $Z_{-n}\coloneqq\cEX[-n]{X}$, and let  $Z_{-\infty} \coloneqq \cEX[-\infty]{X}$. Since $\cF_{-\infty} \subset  \cF_{-n}$ for all $n \in \NN$, it suffices to show that $\lim_{n \to \infty} Z_{-n} = Z_{-\infty}$ $\as{\P}$ on $\{Z_\infty \leq k\}$ for all $k \in \N$ and $\lim_{n \to \infty} Z_{-n} = Z_{-\infty}$ $\as{\P}$ on $\{Z_\infty =\infty\}$. The case of finite $k$ follows from the standard BMCT via
		\begin{equation}
			\lim_{n \to \infty} Z_{-n} \1_{\{Z_\infty \leq k\}} = \lim_{n \to \infty} \cEX[-n]{X\1_{\{Z_\infty \leq k\}}} =  \cEX[-\infty]{X\1_{\{Z_\infty \leq k\}}} = Z_{-\infty} \1_{\{Z_\infty \leq k\}} \quad \as{\P}
		\end{equation}
		For the other case, by the standard BMCT for fixed $k \in \NN$
		\begin{equation}
			\liminf_{n \to \infty} Z_{-n} \geq \lim_{n \to \infty} \cEX[-n]{X \wedge k} =  \cEX[-\infty]{X\wedge k}\quad \as{\P}
		\end{equation}
		Now taking on the right-hand side the monotone limit  as $k \to \infty$ gives $\liminf_{n\to\infty}Z_{-n}\geq \cEX[-\infty]{X}$ $\as{\P}$ Finally, on $\{Z_{-\infty}=\infty\}$ the liminf trivially coincides with the limsup.
	\end{proof}
	
	\begin{lemma}
		\label{lem:hunt}
		Let $(\Omega, \cF, \P)$ be a probability space and $(X_n)_{n \in \NN}$ a nondecreasing sequence of $[0, \infty]$-valued random variables with $\lim_{n \to \infty} X_n  = X$ $\as{\P}$ Let $\cF \supseteq \cF_0 \supseteq\cF_{-1} \supseteq \cF_{-2} \supseteq \cdots$ be a decreasing sequence of sub-$\sigma$-algebras and $\cF_{-\infty} \coloneqq \bigcap_{k = 1}^\infty \cF_{-k}$. Then $\lim_{n \to \infty}\cEX[-n]{X_n} = \cEX[-\infty]{X}$ $\as{\P}$
	\end{lemma}
	
	\begin{proof}
		For $n \in \NN$,        let $Y_n = \cEX[-n]{X_n}$. Then, $\cEX[-n]{X_m}\leq Y_n \leq \cEX[-n]{X}$ for $m\leq n$. Now taking taking the limit as $n \to \infty$ and applying Proposition \ref{prop:GBMCT} gives
		\begin{equation}
			\cEX[-\infty]{X_m}\leq\liminf_{n\to\infty} Y_n\leq\limsup_{n\to\infty} Y_n \leq \cEX[-\infty]{X} \quad \as{P}
		\end{equation}
		Taking the limit as $m\to\infty$, the result follows from the Monotone Convergence Theorem.
	\end{proof}
	
	We may now prove Theorem \ref{thm:approximation by evaluable processes is unique}, the central result of Section \ref{sec:rem bounds}.
	\begin{proof}[\textbf{Proof of Theorem \ref{thm:approximation by evaluable processes is unique}}]
		We only prove the case that $(C^n)_{n\in\N}$ is an increasing sequence and $\VV\subseteq \ol \R_+$. For the case when $(C^n)_{n\in\N}$ is a decreasing sequence and $\VV\subseteq \ol \R_-$, the proof goes through by a symmetric argument. Since $(C^n)_{n\in\N}$ is increasing, so is $(V^n)_{n\in\N}$ by Corollary \ref{cor: associated value process increasing in C and g}. Then, $V^\dagger =\lim_{n\to\infty}V^n$ exists and $V^n \leq V^\dagger$ for each $n\in\N$. Further, for any bounded stopping times ${\tau_1}$ and ${\tau_2}$ with ${\tau_1}\leq{\tau_2}$ $\as{\P}$,
		\begin{align}
			V^\dagger_{\tau_1} = & \lim_{n\to\infty}\cEX[{\tau_1}]{\int_{\tau_1}^{{\tau_2}} g(s,\omega,C^n_s ,V^n_s) \dd s + V^n_{{\tau_2}} }
			\\
			\geq                 & \lim_{n\to\infty}\cEX[{\tau_1}]{\int_{\tau_1}^{{\tau_2}} g(s,\omega,C^n_s ,V^\dagger_s) \dd s + V^n_{{\tau_2}} }
			\\ \label{eq:replace tau with tau_+}
			=                    & ~ \cEX[{\tau_1}]{\int_{\tau_1}^{{\tau_2}} g(s,\omega,C_s ,V^\dagger_s) \dd s + V^\dagger_{{\tau_2}} }
		\end{align}
		It follows that $V^\dagger_{\tau_1}     \geq  \E[ V^\dagger_{\tau_2}|\cF_{\tau_1}]$
		so that $V^\dagger$ is a non-negative generalised optional strong supermartingale. Hence, by Proposition \ref{prop:generalised supermartingale is ladlag}, it is l\`adl\`ag. Since $\E[V^\dagger_{\tau_2}|\cF_{\tau_1}] \geq \E[V^\dagger_{{\tau_2}{+}}|\cF_{\tau_1}]$, \eqref{eq:replace tau with tau_+} becomes
		$V^\dagger_{\tau_1} \geq \cEX[{\tau_1}]{\int_{\tau_1}^{{\tau_2}} g(s,\omega,C_s ,V^\dagger_s) \dd s + V^\dagger_{{\tau_2}{+}} }$.
		Furthermore, since $\VV\subseteq \ol \R_+$, $\liminf_{t\to\infty}V^\dagger_{t+} \geq 0 \text{ a.s.}$ and $V^\dagger$ is a supersolution.
		
		Now, take any other arbitrary monotone sequence $(\widetilde{C}^n)_{n\in\N}$ whose limit is equal to $C$. Let $\widetilde{V}^n$ be the utility process associated to $\widetilde{C}^n$ and $\widetilde{V}^\dagger = \lim_{n\to\infty}\widetilde{V}^n$. Then, since $\widetilde{V}^n\in\UU\II(g,C)$
		is a subsolution associated to $(g,C)$ since $g$ satisfies (c$\uparrow$), we may apply Theorem \ref{thm:comparison} and deduce that $V^\dagger_\tau \geq \widetilde{V}^n_\tau$ for all finite stopping times $\tau$. Taking limits gives that $V^\dagger_\tau \geq \widetilde{V}^\dagger_\tau$. Repeating the argument with the roles of $V^\dagger$ and $\widetilde{V}^\dagger$ reversed, we find that $\widetilde{V}^\dagger_\tau \geq V^\dagger_\tau$ for all finite stopping times $\tau$. Therefore, since $V^\dagger$ and $\widetilde{V}^\dagger$ are optional processes that agree for all finite stopping times, they agree up to indistinguishability (see, for example, \cite[Theorem 3.2]{nikeghbali2006essayon}).
		
		Next, we show that  $V^\dagger$ is the minimal supersolution for $C$.  Let $\overline V$ be any supersolution. Then, since $V^n\in\UU\II(g,C)$ is a subsolution associated to $(g,C)$, $\overline V_t \geq V^n_t$ for all $t\geq0$ by Theorem \ref{thm:comparison}. Taking limits gives $\overline V_t \geq V^\dagger_t$.
		
		Finally, we show that $V^\dagger$ is c\`adl\`ag. To this end, it suffices to show that the right-continuous process $(V^\dagger_{t+})_{t \geq 0}$ is also a supersolution. Then, by the supermartingale property of $V^\dagger$ it follows that $V^\dagger_{\tau+} = \E[V^\dagger_{\tau+}|\cF_{\tau}] \leq \E[V^\dagger_{\tau}|\cF_{\tau}] = V^\dagger_{\tau}$ for each bounded stopping time, and thus by the minimality of $V^\dagger$, $(V^\dagger_t)_{t \geq 0} = (V^\dagger_{t+})_{t \geq 0}$ up to indistinguishability.
		
		To show that $(V^\dagger_{t+})_{t \geq 0}$ is indeed a supersolution, fix bounded stopping times ${\tau_1}$ and ${\tau_2}$ with ${\tau_1}\leq{\tau_2}$. We first assume that there is $\delta > 0$ such that $\tau_1 + \delta \leq \tau_2$. Then for each $\epsilon < \delta$, by the fact that $V^\dagger$ is a supersolution and a generalised optional strong supermartingale,
		\begin{equation}
			V^\dagger_{\tau_1+ \epsilon} \geq \cEX[{\tau_1+\epsilon}]{\int_{\tau_1+\epsilon}^{{\tau_2}} g(s,\omega,C_s ,V^\dagger_{s_+}) \dd s + V^\dagger_{{\tau_2}{+}} }.
		\end{equation}
		Taking the limit as $\epsilon \to 0$, and using the fact that for a.e.~$\omega$, the path $t \mapsto V^\dagger$ is right-continuous outside a countable set by Proposition \ref{prop:generalised supermartingale is ladlag}, we get by Hunt's lemma in the form of Lemma \ref{lem:hunt},
		\begin{equation}
			\label{eq:Vdagger+:delta}
			V^\dagger_{\tau_1+} \geq \cEX[{\tau_1}]{\int_{\tau_1}^{{\tau_2}} g(s,\omega,C_s ,V^\dagger_{s_+}) \dd s + V^\dagger_{{\tau_2}{+}} }.
		\end{equation}
		Now if $\tau_2$ is general, for $\delta > 0$ set $\tau_2^\delta\coloneqq \tau_2 \vee (\tau_1 + \delta)$. Then applying \eqref{eq:Vdagger+:delta} for $\tau_2^\delta$ gives
		\begin{align}
			V^\dagger_{\tau_1+} \1_{\{\tau_2 \geq  \tau_1+\delta\}} & \geq \cEX[{\tau_1}]{\int_{\tau_1}^{{\tau_2^\delta}} g(s,\omega,C_s ,V^\dagger_{s_+}) \dd s + V^\dagger_{{\tau^\delta_2}{+}} }\1_{\{\tau_2 \geq  \tau_1+\delta\}} \notag \\
			& = \cEX[{\tau_1}]{\int_{\tau_1}^{{\tau_2}} g(s,\omega,C_s ,V^\dagger_{s_+}) \dd s + V^\dagger_{{\tau_2}{+}} }\1_{\{\tau_2 \geq  \tau_1+\delta\}}
		\end{align}
		Taking the limit as $\delta \to 0$ gives by monotone convergence,
		\begin{align}
			V^\dagger_{\tau_1+} \1_{\{\tau_2 > \tau_1\}} & \geq \cEX[{\tau_1}]{\int_{\tau_1}^{{\tau_2}} g(s,\omega,C_s ,V^\dagger_{s_+}) \dd s + V^\dagger_{{\tau_2}{+}} }\1_{\{\tau_2 > \tau_1\}}.
		\end{align}
		Since trivially,
		$V^\dagger_{\tau_1+} \1_{\{\tau_2 = \tau_1\}} = \cEX[{\tau_1}]{\int_{\tau_1}^{{\tau_2}} g(s,\omega,C_s ,V^\dagger_{s_+}) \dd s + V^\dagger_{{\tau_2}{+}} }\1_{\{\tau_2 = \tau_1\}}$,
		we conclude
		\begin{align}
			V^\dagger_{\tau_1+} & \geq \cEX[{\tau_1}]{\int_{\tau_1}^{{\tau_2}} g(s,\omega,C_s ,V^\dagger_{s_+}) \dd s + V^\dagger_{{\tau_2}{+}} }.\qedhere
		\end{align}
	\end{proof}
	
	\section{Additional proofs omitted from the main text}
	\label{appendix:additional proofs}
	
	\begin{proof}[\textbf{Proof of Proposition \ref{prop:MMKS utility process is utility process}}]
		Let $V^\Delta$ be a $(\delta \theta, \JJ^{MMS})$-utility process associated to consumption stream $C$ and generator $g^\Delta_{EZ}$. Then, $V^\Delta \in \bS^1_T \cap \II_T(g^\Delta_{EZ}, C)$, $\lim_{t \to \infty} e^{-\delta\theta t}\E[V^\Delta_t] = 0$, and $V^\Delta$ solves \eqref{eq:familyV} with aggregator $g_{EZ}^\Delta$ for all $0\leq t \leq T < \infty$.
		
		Define the process $V = (V_t)_{t \geq 0}$ by $V_t := \exp(-\delta t) V^{\Delta}_t$. Then $V \in \bS^1_T$ and $\lim_{t \to \infty} \EX{V_t}  = 0$ by the transversality condition of $V^\Delta$. We proceed to show that $V \in \II_T(g_{EZ}, C)$ and $V$ satisfies
		\begin{equation}\label{eq:finite time horizon problem `discounted' formulation}
			V_t = \ \cEX[t]{\int_t^T  b e^{-\delta u}\frac{C_u^{1-S}}{1-S}((1-R)V_u)^\rho \dd u + V_T}
		\end{equation}
		for all $T > 0$. So fix $T > 0$. Using that $V^\Delta \in \bS^1_T \cap \II_T(g^\Delta_{EZ}, C)$ and $e^{-\delta t}|V_t|^\rho \leq e^{|\delta\theta| T}|V^\Delta_t|^\rho$ for $t \in [0, T]$, we obtain
		\begin{eqnarray*}
			\lefteqn{\E \int_0^T \left| b e^{-\delta s}\frac{C_s^{1-S}}{1-S}((1-R)V_s)^\rho \right| \dd s} \\
			& \leq &e^{|\delta \theta| T} \EX{\int_0^T \left| b \frac{C_s^{1-S}}{1-S}((1-R)V^\Delta_s)^\rho - \delta\theta V^\Delta_s\right|\dd s}
			+ e^{|\delta \theta| T} T \left|\delta\theta\right| \E\left[\sup_{s \in [0,T]}|V^\Delta_s|\right] ~<~  \infty.
		\end{eqnarray*}
		Thus, $V \in \II_T(g_{EZ}, C)$. Next, define the martingale $M = (M_t)_{t \in [0, T]}$ by
		\begin{align}
			M_t = \ \cEX[t]{\int_0^T \left[ b \frac{C_s^{1-S}}{1-S}((1-R)V_s)^\rho - \delta \theta V_s \right]\dd s +  V_T}
		\end{align}
		As $V^\Delta$ satisfies \eqref{eq:familyV}, it satisfies the BSDE
		\[              V^\Delta_t = V^\Delta_T + \int_t^T \left(b \frac{C_u^{1-S}}{1-S}((1-R)V^\Delta_u)^\rho - \delta \theta V^\Delta_u \right) \dd u - \int_t^T \dd M_u. \]
		Applying the product rule to $V_t = e^{-\delta \theta t}V^\Delta_t$ we find that
		\begin{equation}
			\label{eq:discounted tilde BSDE}
			V_t = V_T + \int_t^T b e^{-\delta u}\frac{C_u^{1-S}}{1-S}((1-R)V_u)^\rho \dd u + \int_t^T e^{-\delta \theta u} \dd M_u.
		\end{equation}
		Since $\E[(1-e^{-\delta \theta T})|M_T| ] < \infty$, $N_t = \int_0^t e^{-\delta \theta s} \dd M_s$ is a martingale by \cite[Lemma A.1.]{herdegen2019sensitivity}.       Now taking expectations gives \eqref{eq:finite time horizon problem `discounted' formulation}.
		
		Next, using that $V$ and the integrand in \eqref{eq:finite time horizon problem `discounted' formulation} have the same sign, it follows from the monotone convergence theorem and $\lim_{T \to \infty} \EX{V_T}  = 0$ that $V$ satisfies \eqref{eq:value process optimal for strategy derivation}. Since $V_0$ is finite, this also gives $V \in \II(g_{EZ}, C)$.
		
		Finally, if $\theta > 1$ then $\delta \theta > \delta$ and any $(\delta, J^{MMS})$-utility process is automatically a $(\delta \theta, J^{MMS})$- utility process. Hence $\sE^{MMS}(g^\Delta_{EZ}) \subseteq \sE(g_{EZ})$.
	\end{proof}
	
	\begin{proof}[\textbf{Proof of Proposition \ref{prop:generalised supersolution increasing in C and g}}]
		Suppose $V \subseteq \ol \R_+$; the case of $V \subseteq \ol \R_-$ follows by a symmetric argument.
		
		Let $C^{2,n}$ be a non-decreasing sequence of processes in $\sE(g)$ with limit $C^2$ and let $C^{1,n}\coloneqq C^{2,n}\wedge C^1$. Then, $C^{1,n}$ is a monotone sequence which approximates $C^1$. Furthermore, let $V^{1,n}\in\UU\II(g,C^{1,n})\subseteq\UU\II(g,C^{2,n})$ and $V^{2,n}\in\UU\II(g,C^{2,n})$ be the utility processes associated to $C^{1,n}$ and $C^{2,n}$ respectively. Then, if $V^{1,\dagger}$ and $V^{2,\dagger}$ are the generalised solutions associated to $C^1$ and $C^2$, it follows from Theorem \ref{thm:approximation by evaluable processes is unique} that $V^{1,\dagger} = \lim_{n\to\infty}V^{1,n}$ and $V^{2,\dagger} = \lim_{n\to\infty}V^{2,n}$.
		
		Since $C^{2,n}\geq C^{1,n}$ and $g$ satisfies $(c \uparrow)$, $g(t,\omega,C^{2,n}_t,V^{2,n}_t) \geq g(t,\omega,C^{1,n}_t,V^{2,n}_t)$ for almost all $(t,\omega)$. Hence, for all finite stopping times ${\tau_1} \leq {\tau_2}$,
		\[ V^{2,n}_{\tau_1} = \cEX[{\tau_1}]{V^{2,n}_{{\tau_2}{+}} +  \int_{\tau_1}^{{\tau_2}} g(s,\omega,C^{2,n}_s,V^{2,n}_s) \dd s }
		\geq \cEX[{\tau_1}]{V^{2,n}_{{\tau_2}{+}} +  \int_{\tau_1}^{{\tau_2}} g(s,\omega,C^{1,n}_s,V^{2,n}_s) \dd s }. \]
		
		Since also $\lim_{t \to \infty} V^{2,n}_{t{+}} = 0 \quad a.s.$, $ V^{2,n}$ satisfies the definition of a supersolution associated to the pair $(g,C^{1,n})$. Hence, by Theorem \ref{thm:comparison} it follows
		that $V^{2,n}_{\tau_1} \geq V^{1,n}_{\tau_1}$ for all finite stopping times ${\tau_1}$. Taking the limit as $n\to\infty$ gives the result.
	\end{proof}
	
\end{document}

%% file: preamble.tex
\usepackage[english]{babel}
\usepackage[utf8]{inputenc}
\usepackage[T1]{fontenc}
\usepackage{mathrsfs}

\usepackage[leqno]{amsmath}
\usepackage{color}

\definecolor{dukeblue}{rgb}{0.0, 0.0, 0.61}
\definecolor{internationalkleinblue}{rgb}{0.0, 0.18, 0.65}
\definecolor{coolblack}{rgb}{0.0, 0.18, 0.39}
\definecolor{darkblue}{rgb}{0.0, 0.0, 0.55}

\usepackage{comment}
\usepackage{stmaryrd} %for \llbracket

\usepackage{geometry} 
\usepackage{amsmath,amssymb}
\usepackage{bm} %nice bold maths
\usepackage{enumerate} %allows us to use roman numerals for lists
\usepackage{amsthm}
\usepackage{mathtools} %gives :=

\usepackage{filemod} %autoupdate last edited date

\usepackage{varwidth}%For enumerating by three letters

\usepackage{pdflscape}
\usepackage{afterpage}
\makeatletter
\newcommand{\leqnos}{\tagsleft@true\let\veqno\@@leqno}
\newcommand{\reqnos}{\tagsleft@false\let\veqno\@@eqno}
\reqnos
\makeatother

\mathtoolsset{showonlyrefs,showmanualtags}%only numbers labelled equations

\newtheoremstyle{boldremark}
{\dimexpr\topsep/2\relax} % space above
{\dimexpr\topsep/2\relax} % space below
{}          % body font
{}          % indent amount
{\bfseries} % theorem head font
{.}         % punctuation after theorem head
{.5em}      % space after theorem head
{}          % theorem hed spec. (empty = "normal")

\theoremstyle{plain}
\newtheorem{thm}{Theorem}[section] % reset theorem numbering for each chapter
\newtheorem{lemma}[thm]{Lemma} % same for Lemmas
\newtheorem{prop}[thm]{Proposition} % same for Propositions
\newtheorem{cor}[thm]{Corollary} % same for Corollaries
\newtheorem{sass}{Standing Assumption}
\newtheorem*{tsass}{Temporary Standing Assumption (for Section~\ref{ssec:trans} only)}
\newtheorem{hypothesis}{Hypothesis} %assumptions
\theoremstyle{definition}
\newtheorem{defn}[thm]{Definition} % definition numbers are dependent on theorem numbers

\numberwithin{equation}{section}

\theoremstyle{remark}
\newtheorem{example}[thm]{Example} % same for example numbers

\theoremstyle{boldremark}
\newtheorem{rem}[thm]{Remark}

\usepackage{dsfont}
\newcommand{\R}{\mathds{R}}
\newcommand{\N}{\mathds{N}}

\newcommand{\B}{\mathscr{B}}

\renewcommand{\theta}{\vartheta}
\renewcommand{\epsilon}{\varepsilon}
\renewcommand{\P}{\mathbb{P}}

\newcommand{\E}{\mathbb{E}}

\newcommand{\II}{\mathds{I}}
\newcommand{\JJ}{\mathds{J}}
\newcommand{\NN}{\mathds{N}}
\newcommand{\OO}{\mathds{O}}
\newcommand{\QQ}{\mathds{Q}}
\newcommand{\RR}{\mathds{R}}
\newcommand{\bS}{\mathds{S}}
\newcommand{\UU}{\mathds{U}}
\newcommand{\VV}{\mathds{V}}

\newcommand{\cB}{\mathcal{B}}

\newcommand{\cE}{\mathcal{E}}
\newcommand{\cF}{\mathcal{F}}

\newcommand{\cP}{\mathcal{P}}

\newcommand{\sA}{\mathscr{A}}
\newcommand{\sB}{\mathscr{B}}
\newcommand{\sC}{\mathscr{C}}
\newcommand{\sD}{\mathscr{D}}
\newcommand{\sE}{\mathscr{E}}

\newcommand{\sL}{\mathscr{L}}
\newcommand{\sP}{\mathscr{P}}
\newcommand{\sS}{\mathscr{S}}

\newcommand{\diff}{\mathrm{d}}
\newcommand{\dd}{\,\mathrm{d}}

\newcommand{\1}{\mathbf{1}}

\newcommand*{\EX}[2][]{\E^{#1}\left [ #2 \right ]}
\newcommand*{\cEX}[2][]{\E\left[ #2 \,\middle\vert\,\cF_{#1} \right]}

\newcommand*{\as}[1]{#1\text{-a.s.}}

\newcommand*{\ol}[1]{\overline{#1}}

\usepackage{lipsum}

\definecolor{pipurple}{RGB}{128,0,128}
\definecolor{vargreen}{RGB}{0,150,0}
\definecolor{varyellow}{RGB}{180,120,0}

\makeatletter
\g@addto@macro\bfseries{\boldmath}
\makeatother

\usepackage{caption}
\usepackage{subcaption}

\usepackage{IEEEtrantools}